\documentclass{lmcs}
\pdfoutput=1

\usepackage{lastpage}
\renewcommand{\dOi}{14(3:15)2018}
\lmcsheading{}{1--\pageref{LastPage}}{}{}%
{Jan.~27,~2016}{Sep.~05,~2018}{}

\usepackage{etex}
\usepackage{mathtools,xspace}
\usepackage{amsmath}
\usepackage{amsfonts}
\usepackage{amssymb}
\usepackage{amsthm}
\usepackage[pdftex,all]{xy}
\SelectTips{cm}{}  % Tips (of arrows) are in accordance with Computer
\xyoption{v2}
\xyoption{curve}
\xyoption{2cell}
\UseAllTwocells
\SilentMatrices

\usepackage{wrapfig}
\usepackage{floatflt}
\usepackage{tikz}

\usepackage{cancel}

\usepackage{stmaryrd}

\newcommand{\prefix}{\text{prefix}}

\newcommand{\pow}{\mathcal{P}}
\newcommand{\dist}{\mathcal{D}}
\newcommand{\lift}{\mathcal{L}}
\newcommand{\giry}{\mathcal{G}}
\newcommand{\place}{\underline{\phantom{n}}\,} % place holder
\newcommand{\Sets}{\mathbf{Sets}}

\newcommand{\Meas}{\mathbf{Meas}}

\newcommand{\lang}{L}
\newcommand{\langinf}{L^{\infty}}
\newcommand{\langfin}{L^{*}}

\newcommand{\sigalg}{\mathfrak{F}}
\newcommand{\FSigma}{F_{\Sigma}}

\newcommand{\cyl}{\text{cyl}}
\newcommand{\op}{\mathrm{op}}

\newcommand{\Reach}{\text{Reach}}

\newcommand{\kto}[0]{\mathrel{\mspace{-4mu} \to \mspace{-16mu} \raisebox{0.1ex}[0ex][0ex]{$\shortmid$} \mspace{6mu}}}
\newcommand{\kleftarrow}[0]{\mathrel{\mspace{-4mu} \leftarrow \mspace{-16mu} \raisebox{0.1ex}[0ex][0ex]{$\shortmid$} \mspace{6mu}}}

\newcommand{\Kl}[0]{\mathcal{K}\mspace{-1mu}\ell}

\newcommand{\id}[0]{\mathrm{id}}
\newcommand{\tr}[0]{{\sf tr}}
\newcommand{\trinf}[0]{{\sf tr^{\infty}}}

\newcommand{\relmiddle}[1]{\mathrel{}\middle#1\mathrel{}}

\newcommand{\Tree}[0]{\text{Tree}}
\newcommand{\Treeinf}[0]{\text{Tree}_{\infty}}

\newcommand{\empseq}[0]{\varepsilon}

\newcommand{\Geq}[0]{\mathcal{G}_{=1}}

\newcommand{\FPE}[0]{{\sf FPE}}

\newcommand{\kar}{\ar|-*\dir{|}}

\newcommand{\treeprefix}{\preceq}

\newcommand{\fwd}[0]{\sqsubseteq_{\bf F}}
\newcommand{\bwd}[0]{\sqsubseteq_{\bf B}}

\newcommand{\tifbwd}[0]{\sqsubseteq_{\bf B}^{\mathrm{TIF}}}
\newcommand{\tbwd}[0]{\sqsubseteq_{\bf B}^{\mathrm{T}}}

\newcommand{\proofmark}{\noindent\textit{Proof.}\,\;}

\makeatletter
\newcommand{\mypushright}[1]{\ifmeasuring@#1\else\omit\hfill$\displaystyle#1$\fi\ignorespaces}
\newcommand{\mypushleft}[1]{\ifmeasuring@#1\else\omit$\hspace*{1mm}\displaystyle#1$\hfill\fi\ignorespaces}
\makeatother

\newlength{\currentparindent}
\newenvironment{minipageparindent}
  {\setlength{\currentparindent}{\parindent}% save the value
   \noindent\begin{minipage}{\hsize}% open the minipage
   \setlength{\parindent}{\currentparindent}% restore the value
  }
  {\end{minipage}}

\newif\ifignore % when set to true, additional text appears containing
\ignorefalse
\newcommand{\auxproof}[1]{
\ifignore\mbox{}\newline
\textbf{BEGIN: AUX-PROOF} \dotfill\newline
{#1}\mbox{}\newline
\textbf{END: AUX-PROOF}\dotfill\newline
\fi}

\theoremstyle{plain}
\newtheorem{sublem}[thm]{Sublemma}

\theoremstyle{definition}

\newtheorem{nota}[thm]{Notation}

\begin{document}
\title{Coalgebraic Infinite Traces and Kleisli Simulations}

\author[Natsuki Urabe]{Natsuki Urabe\rsuper{*}}	%required
\address{Dept.\ Computer Science, The University of Tokyo, Japan}	%required
\email{urabenatsuki@is.s.u-tokyo.ac.jp}  %optional
\thanks{\lsuper{*}JSPS Research Fellow}	%optional

\author[Ichiro Hasuo]{Ichiro Hasuo}	%optional
\address{
 National Institute of Informatics and SOKENDAI,
 Tokyo, Japan
% Dept.\ Computer Science, The University of Tokyo\\ Hongo 7-3-1, Tokyo 113-8656, 
%   Japan
}	%optiona
\email{
i.hasuo@acm.org
}  %optional

%% required for running head on odd and even pages, use suitable
%% abbreviations in case of long titles and many authors:

%% mandatory lists of keywords and classifications:
\titlecomment{An earlier version of this paper~\cite{urabeH15coalgebraicinfinite} 
has been presented at: \emph{6th Conference on Algebra and Coalgebra in Computer Science} (CALCO 2015)}

\begin{abstract}
% 4th ver.
Kleisli simulation is a categorical notion introduced by Hasuo 
to verify finite trace inclusion.
They allow us to give  definitions of \emph{forward and backward simulation} for various types of systems.
A generic categorical theory behind Kleisli simulation has been
 developed and it guarantees the
soundness of those simulations with respect to \emph{finite} trace semantics.
Moreover, those simulations can be aided by \emph{forward partial execution} (FPE)---a 
categorical transformation of systems previously introduced by the authors.

In this paper, we give Kleisli simulation a theoretical foundation that assures 
its soundness also with respect to \emph{infinitary} traces.
%show that Kleisli simulation can be used also to verify \emph{infinite} trace inclusion.
There, following Jacobs' work, infinitary trace semantics is characterized
 as the ``largest homomorphism.'' It turns out that soundness of forward
 simulations is rather straightforward; that of backward simulation
 holds too, although it requires certain additional conditions and its 
proof is more involved.
% It is proved using categorical characterization of the infinite trace introduced by Jacobs intensively for nondeterministic systems.
% There, infinite trace semantics is captured by the ``largest homomorphism''. %to the ``lifted'' final coalgebra.
% For a nondeterministic system that outputs trees, we show that their automata-theoretic infinite language
% can be captured with this framework. %by the largest homomorphism.
%Then we show the soundness of Kleisli simulation wrt.\ finite trace.
%the inclusion between the coalgebraic infinite trace semantics.
We also show that FPE can be successfully employed in the infinitary trace
 setting
to enhance the applicability of Kleisli simulations as witnesses of
 trace inclusion.
Our framework is parameterized in the monad for branching as well as in
 the functor for linear-time behaviors; 
 for the former we mainly use the
 powerset monad (for nondeterminism), the sub-Giry monad
(for probability), and the lift monad (for exception).
\end{abstract}

\maketitle              % typeset the title of the contribution

\section{Introduction}
\subsection{Language Inclusion}
\emph{Language inclusion} of transition systems is an important problem 
in both qualitative and quantitative verification.
In a qualitative setting the problem is concretely as follows:
for given two nondeterministic systems $\mathcal{X}$ and $\mathcal{Y}$,
check if $\lang(\mathcal{X})\subseteq\lang(\mathcal{Y})$---that is, if the set of words 
generated by $\mathcal{X}$ is included in the set of words generated by $\mathcal{Y}$.
In a typical usage scenario, $\mathcal{X}$ is a model of the \emph{implementation} in question while
$\mathcal{Y}$ is a model that represents the \emph{specification} of $\mathcal{X}$.
More concretely, $\mathcal{Y}$ is a system such that
$\lang(\mathcal{Y})$ is easily seen not to contain anything ``dangerous''---therefore the language inclusion $\lang(\mathcal{X})\subseteq\lang(\mathcal{Y})$ immediately implies that
$\lang(\mathcal{X})$ contains no dangerous output, either.
Such a situation can also arise in a \emph{quantitative setting} where 
a specification is about \emph{probability}, \emph{reward}, and so on.
%it is aimed to verify that the system satisfy specifications about
%\emph{probability} or \emph{rewards}.

\begin{exa}\label{ex:firstExample}
%----------------------------------------------------------------
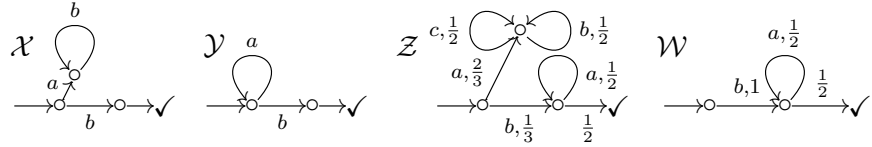
\begin{figure}
 \begin{tabular}{cccc}
\begin{xy}
(-5,8)*{\mathcal{X}}="",
(0,0)*{\circ} = "x",
(8,0)*{\circ} = "y",
(2,4)*{\circ} = "z",
(14,0)*{\checkmark} = "xc",
\ar _{} (-6,0);"x"
\ar ^{a} "x";"z"
\ar _{b} "x";"y"
\ar _{} "y";"xc"
%\ar @(ur,ul)_{a} "x";"x"
\ar @(ur,ul)_{b} "z";"z"
\end{xy}
&
\begin{xy}
(-5,8)*{\mathcal{Y}}="",
(0,0)*{\circ} = "x",
(8,0)*{\circ} = "y",
%(8,8)*{\circ} = "z",
(14,0)*{\checkmark} = "xc",
\ar _{} (-6,0);"x"
%\ar _{a} "x";"z"
\ar _{b} "x";"y"
\ar _{} "y";"xc"
\ar @(ur,ul)_{a} "x";"x"
\end{xy}
&
\begin{xy}
(-10,8)*{\mathcal{Z}}="",
(0,0)*{\circ} = "x",
(10,0)*{\circ} = "y",
(5,10)*{\circ} = "z",
(18,0)*{\checkmark} = "xc",
\ar _{} (-6,0);"x"
\ar ^(.3){a,\frac{2}{3}} "x";"z"
\ar _{b,\frac{1}{3}} "x";"y"
\ar _{\frac{1}{2}} "y";"xc"
%\ar @(ur,ul)_{a} "x";"x"
\ar @(ur,ul)_(.2){a,\frac{1}{2}} "y";"y"
\ar @(dr,ur)_{b,\frac{1}{2}} "z";"z"
\ar @(dl,ul)^{c,\frac{1}{2}} "z";"z"
\end{xy}
&
\begin{xy}
(-5,8)*{\mathcal{W}}="",
(0,0)*{\circ} = "x",
(10,0)*{\circ} = "y",
%(8,8)*{\circ} = "z",
(20,0)*{\checkmark} = "xc",
\ar _{} (-6,0);"x"
%\ar _{a} "x";"z"
\ar ^{b,1} "x";"y"
\ar ^{\frac{1}{2}} "y";"xc"
%\ar @(ur,ul)_{a} "x";"x"
\ar @(ur,ul)_{a,\frac{1}{2}} "y";"y"
\end{xy}
\end{tabular}
\caption{Examples of nondeterministic and probabilistic automata}
\label{fig:exampleautomataintro}
\end{figure}
In Figure~\ref{fig:exampleautomataintro} are four examples of transition
 systems: $\mathcal{X}$ and $\mathcal{Y}$  are
 qualitative/nondeterministic; $\mathcal{Z}$ and $\mathcal{W}$ exhibit
 probabilistic
 branching. We shall denote the \emph{finite} language of a system
 $\mathcal{A}$ 
 by $\langfin(\mathcal{A})$ and the \emph{infinitary}%
 \footnote{\label{footnote:infinitary}Note that in this paper, the word ``\emph{infinitary}'' means ``\emph{possibly infinite}'' and 
 does not mean all the behaviors have an infinite length.
 For example, in Figure~\ref{fig:exampleautomataintro}, $\langinf(\mathcal{X})$ includes a finite-length word 
 $b$ and $\langinf(\mathcal{X})$ assigns a probability $\frac{1}{6}$ to $b$.} 
 one by
 $\langinf(\mathcal{A})$. We define that a generated finite word is one
 with a run that ends with the termination symbol $\checkmark$.

 In the nondeterministic setting, languages are \emph{sets} of words. We
 have
 $\langfin(\mathcal{X})=\{b\}\subseteq
 \{b,ab,aab,\ldots\}=\langfin(\mathcal{Y})$, i.e.\ \emph{finite}
 language inclusion from $\mathcal{X}$ to $\mathcal{Y}$. However 
 $abb\ldots\in \langinf(\mathcal{X})$ while
 $abb\ldots\not\in \langinf(\mathcal{Y})$, hence \emph{infinitary} 
language inclusion fails.

 In the probabilistic setting, languages are naturally \emph{probability
 distributions} over words; and language inclusion refers to the
 pointwise order between probabilities. For example 
$\langfin(\mathcal{Z})=[b\mapsto \frac{1}{6},ba\mapsto \frac{1}{12},baa\mapsto \frac{1}{24},\ldots]$ and
$\langfin(\mathcal{W})=[b\mapsto \frac{1}{2},ba\mapsto
 \frac{1}{4},baa\mapsto \frac{1}{8},\ldots]$; since the former assigns
 no greater probabilities to all the words, we say that the
 finite language of $\mathcal{Z}$ is \emph{included} in that of $\mathcal{W}$.
 This quantitative notion of trace inclusion is also useful in
 verification: it gives e.g.\ an upper bound for the probability for
 something bad.

 Finally, the \emph{infinitary} languages for probabilistic systems call
 for measure-theoretic machinery since, in most of the cases, any
 infinite word gets assigned the probability $0$ (which is also the case
 in
 $\mathcal{Z}$ and $\mathcal{W}$). Here it is standard to assign
 probabilities to \emph{cylinder sets} rather than to individual
 words; see e.g.~\cite{baier08principlesof}. An example of a cylinder
 set is $\{aw\mid w\in\{a,b,c\}^{\omega}\}$. The language
 $\langinf(\mathcal{Z})$ assigns $\frac{2}{3}$ to it, while 
  $\langinf(\mathcal{W})$ assigns $0$; therefore we do not have
 \emph{infinitary} language inclusion from $\mathcal{Z}$ to $\mathcal{W}$.
\end{exa}

\subsection{Simulation}
There are many known algorithms for checking language inclusion. A
well-known one for NFA is a complete one 
that reduces the problem to emptiness check; however it
involves complementation, hence determinization, that
incurs an exponential blowup.

%{\par\sloppy
One of the alternative approaches to  language inclusion is by
\emph{simulation}. 
% It is  an alternative to the well-known algorithms, e.g.\
% for NFA, that reduce the problem to emptiness checking: these algorithms involve complementation, hence determinization, that
% results in an exponential blowup.
%Instead of directly checking language inclusion---that can be
%inefficient---
In the simulation-based verification we look for a simulation, that is, a
witness for \emph{stepwise} language inclusion.
The notion of simulation is commonly defined so that it implies
(proper, global) language inclusion---a property  called
\emph{soundness}. Although its converse (\emph{completeness}) fails 
in many settings, such simulation-based approaches tend to have an
advantage
in computational cost.
% It should be noted that the converse (\emph{completeness}) is not imposed and indeed fails in many cases.
One prototype example of such simulation notions is \emph{forward} and \emph{backward simulation} 
by Lynch and Vaandrager~\cite{lynch95forwardand}, for nondeterministic automata.
%\marginpar{Soundness wrt.\ infinite trace is already mentioned in~\cite{lynch95forwardand}}
They are shown in~\cite{lynch95forwardand} to satisfy soundness
with respect to finite traces: 
explicitly, existence of a forward (or backward) simulation from $\mathcal{X}$ to $\mathcal{Y}$ 
implies 
$\langfin(\mathcal{X})\subseteq\langfin(\mathcal{Y})$ 
%and $\langinf(\mathcal{X})\subseteq\langinf(\mathcal{Y})$, 
%where $\lang(\mathcal{X})$ and $\lang^{\infty}$
where the languages $\langfin(\mathcal{X})$ and $\langfin(\mathcal{Y})$ collects all the \emph{finite} words generated.
The simulations  are also shown in~\cite{lynch95forwardand} to be \emph{partly} sound with respect to \emph{infinite} traces:
i.e.\ existence of a forward (or backward, under the additional assumptions of \emph{image-finite} and \emph{total}) simulation 
from $\mathcal{X}$ to $\mathcal{Y}$ implies 
$\langinf(\mathcal{X})\subseteq\langinf(\mathcal{Y})$ where the languages $\langinf(\mathcal{X})$ and $\langinf(\mathcal{Y})$
 collects all the \emph{infinitary} words.
%it is also shown that forward and backward simulation can be also used to check \emph{infinite} trace inclusion:
%namely, 
%\par}

% that all finite words that $\mathcal{X}$ outputs are also output by $\mathcal{Y}$.

\subsection{Kleisli Simulation and Finite Trace}
\emph{Kleisli
simulation}~\cite{hasuo06genericforward,hasuo10genericforward,urabeH14genericforward,urabeH17matSim}
is a categorical generalization of these notions of forward and backward simulation by Lynch and Vaandrager.
It builds upon the use of coalgebras in a \emph{Kleisli category},
in~\cite{hasuo07generictrace,jacobs04tracesemantics,power99categorytheory}, where they are used to characterize
finite traces.
% for characterizing finite traces.
 % coalgebraic characterization of \emph{finite} trace semantics
 % studied in~\cite{hasuo07generictrace}.
Specifically:
\begin{itemize}
 \item A branching system $\mathcal{X}$ is represented as  an
       \emph{$F$-coalgebra} $c:X\kto FX$ in the Kleisli category $\Kl(T)$, for
       a suitable choice of a functor $F$ and a monad $T$. 
Here $F$ and $T$ are  parameters that determine
the (linear-time) \emph{transition type} and the \emph{branching type},
       respectively, of the  system $\mathcal{X}$. Examples are:
       \begin{itemize}
	\item $F=1+\Sigma\times(\place)$ (terminate, or (output and
	      continue)) and the
\emph{powerset monad} $T=\pow$ on $\Sets$ (nondeterminism), if
	      $\mathcal{X}$ is
	      a nondeterministic automaton (with explicit termination); 
	\item the same functor $F=1+\Sigma\times(\place)$ and the
\emph{sub-Giry monad} 
$T=\giry$~\cite{giry82categoricalapproach} on the category  $\Meas$  of
measurable spaces and measurable functions, for their probabilistic variant; and
	\item the  same functor $F=1+\Sigma\times(\place)$ and the
	\emph{lift monad} $T=\lift$ on $\Sets$ for automata with exception.
       \end{itemize}
 \item In~\cite{hasuo07generictrace}, under certain conditions
on $F$ and $T$,  it is shown that a \emph{final
       $F$-coalgebra} in $\Kl(T)$ arises as a lifting of an initial 
       $F$-algebra in $\Sets$. Moreover, it is observed that the natural
       notion of
       ``finite trace semantics'' or ``(finite) languages''  is  captured
       by a unique homomorphism via finality. This works uniformly for 
       a wide variety of systems, by changing $F$ and $T$.
\end{itemize}

It is shown in~\cite{hasuo06genericforward} that,
with respect to this categorical modeling of finite traces~\cite{hasuo07generictrace},
both forward and backward Kleisli simulation are indeed sound.
This categorical background allows us to instantiate Kleisli simulation
for various concrete systems---including both qualitative and
quantitative ones---and obtain simulation notions whose soundness with respect to \emph{finite} traces comes for free~\cite{hasuo06genericforward,hasuo10genericforward}.
%define simulation for various concrete systems---including both qualitative and quantitative ones---where
%its soundness wrt.\ \emph{finite} trace is assured while completeness is not.
%By It allows us to define a notion of forward and backward simulation for various concrete systems 
%including both qualitative and quantitative systems.
%the defined simulation is assured to be sound wrt.\ \emph{finite} trace.
%In contrast its completeness is not assured and fails for many systems.
Like many other notions of simulation, the resulting simulation notion sometimes fails to be complete.
This drawback of incompleteness 
%feature of forward and backward Kleisli simulation
with respect to finite traces can be partly mended by 
%to some extent by 
\emph{forward partial execution} (FPE), a transformation of %coalgebraic
systems introduced in~\cite{urabeH14genericforward} 
(and its extended version~\cite{urabeH17matSim})
that potentially
increases the likelihood of existence of simulations.
% The property of FPE that it aids Kleisli simulation 
% can be expressed using two terms---\emph{soundness} and \emph{adequacy}.

\subsection{Contributions}
While the automata-theoretic simulations in~\cite{lynch95forwardand} are known to be
useful for checking both finite and infinitary trace inclusion, 
their coalgebraic generalization (Kleisli simulation) 
has been applied only to
%is known to be useful only for 
the finite trace setting.
In this paper we continue our series of
work~\cite{hasuo06genericforward,hasuo10genericforward,urabeH14genericforward,urabeH17matSim}
and study the relationship between Kleisli simulations and
\emph{infinitary} traces. This turns out to be more complicated than 
we had expected, a principal reason being that \emph{infinitary} traces
are less well-behaved than finite traces (the latter being characterized simply
by finality).
% Using Kleisli simulation to check \emph{infinite} trace inclusion is
% the main aim of this paper.

For a suitable coalgebraic modeling of infinitary traces we
principally follow~\cite{jacobs04tracesemantics}---also relying on 
observations
in~\cite{cirstea10genericinfinite,kerstan13coalgebraictrace}---and
characterize infinitary traces in terms of \emph{largest homomorphisms}.
% To this end, we need a categorical characterization of infinite trace;
% we follow the framework introduced in~\cite{jacobs04tracesemantics}.
%where nondeterministic systems are intensively studied.
%With respect to \emph{infinite} trace,
%similar study is done in~\cite{jacobs04tracesemantics} 
%intensively for nondeterministic systems.
More specifically, we lift 
a final $F$-coalgebra
% $\zeta:Z\to FZ$ 
in $\Sets$ to
%  the coalgebra
% $J\zeta:Z\kto FZ$ in
the Kleisli category $\Kl(T)$ 
and prove that the latter admits a largest homomorphism. 
In this paper
we (principally) work with: the powerset monad $\pow$ (on $\Sets$),
 the sub-Giry monad $\giry$ (on $\Meas$), and the lift monad $\lift$ (on $\Sets$) 
 as a monad $T$ for branching (see Table~\ref{table:settings}); and
a polynomial functor $F$ for linear-time behaviors.

\begin{table}
\begin{tabular}{|l|c||c|c|}\hline
 \shortstack{branching type} & \shortstack{monad\\$T$} & 
 \shortstack{finite trace \\ \cite{hasuo06genericforward}} & 
 \shortstack{infinitary trace \\ {[current work]}} \\ \hline\hline
 nondeterministic & $\pow$  & \shortstack{fwd.\ sim.\\bwd.\ sim.} & \shortstack{fwd.\ sim.\\TIF-bwd.\ sim.} \\ \hline
 probabilistic    & $\giry$ & \shortstack{fwd.\ sim.\\bwd.\ sim.} & \shortstack{fwd.\ sim.\\total bwd.\ sim.} \\ \hline
 with exception   & $\lift$ & \shortstack{fwd.\ sim.\\bwd.\ sim.} & \shortstack{fwd.\ sim.\\total bwd.\ sim.} \\ \hline
\end{tabular}
\caption{Three different settings}
\label{table:settings}
\end{table}

Here are our concrete contributions.
For each of the above combinations of $T$ and $F$:
\begin{itemize}
 \item We show that
       forward Kleisli simulations are sound with respect to inclusion of
       \emph{infinitary}
       languages. The proof of this general result is not hard,
       exploiting the above
       coalgebraic modeling of infinitary languages as largest homomorphisms.
 \item We show that backward simulations are sound too, although here we have
       to impose 
       suitable restrictions, namely \emph{totality} and
       \emph{image-finiteness}. The soundness
       proofs are much more involved, too, and call for careful inspection of the 
       construction of infinitary trace semantics. The proofs are given
       separately for $T=\pow$, and for $T=\giry$ and $\lift$, because of differences 
       in the relevant constructions
       %in property between them 
       (see Remark~\ref{rem:differencePandG}).
 \item We show that \emph{forward partial execution} (FPE)---a
       transformation from~\cite{urabeH14genericforward,urabeH17matSim} that aids 
       discovery of forward or backward simulations---is applicable also to
       the current setting of \emph{infinitary} trace inclusion. More
       specifically we prove: \emph{soundness} of FPE (discovery of a
       simulation after FPE indeed witnesses infinitary language
       inclusion); and its \emph{adequacy} (FPE does not destroy simulations that
       are already there). 
       Suitable restrictions, \emph{totality} and \emph{image-finiteness}, are imposed on the simulating system in order to
       ensure the adequacy with respect to backward simulation.
\end{itemize}

\subsection*{Organization}
 Section~\ref{sec:preliminaries} is devoted to categorical preliminaries; we
 fix notations there.
In Section~\ref{sec:weaklyfinalcoalg} we review the previous works that we
 rely on, namely coalgebraic infinitary trace
 semantics~\cite{jacobs04tracesemantics}, Kleisli
 simulation~\cite{hasuo06genericforward,hasuo10genericforward,urabeH14genericforward,urabeH17matSim},
and FPE~\cite{urabeH14genericforward,urabeH17matSim}.
Our technical contributions are in the subsequent sections: in
Section~\ref{sec:powersetmonad} we study the nondeterministic setting (i.e.\
 the powerset monad $\pow$ on $\Sets$ and a polynomial functor $F$); 
 Section~\ref{sec:girymonad} is for the probabilistic setting (where the monad
 $T$ is the sub-Giry monad $\giry$); and
 Section~\ref{sec:liftmonad} is for systems with exception (where $T$ is the lift monad $\lift$).

Some definitions and results
in Sections~\ref{sec:powersetmonad}--\ref{sec:girymonad} are marked with
$\dagger$. Those marked ones are essentially %proofs of the 
results for specific
settings (namely $T=\pow$ and $T=\giry$) but formulated in general
terms with an arbitrary $T$ subject to certain axioms. We do so in the hope that the axioms thus
identified
will help to discover new instances. 
Indeed, 
%For example,
the results 
%as we will see 
in Section~\ref{sec:liftmonad} for $T=\lift$ are derived from the general results 
developed in Section~\ref{sec:girymonad}.
%are
% applicable to 
%another monad (namely the lift monad $T=\lift$).

Compared to the earlier version~\cite{urabeH15coalgebraicinfinite} of this paper, 
the current version additionally contains the following materials.
\begin{itemize}
\item Section~\ref{subsec:rankedAlphabetAndInfiniteTrees} is added, where preliminaries about ranked alphabet and
infinitary trees are given.
\item 
We added sections where coincidence between
the coalgebraic definition and the automata-theoretic definition of infinitary language is presented.
Namely: Section~\ref{subsec:automcharapow} is for nondeterministic setting,  Section~\ref{subsec:automcharagiry} is
for probabilistic setting, and Section~\ref{subsubsec:automcharaexception} for the
setting where the system can abort with an exception.
%systems with exception.
\item In~\cite{urabeH15coalgebraicinfinite} we mainly used two monads---$\pow$ and $\giry$.
Now one additional monad---the \emph{lift monad} $\lift$---is also discussed in this paper. 
Moreover, a brief discussion about the \emph{subdistribution monad} $\dist$ is added, too.
%Now two additional monads---the \emph{subdistribution monad} $\dist$ and the lift monad $\lift$ are 
%also discussed in this paper.
%in Section~\ref{subsec:subdistmonad} investigates the \emph{subdistribution monad} $\dist$, 
%and Section~\ref{sec:liftmonad} investigates 
\item We added some examples that are absent in~\cite{urabeH15coalgebraicinfinite}.
\item We added proofs that are omitted in~\cite{urabeH15coalgebraicinfinite}. % because of space limitation.
\end{itemize}

\section{Preliminaries}
\label{sec:preliminaries}
\subsection{Categorical Preliminaries}
\label{subsec:categoricalPreliminaries}
%\subsection{Categorical Notions}
\begin{defi}
 A  \emph{polynomial functor} $F$ on $\Sets$  is defined by the following BNF notation:
 $F::=\id\mid A\mid F_1\times F_2\mid \coprod_{i\in I} F_i$. 
 Here $A\in\Sets$ and $I$ is a countable set.
\end{defi}
The notion of polynomial functor can be also defined  for $\Meas$---the category 
of measurable spaces and measurable functions between them.

\begin{defi}\label{def:polyfuncmeas}
A \emph{(standard Borel) polynomial functor} $F$ on $\Meas$ is defined by the following BNF notation:
$%\begin{equation*}
F\;::=\; \id\,\mid\, (A,\sigalg_A) \,\mid\, F_1\times F_2 \,\mid\, \coprod_{i\in I} F_i\, .
$ %\end{equation*} 
%Here, $A\in\Meas$.
%Here, $A$ is a countable set.
Here
 $I$ is a countable set; and
 we require that $(A,\sigalg_A)\in\Meas$  is a \emph{standard Borel
 space}
 (see e.g.~\cite{doob94measuretheory}). 
The  $\sigma$-algebra $\sigalg_{FX}$ associated to $FX$ is defined in the obvious
 manner. Namely: 
%We have to specify $\sigma$-algebras.
% For $(X,\sigalg_X)\in\Meas$, $F(X,\sigalg_X)=(FX,\sigalg_{FX})$ is given as follows:
% the underling set $FX$ is given in the same manner as that for the  polynomial functors on $\Sets$.
% The $\sigma$-algebra $\sigalg_{FX}$ is defined inductively as follows.
for $F=\id$, $\sigalg_{FX}=\sigalg_X$; for $F=(A,\sigalg_A)$, $\sigalg_{FX}=\sigalg_A$;
for $F=F_1\times F_2$, $\sigalg_{FX}$ is the smallest $\sigma$-algebra that contains $A_1\times A_2$ for 
all $A_1\in\sigalg_{F_1X}$ and $A_2\in\sigalg_{F_2X}$;
for for $F=\coprod_{i\in I} F_i$, 
$\sigalg_{FX}=\{\coprod_{i\in I} A_i\mid A_i\in \sigalg_{F_iX}\}$.

On arrows, $F$ acts in the same manner as a polynomial functor on
 $\Sets$.

In what follows, a standard Borel polynomial functor is often called
 simply a \emph{polynomial functor}. 
\end{defi}
%We only give their action on objects $(X,\sigalg_X)$ in $\Meas$.
 The technical  requirement of being standard Borel in the above will be used in the probabilistic
setting of~Section~\ref{sec:girymonad} (it is also assumed and exploited in~\cite{cirstea10genericinfinite}
and in~\cite{schubert09terminalcoalgebras} that we rely on).  A standard Borel space is a measurable
space induced by a Polish space; for further details see
e.g.~\cite{doob94measuretheory}.

 We go on to introduce monads $T$ for branching.
We principally use three monads---the \emph{powerset monad} $\pow$ on
$\Sets$, the \emph{sub-Giry} monad $\giry$ on $\Meas$, 
and the \emph{lift monad} $\lift$ on $\Sets$. 
%The latter is
The monad $\giry$ is
an adaptation of the \emph{Giry monad}~\cite{giry82categoricalapproach}
and inherits most of its structure from the Giry monad;
see Remark~\ref{rem:giryAndSubGiry}.
%Giry monad is defined in \cite{giry82categoricalapproach}
\begin{defi}[monads $\pow$, $\giry$ and $\lift$]\label{def:variousmonads}
The \emph{powerset monad} is the monad $(\pow,\eta^{\pow},\mu^{\pow})$ on $\Sets$ such that 
$\pow X=\{A\subseteq X\}$ and $\pow f(A)=\{f(x) \mid x\in A\}$.
Its unit is given by the singleton set $\eta^{\pow}_{X}(x)=\{x\}$ and its multiplication is given by $\mu^{\pow}_{X}(M)=\bigcup_{A\in M}A$.

The \emph{sub-Giry monad} is the monad $(\giry,\eta^{\giry},\mu^{\giry})$ on $\Meas$ such that
\begin{itemize}
\item $\giry(X,\sigalg_X)=(\giry X, \sigalg_{\giry X})$, where 
%\begin{itemize}
%\item 
the underling set $\giry X$ is the set of all \emph{subprobability
      measures} on $(X,\sigalg_X)$. The latter means those measures which assign to the whole
      space $X$ a value in the unit interval $[0,1]$.
 \item 
%\item 
The $\sigma$-algebra $\sigalg_{\giry X}$ on $\giry X$ is the smallest $\sigma$-algebra such that, for all $S\in\sigalg_X$, 
the function $\text{ev}_S:\giry X\to[0,1]$ defined by $\text{ev}_S(P)=P(S)$ is measurable.
%\end{itemize}
\item $\giry f(\nu)(S)=\nu(f^{-1}(S))$ where $f:(X,\sigalg_X)\to(Y,\sigalg_Y)$ is measurable, $\nu\in\giry X$, and $S\in\sigalg_Y$.
\item $\eta^{\giry}_{(X,\sigalg_X)(x)}$ is given by the \emph{Dirac measure}: $\eta^{\giry}_{(X,\sigalg_X)}(x)(S)$ is $1$ if $x\in S$ and $0$ otherwise.
%$\eta^{\giry}_{(X,\sigalg_X)}(x)(S)=\begin{cases} 1 & (x\in S) \\ 0 & (x\notin S)\end{cases}$ 
%where $x\in X$ and $S\in\sigalg_X$, and
\item $\mu^{\giry}_{(X,\sigalg_X)}(\Psi)(S)=\int_{\giry (X,\sigalg_X)} \text{ev}_S \,d\Psi$ where 
$\Psi\in\giry^2 X$, $S\in\sigalg_X$ and $\text{ev}_S$ is defined as above.
\end{itemize}

The \emph{lift monad} is a monad $(\lift,\eta^{\lift},\mu^{\lift})$ on $\Sets$  such that
%\begin{itemize}
%\item 
$\lift X=\{\bot\}+X$,
%\item 
%$\lift f(y)=\begin{cases} f(y) & (y\neq \bot) \\ \bot & (\text{otherwise}),\end{cases} $
$\lift f(y)= f(y)$ if  $y\neq \bot$ and $\lift f(y)=\bot$ otherwise.
Its unit is given by
%\item 
$\eta^{\lift}_{X}(x)=x$ and its multiplication is given by
%\item 
$\mu^{\lift}_{X}(z)=z$.
%\end{itemize}
\end{defi}

A monad gives rise to a category called its \emph{Kleisli category} (see
e.g.~\cite{mac98categoriesfor}).

\begin{defi}[Kleisli category $\Kl(T)$]\label{def:KleisliCategory}
Given
%a category $\mathbb{C}$ and 
a monad $(T,\eta,\mu)$ on a category $\mathbb{C}$,
the \emph{Kleisli category} for $T$ is the category $\Kl(T)$ whose objects are the same as those of $\mathbb{C}$, 
and for each pair of objects $X,Y$, the homset $\Kl(T)(X,Y)$ is given by 
$\mathbb{C}(X,TY)$. 
An arrow  in $\Kl(T)$ is referred to as a \emph{Kleisli arrow}, and
 depicted by $X\kto Y$ for distinction from $\mathbb{C}$. Note that it is nothing but
 an arrow $X\to TY$ in the base category $\mathbb{C}$.
% a set of \emph{Kleisli arrows}: namely,  
% $\Kl(T)(X,Y)=\mathbb{C}(X,TY)$.
% An arrow $f$ from $X$ to $Y$ in the Kleisli category is denoted by
 % $f:X\kto Y$ for distinction.

Moreover, for two sequential Kleisli arrows $f:X\kto Y$ and $g:Y\kto Z$,
their composition is given by $\mu_Z\circ Tg\circ f$ and denoted by $g\odot f$.
The \emph{Kleisli inclusion functor} is the functor $J:\mathbb{C}\to\Kl(T)$ 
%that is defined by
such that
$JX=X$ and $Jf=\eta_Y\circ f$ for $f:X\to Y$ in $\mathbb{C}$.
\end{defi}

%Let $F$ be an endofunctor and $T$ be a monad on a category $\mathbb{C}$.
%For a monad $T$, 
It is known that a functor $F:\mathbb{C}\to\mathbb{C}$ canonically lifts
to a functor $\overline{F}:\Kl(T)\to\Kl(T)$, given that 
there exists a natural transformation $\lambda:FT\Rightarrow TF$ that is compatible 
with the unit and the multiplication of $T$.
%Such a natural transformation is called a \emph{distributive law}. 
%Namely, for $X\in\Kl(T)$ (hence $X\in\mathbb{C}$), $\overline{F}X$ is given by $\overline{F}X=FX$ and
%for $f\in\Kl(T)(X,Y)=\mathbb{C}(X,TY)$, $\overline{F}f\in\Kl(T)(\overline{F}X,\overline{F}Y)=\mathbb(C)(FX,TFY)$ is given by 
%$\overline{F}f=\lambda_{Y}\circ Ff$.
%For more details, see e.g.~\cite{mulry93liftingtheorems}.
%
\begin{lem}[distributive law, \cite{mulry93liftingtheorems}]
\label{lem:distributiveLawLift}
Let $T$ be a monad and $F$ be an endofunctor on a category $\mathbb{C}$.
The following conditions are equivalent
\begin{enumerate}
\item The functor $F$ can be lifted to the Kleisli category $\Kl(T)$: namely,
there exists a functor $\overline{F}:\Kl(T)\to\Kl(T)$ such that $\overline{F}\circ J=J\circ F$.

\item There exists a natural transformation $\lambda:FT\Rightarrow TF$ such that the following diagrams commute for all $X\in\mathbb{C}$.
\begin{displaymath}
 \begin{xy}
 \xymatrix@R=1.6em@C=2.4em{
 {F X} \ar[r]^{F\eta_{X}} \ar[dr]_{\eta_{FX}}  & {FTX} \ar[d]^{\lambda_X} \\
 {} & {TFX} 
 }
 \end{xy}
 \;\;\;\;\;\;\;\;\;\;
 \begin{xy}
 \xymatrix@R=1.6em@C=2.4em{
 {FT^2X} \ar[d]_{F\mu_{X}} \ar[r]^{\lambda_{TX}}  & {TFTX} \ar[r]_{T\lambda_X} & {T^2FX} \ar[d]^{\mu_{FX}} \\
 {FTX} \ar[rr]^{\lambda_{X}} & & {TFX} 
 }
 \end{xy}
 \end{displaymath}
\end{enumerate}
Such $\lambda$ is called a \emph{distributive law}.   \qed
\end{lem}

%We  define orders on each of the homsets in $\Kl(\pow)$ and $\Kl(\giry)$ as follows.
Throughout this paper, we fix the orders on  homsets of $\Kl(\pow)$, $\Kl(\giry)$ and $\Kl(\lift)$ as follows.
\begin{defi}[order enrichment of $\Kl(\pow)$, $\Kl(\giry)$ and $\Kl(\lift)$]\label{def:orderenrichment}
We define an order on $\Kl(\pow)(X,Y)$ by: $f\sqsubseteq g$ if and only if $\forall x\in X.\, f(x)\subseteq g(x)$.
%We define an order on 
For
$\Kl(\giry)(X,Y)$ we define: $f\sqsubseteq g$ if and only if
 $\forall x\in X.\,\forall A\in \sigalg_Y.\, f(x)(A)\leq g(x)(A)$. Here
 the last $\le$ is the usual order in the unit interval $[0,1]$.
 We define an order on $\Kl(\lift)(X,Y)$ by 
$f\sqsubseteq g$ if and only if $\forall x\in X.\, f(x)=\bot \,\text{or}\, f(x)=g(x)$.
%
%For $T=\pow$ and $\lift$, we define an order on $\Kl(T)(X,Y)$  as follows.
%\begin{itemize}
%\item For $T=\pow$, $f\sqsubseteq g \;\defarrow\; \forall x\in X.\, f(x)\subseteq g(x)$.
%\item For $T=\lift$, $f\sqsubseteq g \;\defarrow\; \forall x\in X.\, g(x)=\bot \Rightarrow f(x)=\bot$.
%\end{itemize}
%For $T=\giry$, we define an order on $\Kl(T)((X,\sigalg_X),(Y,\sigalg_Y))$ as follows.
%\begin{itemize}
%\item $f\sqsubseteq g \;\defarrow\; \forall x\in X.\,\forall A\in \sigalg_Y.\, f(x)(A)\leq g(x)(A)$.
%\end{itemize}
\end{defi}

\begin{rem}\label{rem:giryAndSubGiry}
 The sub-Giry monad $\giry$ is an adaptation of the \emph{Giry monad}
from~\cite{giry82categoricalapproach}; in the original Giry monad one only allows
(proper) \emph{probability measures}, i.e.\  measures that map the
 whole space to $1$. We work with the sub-Giry monad because, without
 this relaxation from probability to subprobability, the order
 structure in Definition~\ref{def:orderenrichment} is reduced to the equality.
\end{rem}

\subsection{Ranked Alphabet and Infinitary Trees}
\label{subsec:rankedAlphabetAndInfiniteTrees}
There is a natural correspondence between polynomial functors and
\emph{ranked alphabets}. In this paper  a functor $F$ for the (linear-time)
transition type is restricted to a polynomial one; this means that we are dealing with
($T$-branching) systems that generate \emph{trees}
over some ranked alphabet. 
%All  the systems that we will be using in this paper are those which
%generate trees.
Here we collect some standard notions and notations on (the conventional
presentation of) such finite/infinite trees;
they will be used later in showing that our coalgebraic infinitary traces indeed
capture
infinitary tree languages of such systems.

%The generated 
Trees are labeled with 
letters from an alphabet.
%a family of alphabets.
%Before focusing on specific functors and monads, we give a formal definition of trees.
\begin{defi}
A \emph{ranked alphabet} is a family $\Sigma=(\Sigma_n)_{n\in\omega}$ of sets.
A \emph{standard Borel ranked alphabet} is a family $\Sigma=\bigl((\Sigma_n,\sigalg_n)\bigr)_{n\in\omega}$ of
standard Borel spaces.
%In the above, 
The index $n\in\omega$ is called an \emph{arity}.
\end{defi}

For the definition of infinitary trees we follow~\cite{courcelle83fundamentalproperties}.
%The defined tree is an infinite tree whose nodes are labeled with letters in a ranked alphabet.
Each node 
labeled with a letter $a\in\Sigma_n$ of arity $n$
%whose arity is $n$ 
has $n$ children.
The idea of $k$-prefix trees, presented below, is introduced in~\cite{ellis71probabilistictree}.
It can be regarded as a finite tree of depth $k$ that is obtained by 
%extracting nodes of 
truncating
an infinitary tree.

\begin{defi}\label{def:treeNotions}
Let $\Sigma=(\Sigma_n)_{n\in\omega}$ be a ranked alphabet.
A \emph{$\Sigma$-labeled  infinitary tree} is a pair $(D,l)$ of a \emph{domain} $D\subseteq \mathbb{N}^*$ and 
a \emph{labeling function} $l:D\to\bigcup_{n\in\omega}\Sigma_n$ such that: %satisfying the following conditions:
\begin{itemize}
\item the domain $D$ is
%\begin{itemize}
%\item 
prefix-closed 
(i.e.\ $\forall \alpha\in\mathbb{N}^*.\,\forall i\in\mathbb{N}.\; \alpha i\in D\,\Rightarrow\, \alpha \in D$),
%(i.e.\ for  $\alpha\in\mathbb{N}^*$ and $i\in\mathbb{N}$, $\alpha i\in D$ implies $\alpha \in D$),
%\item 
nonempty, and
%(equivalently, empty sequence $\empseq$ is in $D$), and
%\item 
downward-closed 
(i.e.\ $\forall \alpha\in\mathbb{N}^*.\,\forall i\in\mathbb{N}.\; \alpha
      i\in D\,\Rightarrow\, \forall j\leq i.\, \alpha j\in D$); and
%(i.e.\ for  $\alpha\in\mathbb{N}^*$ and $i\in\mathbb{N}$, $\alpha i\in D$ implies $\alpha j\in D$ for all $j\leq i$). 
%\end{itemize}
\item %The labelling function $l$ satisfies the arity condition: namely,
%\begin{itemize}
%moreover, 
for all $\alpha\in D$  such that $l(\alpha)\in \Sigma_n$ and $i\in\mathbb{N}$, $\alpha i\in D$ if and only if $0\leq i\leq n-1$. 
%\end{itemize}
\end{itemize}
For 
%a ranked alphabet $\Sigma=(\Sigma_i)_{i\in\omega}$ and 
%a natural number 
$k\in\omega$,  
a \emph{$\Sigma$-labeled $k$-prefix tree} is 
a pair $(D,l)$ of a \emph{domain} $D$ and a \emph{labeling function} $l:D\to\bigcup_{n\in\omega}\Sigma_n$ 
such that:
%that satisfy the following conditions.
\begin{itemize}
\item
if $k=0$ then $D$ is an empty set and
if $k>0$ then $D\subseteq \bigcup_{i\leq k-1} \mathbb{N}^{i}$;
\item the domain $D$ is prefix-closed, nonempty, and downward-closed; and
\item
%Moreover, 
for $\alpha\in D$ such that $|\alpha| <k-1$ and $l(\alpha)\in \Sigma_n$ and $i\in\mathbb{N}$, $\alpha i\in D$ if and only if $0\leq i\leq n-1$. 
\end{itemize}
%defined in the similar manner to $\Sigma$-labeled  infinite tree, except for the following points.
%\begin{itemize}
%\item The domain $D$ is restricted to $D\subseteq \mathbb{N}^{k-1}$ if $k>0$ and $\emptyset$ if $k=0$.
%\item The empty sequence $\empseq$ does not have to be in  $D$.
%\item The condition for the labelling function is replaced by the following.
%\begin{itemize}
%\item for each $\alpha\in D$ such that $|\alpha| <n$ and $l(\alpha)\in \Sigma_i$, and each $n\in\mathbb{N}$, $\alpha n\in D$ iff $0\leq n\leq i-1$. 
%\end{itemize}
%\end{itemize}
Here $k$ is called the \emph{depth} of $(D,l)$.
%
%For a given ranked alphabet $\Sigma$, 
%An \emph{empty tree} $\empseq$ is the unique $0$-prefix tree.
We write $\Treeinf(\Sigma)$ and $\Tree^k(\Sigma)$ for
the sets of all $\Sigma$-labeled  infinitary trees and 
$\Sigma$-labeled  $k$-prefix trees, respectively. 
%\end{defi}
%
%
%By dropping or extracting nodes of a tree, we can obtain a new tree.
%Conversely, we can construct a new tree by combining multiple trees with a new node.
%%A $k$-prefix tree can be 
%%%regarded as a finite tree of depth $k$ that is 
%%obtained by extracting nodes of an infinite tree in several ways.
%%Conversely, we can construct an 

%\begin{defi}
%A $\Sigma$-labeled $k$-prefix tree $t=(D,l)$ is 
A $k$-prefix tree $t=(D,l)\in\Tree^k(\Sigma)$ is
%said to be 
called
a \emph{prefix} of 
a tree 
$t'=(D',l')\in\Treeinf(\Sigma)\cup\bigcup_{k'\geq k}\Tree^{k'}(\Sigma)$
%$\Sigma$-labeled infinite tree 
%(or $k'$-prefix tree for some $k'\geq k$) $t'=(D',I')$  
if $D\subseteq D'$ and 
%restriction of $l'$ on $D$ coincides with $l$ (i.e.\ 
%for each $\alpha\in D$, 
$l(\alpha)=l'(\alpha)$ holds for all $\alpha\in D$.
We write $t\treeprefix t'$ if $t$ is a prefix of $t'$.
For 
%a tree
%$\Sigma$-labeled  infinite $k$-prefix tree 
$t\in\Tree^k(\Sigma)$, a \emph{cylinder set induced by $t$} 
is the set $\cyl(t)\subseteq\Treeinf(\Sigma)$ that is defined by
%$\cyl(t)\subseteq\Treeinf(\Sigma)$ that is defined by
$%\begin{equation*}
\cyl(t)=\{t'\in\Treeinf(\Sigma)\mid \text{$t\treeprefix t'$}\}\,.
$%\end{equation*}

For %a tree %$\Sigma$-labeled infinite tree 
$t=(D,l)\in\Treeinf(\Sigma)$ and $\alpha\in D$, %such that $l(\empseq)\in\Sigma_n$ and $i\leq n-1$, %$\alpha\in D$,
\emph{the $\alpha$-th subtree of $t$} is a 
$\Sigma$-labeled  infinitary tree 
$t_{\alpha}=(D_{\alpha},l_{\alpha})\in\Treeinf(\Sigma)$ where
$D_{\alpha}=\{\beta\in \mathbb{N}^* \mid \alpha\beta\in D\}$ and 
$l_{\alpha}(\beta)=l(\alpha\beta)$.
%
%For $a\in\Sigma_n$ and a family of trees 
%$\Sigma$-labeled infinite trees 
%$t_0=(D_0,l_0),\ldots,t_{n-1}=(D_{n-1},l_{n-1})\in\Treeinf(\Sigma)$,
% %$(t_i=(D_i,l_i))_{0\leq i\leq n-1}$,
%we write $(a,t_0,\ldots,t_{n-1})$ for a $\Sigma$-labeled infinite tree $t=(D,l)\in\Treeinf(\Sigma)$ where
%$D=\{i \alpha_i\mid 0\leq i\leq n-1, \alpha_i\in D_i\}$ and $l(i\alpha)=l_i(\alpha)$ for  $i\in\{0,\ldots,n-1\}$.
\end{defi}

In later sections we will use an \emph{infinitary tree automaton}---an automaton that generates $\Sigma$-labeled trees.
Note that an infinitary tree automaton might output a finite-depth tree whose leaves are labeled with $0$-ary letters $a\in\Sigma_0$
(cf.\ footnote~\ref{footnote:infinitary} on page~\pageref{footnote:infinitary}).
Especially when $\Sigma_0=\{\checkmark\}$ and $\Sigma_i=\emptyset$ for all $i\geq 2$, 
an infinitary tree automaton can be regarded as 
%an infinite \emph{word} automaton.
an automaton that generates \emph{words} instead of trees. %---a special tree.
We call  such an automaton \emph{infinitary automaton} (suppressing the word ``tree'').
%Note that an infinitary automaton might output a finite-length word terminating with $\checkmark$ 
%(c.f.\ footnote~\ref{footnote:infinitary} on page~\pageref{footnote:infinitary}).

\begin{rem}\label{rem:1807081506}
As described above, we regard infinitary tree automata as \emph{generative} ones that \emph{output} trees throughout this paper.
Another characterization of behaviors of infinitary tree automata is to regard them as \emph{reactive} ones that \emph{accept} trees.
These two characterizations coincide 
in the nondeterministic setting.
%generative infinitary tree automata %nondeterministically outputs trees in a \emph{generative} manner.
%can be identified with reactive ones.
%We can identify such automata with a \emph{reactive} automata that nondeterministically \emph{accepts} trees. 
%that nondeterministically outputs a tree can be identified with an automaton that \emph{accepts} trees by considering that 
In contrast,  they are distinguished in the probabilistic setting:
in the former case, an automaton randomly outputs a tree while in the latter case,  an automaton assigns a probability
where the tree is accepted to each tree.
%Probabilistic infinitary tree automata discussed in this paper are generative ones.
\end{rem}

A system that generates $\Sigma$-labeled infinitary trees will later be represented as an $\overline{\FSigma}$-coalgebra 
on the Kleisli category of some monad where $\overline{\FSigma}$ is a lifting of $\FSigma$.
Here the functor $\FSigma$ is a polynomial functor defined as follows.
%A ranked alphabet and standard Borel ranked alphabet give rise to polynomial functors on $\Sets$ and $\Meas$ respectively.
%We will see later that they represent a branching type of systems that generate infinite trees.
\begin{defi}\label{def:FSigma}
For a ranked alphabet $\Sigma=(\Sigma_n)_{n\in\omega}$,
we define 
%a polynomial functor 
$\FSigma:\Sets\to\Sets$ by $\FSigma=\coprod_{n\in\omega} \Sigma_n\times (\place)^n$.
For a standard Borel ranked alphabet $\Sigma=\bigl((\Sigma_n,\sigalg_n)\bigr)_{n\in\omega}$,
we define 
%a polynomial functor 
$\FSigma:\Meas\to\Meas$ by $\FSigma=\coprod_{n\in\omega} (\Sigma_n,\sigalg_n)\times (\place)^n$.
%Let $\Sigma=(\Sigma_n)_{n\in\omega}$ be a ranked alphabet and
%$\mathbb{C}$ be a category where the notion of polynomial functors is defined.
%We define a polynomial functor $F_{\Sigma}:\mathbb{C}\to\mathbb{C}$ by 
%$F_{\Sigma}=\coprod_{n\in\omega} \Sigma_n\times (\place)^n$.
\end{defi}

\section{Infinite Traces, Kleisli Simulations and Coalgebras in $\Kl(T)$}\label{sec:weaklyfinalcoalg}
In this section we review some categorical constructs, the relationship
among which lies at the heart of this paper. They are namely: coalgebraic
infinitary trace semantics~\cite{jacobs04tracesemantics}, Kleisli
simulation~\cite{hasuo06genericforward,hasuo10genericforward,urabeH14genericforward,urabeH17matSim}
and forward partial execution (FPE)~\cite{urabeH14genericforward,urabeH17matSim}.

%The following situation is identified in~\cite{jacobs04tracesemantics}.
The following situation is identified in~\cite{jacobs04tracesemantics} 
(see also Section~\ref{subsec:automcharapow}, Section~\ref{subsec:automcharagiry}
and Section~\ref{subsubsec:automcharaexception}):
the largest homomorphism 
to a certain coalgebra (that we describe below)
%to an $\overline{F}$-coalgebra $J\zeta$ in $\Kl(T)$ 
%where $\zeta$ is a final $F$-coalgebra in $\mathbb{C}$
%is observed to 
coincides with the standard, conventionally defined notion of infinitary language, for a variety of systems.
%This situation allows us to characterize the infinite trace semantics of a system as 
%the largest homomorphism.
% to an $\overline{F}$-coalgebra in $\Kl(T)$ that is obtained by
%lifting a final $F$-coalgebra in $\mathbb{C}$.
An instance of it is shown to arise, in~\cite{jacobs04tracesemantics}, when $\mathbb{C}=\Sets$, $T=\pow$ and $F$ is a
polynomial functor.
In Section~\ref{sec:powersetmonad} we will give another proof for this fact;
the new proof will
 serve our goal of showing soundness of backward simulations.
\begin{defi}[infinitary trace situation]\label{def:infiniteTraceSituation}
%Let $\mathbb{C}$ be a category, let $F$ be an endofunctor on $\mathbb{C}$ and $T$ be a monad on $\mathbb{C}$.
Let $F$ be an endofunctor and $T$ be a monad on a category $\mathbb{C}$.
We assume that each homset of the Kleisli category $\Kl(T)$ carries an order $\sqsubseteq$.
The functor $F$ and the monad $T$ constitute an \emph{infinitary trace situation} with respect to $\sqsubseteq$ if they satisfy the following conditions.
\begin{itemize}
\item There exists a final $F$-coalgebra $\zeta:Z\to FZ$ in $\mathbb{C}$.
\item There exists a distributive law $\lambda:FT\Rightarrow TF$,
      yielding a lifting $\overline{F}$ on $\Kl(T)$ of $F$ by Lemma~\ref{lem:distributiveLawLift}.
%, and hence $F$ can be lifted to an endofunctor $\overline{F}$ on $\Kl(T)$ (see~\cite{mulry93liftingtheorems} for more details).
\item For each coalgebra $c:X\kto \overline{F}X$ in $\Kl(T)$, the
      lifting $J\zeta:Z\kto \overline{F}Z$ of $\zeta$ admits the largest homomorphism.
That is, there exists a homomorphism $\trinf(c):X\kto Z$ from $c$ to
      $J\zeta$ such that, for any homomorphism $f$ from $c$ to $J\zeta$, $f\sqsubseteq\trinf(c)$ holds.
%(i.e.\ $\trinf(c)$ is a A coalgebra $J\zeta:Z\kto \overline{F}Z$ is weakly final.
\end{itemize}
%In this setting, for an $\overline{F}$-coalgebra $c:X\kto\overline{F}X$, 
%its \emph{coalgebraic infinite trace semantics} is the largest homomorphism  from $c$ to $J\zeta$ and denoted by $\trinf(c)$.
%The largest homomorphism $\trinf(c)$ is 
%Namely, for every homomorphism $f:X\kto Z$ from $c$ to $J\zeta$, $f\sqsubseteq \trinf(c)$ holds.
\end{defi}

In~\cite{hasuo06genericforward,hasuo10genericforward,urabeH14genericforward,urabeH17matSim}
we augment a coalgebra with an explicit arrow for initial states. The
resulting notion is called a $(T,F)$-system.

\vspace*{1mm}
\noindent\begin{minipage}{0.75\hsize}
\begin{defi}[$(T,F)$-systems and their infinitary trace semantics, \cite{hasuo07generictrace,jacobs04tracesemantics}]\label{def:coalginftr}
Let $\mathbb{C}$ be a category with a final object $1\in\mathbb{C}$.
A \emph{$(T,F)$-system} is a triple $\mathcal{X}=(X,s,c)$ consisting of
a \emph{state space} $X\in\mathbb{C}$,
a Kleisli arrow $s:1\kto X$ for  \emph{initial states}, and
%a Kleisli arrow 
$c :X\kto \overline{F}X$ for \emph{transition}.
%
%We further assume that each homset of the Kleisli category $\Kl(T)$ carries an order $\sqsubseteq$
%and an endofunctor $F$ and a monad $T$ on $\mathbb{C}$ constitute an
 %infinite trace situation with respect to the order.

Let us assume that the endofunctor $F$ and the monad $T$ on $\mathbb{C}$
 constitute an infinitary trace situation.
%  with respect to 
% an order $\sqsubseteq$.
The \emph{coalgebraic infinitary trace semantics} of a $(T,F)$-system
 $\mathcal{X}=(X,s,c)$ is the Kleisli arrow $\trinf(c)\odot s:1\kto Z$
 where $\trinf(c)$ is the largest homomorphism in Definition~\ref{def:infiniteTraceSituation}
 (see the diagram, in $\Kl(T)$, on the right).
Recall that $\odot$ denotes composition in $\Kl(T)$ (Definition~\ref{def:KleisliCategory}).
\end{defi}
\end{minipage}
%\begin{wrapfigure}{r}{3cm}
%\vspace*{-\intextsep}
\begin{minipage}{0.25\hsize}
\vspace{-5mm}
\begin{xy}
 \xymatrix@R=2.4em@C=1.8em{
 {\overline{F} X} \ar@{}[drr]|{=} \kar@{~>}[rr]^{\overline{F}(\trinf(c))} & & {\overline{F}Z}  \\
 {X} \kar[u]_{c}  \kar@{~>}^{\trinf(c)}[rr]  & & {Z} \kar[u]_{J\zeta}  \\
1 \kar[u]_{s} &   & }
\end{xy}
\end{minipage}\vspace{3mm}
%\end{wrapfigure}

Suppose that we are given two $(T,F)$-systems $\mathcal{X}=(X,s,c)$ and $\mathcal{Y}=(Y,t,d)$.
Let us say we aim to prove the inclusion between infinitary trace
semantics, that is, to show
% that the coalgebraic infinite trace semantics of $\mathcal{X}$ is included in that of $\mathcal{Y}$.
%Namely, we want to verify that
 $\trinf(c)\odot s\sqsubseteq \trinf(d)\odot t$
with respect to the order in the homset $\Kl(T)(1,Z)$.
%This situation is called \emph{(coalgebraic) infinite trace inclusion}.
Our goal in this paper is to offer \emph{Kleisli simulations} as a sound means
to do so.

The notions of \emph{forward} and \emph{backward Kleisli simulation} are introduced in~\cite{hasuo06genericforward}
as a categorical generalization of forward or backward simulations in~\cite{lynch95forwardand}.
They are defined as Kleisli arrows between (the state spaces of) two
$(T,F)$-system that are subject to certain inequalities---in short they
are \emph{lax/oplax coalgebra
homomorphisms}. In~\cite{hasuo06genericforward} they are
shown to be sound with respect to \emph{finite} trace semantics---the
languages of finite words, concretely, and the unique arrow to a
lifted initial algebra (that is a final coalgebra,
see~\cite{hasuo07generictrace} and the introduction), abstractly. In this
paper we are interested in their relation to \emph{infinitary} trace semantics.
% They satisfy soundness, namely, 
% the existence of forward or backward Kleisli simulation witness coalgebraic \emph{finite} trace inclusion.
% In the later sections, we will see that forward and backward Kleisli simulation can be also used to witness infinite trace inclusion defined above.

\vspace{5mm}\noindent\begin{minipage}{0.8\hsize}
\vspace{-2mm}
\begin{defi}[forward and backward Kleisli simulation, \cite{hasuo06genericforward}]\label{def:fwdbwdsim}
Let $F$ be an endofunctor and $T$ be a monad on $\mathbb{C}$ 
such that each homset of $\Kl(T)$ carries an order $\sqsubseteq$.
%that constitute an infinite trace situation with respect to an order $\sqsubseteq$.
Let $\mathcal{X}=(X,s,c)$ and $\mathcal{Y}=(Y,t,d)$ be $(T,F)$-systems.

A \emph{forward Kleisli simulation} from $\mathcal{X}$ to $\mathcal{Y}$
 is a Kleisli arrow $f:Y\kto X$ that satisfies the following conditions
 (see the diagram):
\begin{equation*}
s\sqsubseteq f\odot t, \;\;\;\text{and}\;\;\;
c\odot f \sqsubseteq \overline{F}f\odot d.
\end{equation*}
We write $\mathcal{X}\fwd\mathcal{Y}$ if there exists a forward simulation from $\mathcal{X}$ to $\mathcal{Y}$.

A \emph{backward Kleisli simulation} from $\mathcal{X}$ to $\mathcal{Y}$
 is a Kleisli arrow 
 $b:X\kto Y$ that satisfies the following conditions
 (see the diagram):
\begin{equation*}
b\odot s\sqsubseteq t, \;\;\;\text{and}\;\;\;
\overline{F}b\odot c \sqsubseteq d\odot b.
\end{equation*}
We write $\mathcal{X}\bwd\mathcal{Y}$ if there exists a backward simulation from $\mathcal{X}$ to $\mathcal{Y}$.
\end{defi}
\end{minipage}
\begin{minipage}{0.2\hsize}
\vspace*{-\intextsep}
 \begin{xy}
 \xymatrix@R=1.4em@C=1.2em{
 {\overline{F} X} \ar@{}[drr]|{\sqsubseteq}  & & {\overline{F} Y} \kar[ll]_{\overline{F} f } \\
 {X} \kar[u]_{c} \ar@{}[drr]|{\sqsubseteq} & & {Y} \kar[u]_{d} \ar_(.3){f}|-*\dir{|}[ll]  \\
 {} \ar@{}[ru]|(.10)*\dir{-} & {1} \ar`l[lu][lu]^(0){s}  \ar`r[ru][ru]_(0){t}  & \ar@{}[lu]|(.10)*\dir{-} }
 \end{xy}
 \vspace{1mm}
 \begin{xy}
 \xymatrix@R=1.4em@C=1.2em{
 {\overline{F} X} \ar@{}[drr]|{\sqsubseteq} \kar[rr]^{\overline{F} b } & & {\overline{F} Y}  \\
 {X} \kar[u]_{c} \ar@{}[drr]|{\sqsubseteq} \ar^(.3){b}|-*\dir{|}[rr]  & & {Y} \kar[u]_{d}  \\
 {} \ar@{}[ru]|(.10)*\dir{-} & {1} \ar`l[lu][lu]^(0)s \ar`r[ru][ru]_(0)t  & \ar@{}[lu]|(.10)*\dir{-} }
 \end{xy}
\end{minipage}
\vspace{1mm}

\emph{Forward partial execution} (FPE) is a transformation of 
$(T,F)$-systems introduced in~\cite{urabeH14genericforward} 
(and its extended version~\cite{urabeH17matSim})
for the
purpose of aiding discovery of Kleisli simulations.
Intuitively, it ``executes'' the given system by one step.

\begin{defi}[FPE, \cite{urabeH14genericforward,urabeH17matSim}]\label{def:fpe}
Let $F$ be an endofunctor and $T$ be a monad on $\mathbb{C}$.
\emph{Forward partial execution} (FPE) is a transformation that takes 
a $(T,F)$-system 
$\mathcal{X}=(X,s:1\kto X,c:X\kto\overline{F}X)$ 
%and $X_1\in X$ such that there exists $X_2\in X$ satisfying $X=X_1+X_2$
as input,
and returns a $(T,F)$-system 
$\mathcal{X}_{\FPE}=(\overline{F}X,c\odot s:1\kto\overline{F}X,\overline{F}c:\overline{F}X\kto\overline{F}^2X)$ 
as output.
%Here, $c_i:X_i\kto \overline{F}X$ ($i\in\{1,2\}$) is defined by 
%$c_i=c\odot \kappa_i$ where $\kappa_i:X_i\kto X=X_1+X_2$ is an injection.
\end{defi}
\noindent
It is shown in~\cite{urabeH17matSim} that FPE is a valid technique for
establishing  inclusion of \emph{finite} trace semantics, in the
 technical senses of \emph{soundness} and \emph{adequacy}. Soundness 
asserts that  discovery of a Kleisli simulation after applying FPE
indeed witnesses trace inclusion between the original systems; adequacy
asserts that if there is a Kleisli simulation between the original
systems,
then there is one, too, between the transformed systems. In this paper, naturally, we wish to 
establish the same results for \emph{infinitary} trace semantics.

\section{Systems with Nondeterministic Branching}\label{sec:powersetmonad}
%We call a \emph{nondeterministic tree automaton} for a
%$(\pow,\FSigma)$-system.
In the rest of the paper we develop a coalgebraic theory of infinitary
traces and (Kleisli) simulations---the main contribution of the paper. We do so separately for the
nondeterministic
setting ($T=\pow$), 
for the probabilistic setting ($T=\giry$),
and for the setting where the system can abort with an exception ($T=\lift$). 
This is
because of the difference in the constructions of infinitary traces, and
consequently in the soundness proofs. 

In this section we focus on the
nondeterministic setting; we assume that $F$ is a polynomial functor on $\Sets$.

\subsection{Construction of Infinitary Traces
%the Largest Homomorphism for  $(\pow,\FSigma)$-systems
}\label{subsec:constructlargesthompow}

We start with showing
that the combination of  polynomial $F$  and
$T=\pow$  constitute an
infinitary trace situation (Definition~\ref{def:infiniteTraceSituation}).
This is already known from~\cite{jacobs04tracesemantics}. 
The proof in~\cite{jacobs04tracesemantics} combines
fibrational intuitions with some constructions
that are
specific to $\Sets$. Here we present a
different proof. It exploits an order-theoretic structure of the
Kleisli category $\Kl(\pow)$; this will be useful later in showing
soundness of (restricted) backward simulations.
Our proof also paves the way to the probabilistic case in Section~\ref{sec:girymonad}.

In fact, our proof for %the constitution of 
infinitary trace situation
%Theorem~\ref{thm:inftrsituationpow} 
is stated
axiomatically, in the form of the following proposition. % marked with $\dagger$.
%that is 
(Recall that statements marked with $\dagger$ are axiomatic ones.)
This is potentially useful in identifying new examples other than the
combination of polynomial $F$ and $T=\pow$ (although we have not yet
managed to do so). 
% (We note,
% however, that we are yet to find examples other than polynomial $F$ and $T=\pow$.) 
%It
%The proposition
Its proof
 is essentially  the construction of a greatest fixed
point by transfinite induction~\cite{cousotC79constructiveversions}.
%Note that the following proposition is 
% is marked with $\dagger$ because it is formulated for a general $T$.
 % we start from the largest element in $\Kl(T)(X,Z)$, and calculate the largest homomorphism by iteration. 
\begin{prop}\label{prop:generalconstructiontop}$\!\!\!{}^\dagger$
Let $\mathbb{C}$ be a category, $F$ be an endofunctor on $\mathbb{C}$, 
and $T$ be a monad on $\mathbb{C}$.
%such that each homset of $\Kl(T)$ carries an order $\sqsubseteq$. 
%Let $J:\mathbb{C}\to\Kl(T)$ be a Kleisli inclusion functor.
 Assume the following conditions.
\begin{enumerate}
\item There exists a final $F$-coalgebra $\zeta:Z\to FZ$ in $\mathbb{C}$.
 \label{asm:generalconstructiontopFinalCoalg}
\item There exists a distributive law $\lambda:FT\Rightarrow TF$,
      yielding a lifting $\overline{F}$ on $\Kl(T)$ of $F$.
%, and hence $F$ can be lifted to an endofunctor $\overline{F}$ on $\Kl(T)$. 
\label{asm:generalconstructiontopDistLaw}
\item For each 
%pair of objects $X$ and $Y$ in $\Kl(T)$, 
$X,Y\in\Kl(T)$,
the homset $\Kl(T)(X,Y)$ carries a partial order $\sqsubseteq$.
Moreover, $\overline{F}$'s action on arrows, as well as composition of
      arrows in $\Kl(T)$, is monotone with respect to this order.
%and continuous (i.e.\ preserves the greatest lower bound). 
\label{asm:generalconstructiontopMonotone}
%\item One of the following  is satisfied.
%\begin{enumerate}
%\item  For each pair of objects $X$ and $Y$ in $\Kl(T)$,  
%every decreasing sequence $f_0\sqsupseteq f_1\sqsubseteq \ldots$ in $\Kl(T)(X,Y)$ has the greatest lower bound $\bigsqcap_{i\in\omega}f_i$.
%%The Kleisli category $\Kl(T)$ is $\omega\mathbf{Cpo}$-enriched in downward manner
%Moreover, composition in $\Kl(T)$ and a lifted functor $\overline{F}$ are  locally $\omega^{\text{op}}$-continuous, respectively.
%(i.e.\ for each $g:Z\kto X$ and $h:Y\kto W$, we have 
%$g\odot (\bigsqcap_{i\in\omega}f_i)=\bigsqcap_{i\in\omega}(g\odot f_i)$, 
%$(\bigsqcap_{i\in\omega}f_i)\odot h=\bigsqcap_{i\in\omega}(f_i\odot h)$, and
%$F(\bigsqcap_{i\in\omega}f_i)=\bigsqcap_{i\in\omega}(F f_i)$).
%\label{asm:generalconstructiontopGFPcont}
\item For each 
$X,Y\in \Kl(T)$,
%pair of objects $X$ and $Y$ in $\Kl(T)$,  
every  (possibly transfinite) decreasing sequence in $\Kl(T)(X,Y)$ has the greatest lower bound. 
That is: let $\mathfrak{a}$ be a ordinal and
      $(g_{\mathfrak{i}}:X\kto Y)_{\mathfrak{i}<\mathfrak{a}}$  be 
a family of arrows such that $\mathfrak{i}\leq \mathfrak{j}$ implies
      $g_{\mathfrak{i}}\sqsupseteq g_{\mathfrak{j}}$. 
      Then their infimum
$\bigsqcap_{\mathfrak{i}<\mathfrak{a}}g_{\mathfrak{i}}$ exists.
 
%  given a family of arrows $(g_{\mathfrak{i}}:X\kto Y)_{\mathfrak{i}<\mathfrak{a}}$ where $\mathfrak{a}$ is any limit ordinal and 
% $\mathfrak{i}\leq \mathfrak{j}$ implies $g_{\mathfrak{i}}\sqsupseteq g_{\mathfrak{j}}$,
% the greatest lower bound $\bigsqcap_{\mathfrak{i}<\mathfrak{a}}g_{\mathfrak{i}}$ exists.
%The Kleisli category $\Kl(T)$ is $\mathbf{Dcpo}$-enriched in downward manner
\label{asm:generalconstructiontopGFPtransfinite}
%\end{enumerate}
%\label{asm:generalconstructiontopGFP}
%(We denote this order by $\sqsubseteq$.)
\item For each  $X\in\mathbb{C}$, 
the homset $\Kl(T)(X,Z)$ has the largest element $\top_{X,Z}$.
%the lifting $J(!_X)$ of the unique arrow to the final object is the largest element of $\Kl(T)(X,1)$.
 \label{asm:generalconstructiontopTop}
\end{enumerate}
Then $T$ and $F$ constitute an infinitary trace situation with respect to $\sqsubseteq$.%$J\zeta:X\kto\overline{F}Z$ is a weakly final $\overline{F}$-coalgebra.
\end{prop}

%\begin{myproof}
\proofmark
%\proof
Let $c:X\kto \overline{F}X$ be an $\overline{F}$-coalgebra in $\Kl(T)$.
We shall construct the largest homomorphism $\trinf(c):X\kto Z$ from $c$
to $J\zeta$,
by transfinite induction.

%\begin{wrapfigure}[4]{r}{0pt}
%\vspace*{-\intextsep}
%\begin{xy}
% \xymatrix@R=1.5em@C=3.2em{
% {\overline{F}X} \ar@{}[dr]|{\sqsubseteq} \kar[r]^{\overline{F}\top_{X,Z}} & \overline{F}Z \\
% {X} \kar[u]^{c} \kar[r]_(.47){\top_{X,Z}} & {Z} \kar[u]^{J\zeta}_{\cong} }
%\end{xy}
%\end{wrapfigure}
%\vspace{1mm}
%\noindent\begin{minipage}{0.8\hsize}
We define an endofunction $\Phi_c:\Kl(T)(X,Z)\to\Kl(T)(X,Z)$ by $\Phi_c(f)=J\zeta^{-1}\odot \overline{F}f\odot c$.
By the monotonicity of Kleisli composition $\odot$ and the functor $\overline{F}$
 (Assumption~(\ref{asm:generalconstructiontopMonotone})), 
$\Phi_c$ is also monotone.
%Moreover, 
For each ordinal $\mathfrak{a}$, we define
$\Phi_c^{\mathfrak{a}}(\top_{X,Z})\in\Kl(T)(X,Z)$ by %the following 
transfinite induction
 on $\mathfrak{a}$ 
as follows:
\begin{itemize}
\item $\Phi_c^0(\top_{X,Z})= \top_{X,Z}$;
\item For a successor ordinal $\mathfrak{a}$, $\Phi_c^{\mathfrak{a}}(\top_{X,Z})=\Phi_c(\Phi_c^{\mathfrak{a}-1}(\top_{X,Z}))$; and
\item For a limit ordinal $\mathfrak{a}$,
      $\Phi_c^{\mathfrak{a}}(\top_{X,Z})=\bigsqcap_{\mathfrak{i}<\mathfrak{a}}\Phi^{\mathfrak{i}}_c(\top_{X,Z})$ (cf.\
      Assumption~(\ref{asm:generalconstructiontopGFPtransfinite})).
\end{itemize}
%\end{minipage}
%\begin{minipage}{0.2\hsize}
%
%\end{minipage}\vspace{1mm}
%We define an ordinal $\mathfrak{l}$ as follows.
%
%If Assumption~\ref{asm:generalconstructiontopGFPcont} is satisfied, then $\Phi_X^{\mathfrak{a}}$ is well-defined for all $\mathfrak{a}\leq\omega$.
%We define $\mathfrak{l}$ by $\mathfrak{l}=\omega$.
%By the $\omega^{\text{op}}$-continuity of composition in $\Kl(T)$ and $\overline{F}$, $\Phi_X$ is also $\omega$-continuous.
%Therefore $\Phi_X^{\mathfrak{l}}(f)$ is the greatest fixed point of $\Phi_X$.
%
%If Assumption~\ref{asm:generalconstructiontopGFPtransfinite} is satisfied, then $\Phi_X^{\mathfrak{a}}$ is well-defined for all ordinal $\mathfrak{a}$.
%
We define $\mathfrak{l}$ to be the smallest ordinal such that the cardinality of $\mathfrak{l}$ is greater than that of $\Kl(T)(X,Z)$.
Then from~\cite{cousotC79constructiveversions}, $\Phi_c^{\mathfrak{l}}(\top_{X,Z})$ is the greatest fixed point of $\Phi_c$.
%
%As a consequence, $\Phi_X^{\mathfrak{l}}(f)$ is the greatest fixed point of $\Phi_X$ in both cases.
Note that a Kleisli arrow is a homomorphism from $c$ to $J\zeta$ if and only if it is a fixed point of $\Phi_c$.
%This immediately implies that
Therefore
 $\Phi_c^{\mathfrak{l}}(\top_{X,Z})$ is the largest homomorphism from $c$ to $J\zeta$.
\qed
%\end{myproof}
\vspace{2mm}

Note that the $\omega^{\op}$-continuity---preservation of the 
greatest lower bound of a decreasing sequence---of composition $\odot$ in $\Kl(T)$ is not assumed.
%Note that it is not assumed that the Kleisli composition $\odot$ 
%preserves the greatest lower bound of a decreasing sequence.
This is because $\pow$---our choice for $T$ in this section---does not satisfy it,
while it satisfies $\omega$-continuity---preservation of
the least upper bound of an increasing sequence.
Indeed,  consider $f:X\kto Y$ and a decreasing sequence $(g_i:Y\kto
Z)_{i\in\omega}$, both
in $\Kl(\pow)$.
Then we have
$\left(\left(\bigsqcap_{i\in\omega}g_i\right)\odot f\right)(x)=\bigcup_{y\in f(x)}\bigcap_{i\in\omega} g_i(y)$
while $\left(\bigsqcap_{i\in\omega}(g_i\odot
f)\right)(x)=\bigcap_{i\in\omega}\bigcup_{y\in f(x)} g_i(y)$, and these two are
not equal in general. 
% (e.g.\ Example~\ref{example:needtransfinite}).
 This failure of $\omega^{\op}$-continuity prevents us from applying the (simpler)
 \emph{Kleene fixed-point theorem}, in which induction terminates after
 $\omega$ steps.
% This restriction prevents us from stopping the iteration at $\mathfrak{i}=\omega$ and require going further using transfinite induction.
The following examples show that
there does exist a nondeterministic automaton for which the largest
homomorphism is obtained after steps bigger than $\omega$.
%; see Example~\ref{example:needtransfinite}.
% where we need transfinite induction to obtain the largest homomorphism (e.g.\ Example~\ref{example:needtransfinite}).

\begin{exa}\label{example:needtransfinite}
In the construction of the largest homomorphism in Proposition~\ref{prop:generalconstructiontop},
we need $\omega+1$ steps %times of iterations 
%in order to construct the homomorphism 
for the nondeterministic automaton $\mathcal{X}$ on the left below. 
 We need $2\omega+1$ steps %times of iterations
% in order to construct the homomorphism
%a homomorphism 
for $\mathcal{Y}$ on the right below. 
%In the proof of Prop.~\ref{prop:generalconstructionnotop}, we used transfinite induction 
%twice---once for constructing a cone over a lifted final sequence, and once for constructing a homomorphism.
%We need $\omega+1$ times of iterations in order to construct a cone for a nondeterministic automaton $\mathcal{X}$ on the left below. 
%We need $2\omega+1$ times of iterations in order to construct a homomorphism for a nondeterministic automaton $\mathcal{Y}$ on the right below. 
In a similar manner, for an arbitrary ordinal $\mathfrak{a}$,
%by the transfinite induction on the ordinal $\mathfrak{a}$,
 we can construct an automaton where  $\mathfrak{a}$ steps %times of iteration 
 are needed.
% or a homomorphism for it.

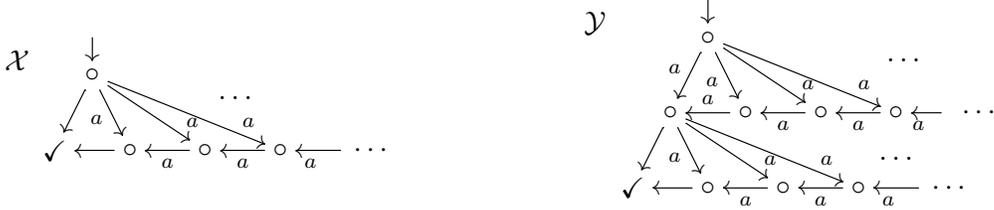
\begin{minipage}{0.5\hsize}
\begin{center}
\begin{xy}
(5,12)*{\mathcal{X}} = "",
%(10,0)*+{} = "xc",
(10,0)*+{\checkmark} = "x0",
(20,0)*+{\circ} = "x1",
(30,0)*+{\circ} = "x2",
(40,0)*+{\circ} = "x3",
(48,0)*{} = "x4",
(56,0)*{} = "x5",
(52,0)*{\cdots}="",
(34,7)*{\dots}="",
(15,10)*+{\circ} = "xo",
%\ar ^{} "x0";"xc"
\ar ^{} "x1";"x0"
\ar ^{a} "x2";"x1"
\ar ^{a} "x3";"x2"
\ar ^{a} "x4";"x3"
%\ar @{.}^{} "x5";"x4"
\ar _{} "xo";"x0"
\ar _{a} "xo";"x1"
\ar ^(.8){a} "xo";"x2"
\ar ^(.8){a} "xo";"x3"
\ar (15,15);"xo"
\end{xy}
\end{center}
\end{minipage}
\begin{minipage}{0.5\hsize}
\begin{center}
\begin{xy}
(5,22)*{\mathcal{Y}} = "",
%(0,0)*+{\checkmark} = "xc",
(10,0)*+{\checkmark} = "x0",
(20,0)*+{\circ} = "x1",
(30,0)*+{\circ} = "x2",
(40,0)*+{\circ} = "x3",
(48,0)*{} = "x4",
(56,0)*{} = "x5",
(15,10)*+{\circ} = "xo",
(25,10)*+{\circ} = "xo1",
(35,10)*+{\circ} = "xo2",
(45,10)*+{\circ} = "xo3",
(52,0)*{\cdots}="",
(56,10)*{\cdots}="",
(51,10)*{} = "xo4",
(61,10)*{} = "xo5",
(20,20)*+{\circ} = "x2o",
(45,4)*{\dots}="",
(46,17)*{\dots}="",
%\ar ^{} "x0";"xc"
\ar ^{} "x1";"x0"
\ar ^{a} "x2";"x1"
\ar ^{a} "x3";"x2"
\ar ^{a} "x4";"x3"
%\ar @{.}^{} "x5";"x4"
\ar _{} "xo";"x0"
\ar _{a} "xo";"x1"
\ar ^(.8){a} "xo";"x2"
\ar ^(.8){a} "xo";"x3"
\ar _{a} "xo1";"xo"
\ar ^{a} "xo2";"xo1"
\ar ^{a} "xo3";"xo2"
\ar ^{a} "xo4";"xo3"
%\ar @{.}^{} "xo5";"xo4"
\ar _{a} "x2o";"xo"
\ar _{a} "x2o";"xo1"
\ar ^(.8){a} "x2o";"xo2"
\ar ^(.8){a} "x2o";"xo3"
\ar (20,25);"x2o"
\end{xy}
\end{center}
\end{minipage}
\end{exa}

%As a result of employing transfinite induction,
%The functor and the monad we use in this section---$\FSigma$ and $\pow$---satisfy the Assumption of the above proposition.
%The assumption of the above proposition are satisfied by $\FSigma$ and $\pow$.

It is easy to check that all the assumptions in
Proposition~\ref{prop:generalconstructiontop} are satisfied by polynomial $F$
and $T=\pow$. 
Therefore we have the following result that is the main theorem of this section.

\begin{thm}\label{thm:inftrsituationpow}
The combination of  polynomial $F$  and
$T=\pow$  constitute an
infinitary trace situation. % (Definition~\ref{def:infiniteTraceSituation}).
% The functor $\FSigma$ and $\pow$ constitute an infinite trace situation.
%with respect to the order in Def.~\ref{def:orderenrichment}. 
%\qed
\end{thm}

\proof
We show that 
%a polynomial functor 
$F$ and $\pow$ satisfy Assumptions~(\ref{asm:generalconstructiontopFinalCoalg})--(\ref{asm:generalconstructiontopTop}) in Proposition~\ref{prop:generalconstructiontop}.

It is known that Assumption~(\ref{asm:generalconstructiontopFinalCoalg}) is satisfied~\cite{adamekK79leastfixed}.
It is  known that Assumption~(\ref{asm:generalconstructiontopDistLaw}) is also satisfied~\cite[Lemma~2.4]{hasuo07generictrace}.
It is easy to see that $F$ and $\pow$ on $\Sets$ satisfy the Assumptions~(\ref{asm:generalconstructiontopMonotone}) and 
(\ref{asm:generalconstructiontopGFPtransfinite}) where the infimum is given by intersection.
It is also easy to see that Assumption~(\ref{asm:generalconstructiontopTop}) is satisfied; $\top_{X,Z}:X\kto Z$ is given by $\top_{X,Z}(x)=Z$ for all $x\in X$.

Therefore by Proposition~\ref{prop:generalconstructiontop}, 
%polynomial 
$F$  and
$\pow$  constitute an
infinitary trace situation.
\qed

\subsection{Kleisli Simulations for Nondeterministic Systems}\label{subsec:Klsimpow}
In this section we prove that forward and backward Kleisli simulations 
%(Def.~\ref{def:fwdbwdsim}) 
can be used to witness \emph{infinitary} trace inclusion. 
This fact is already shown in~\cite{lynch95forwardand} 
for
%that forward and backward simulations 
%%forward or (total and image-finite) backward simulation between nondeterministic word automata
%are useful for checking 
%%witnesses 
%infinite trace inclusion between 
nondeterministic word automata. 
The coalgebraic theory developed in this section is a generalization of the results in~\cite{lynch95forwardand} 
because %the functor $F$ can be an arbitrary polynomial functor, therefore 
it is 
%it is 
applicable  
%not only to nondeterministic word automata but
 also to nondeterministic \emph{tree} automata.
  
\subsubsection{Forward Simulations}
Soundness of forward simulation is not hard; we do not have to go
into the  construction in Proposition~\ref{prop:generalconstructiontop}.
% because it can be proved without going through a general construction in Prop.~\ref{prop:generalconstructiontop}.

\begin{thm}\label{thm:soundnessFwdpow}
Given  two $(\pow,F)$-systems $\mathcal{X}=(X,s,c)$ and $\mathcal{Y}=(Y,t,d)$,
$\mathcal{X}\fwd\mathcal{Y}$ implies $\trinf(c)\odot s\sqsubseteq \trinf(d)\odot t$.
\qed
\end{thm}

The proof, 
much like Proposition~\ref{prop:generalconstructiontop},
%again, 
is formulated as a general result, singling out some
sufficient axioms. 
%The proof is based on Knaster-Tarski theorem~\cite{tarski55latticetheoretical}.

% \begin{myproof}
% It is known that $\FSigma$ and $\pow$ satisfy the assumptions in Lem.~\ref{lem:soundnessFwd}~\cite{hasuo07generictrace} below.
% Therefore immediate from Thm.~\ref{thm:soundnessFwd}.
% \qed
% \end{myproof}

%

% Rem: assumption (2) is not necessary
\begin{lem}\label{lem:soundnessFwd}$\!\!\!{}^\dagger$
Let $F$ be an endofunctor and $T$ be a monad on $\mathbb{C}$; assume
 further that they constitute an infinitary trace situation (with respect to $\sqsubseteq$).
We assume the following conditions.
\begin{enumerate}
\item Each homset of $\Kl(T)$ is $\omega$-complete, that is, each
      increasing $\omega$-sequence in it has the least upper bound. 
\label{asm:soundnessFwd1}
% \item Composition of arrows in $\Kl(T)$ and $\overline{F}$ are respectively monotone. %(i.e.\ composition of Kleisli arrows and $\overline{F}$ preserve the order).
% \label{asm:soundnessFwd2}
\item Composition $\odot$ of arrows in $\Kl(T)$ and $\overline{F}$'s
      action on arrows are both  $\omega$-continuous (i.e.\ they
      preserve the least upper bound of an increasing $\omega$-sequence). 
%      It follows that they are both  monotone.
\label{asm:soundnessFwd3}
\end{enumerate}
For two $(T,F)$-systems  $\mathcal{X}=(X,s,c)$ and $\mathcal{Y}=(Y,t,d)$,
if $f:Y\kto X$ is a forward simulation from $\mathcal{X}$ to
 $\mathcal{Y}$, then $\trinf(c)\odot f\sqsubseteq \trinf(d)$.
As a consequence we have
% We assume the conditions in Lem.~\ref{lem:soundnessFwd}.
% For  two $(T,F)$-system $\mathcal{X}=(X,s,c)$ and $\mathcal{Y}=(Y,t,d)$,
% $\mathcal{X}\fwd\mathcal{Y}$ implies
 $\trinf(c)\odot s\sqsubseteq \trinf(d)\odot t$.
\end{lem}

%\noindent\begin{minipage}{0.68\hsize}
%\begin{myproof}

\proofmark
%\proof
Let $\zeta:Z\to FZ$ be a final $F$-coalgebra in $\mathbb{C}$.
%Let $f:Y\kto X$ be a forward Kleisli simulation from $\mathcal{X}$ to $\mathcal{Y}$. %(see the above diagram). 
We define a function $\Phi_d:\Kl(T)(Y,Z)\to\Kl(T)(Y,Z)$ by
$\Phi_d(g)=J\zeta^{-1}\odot \overline{F}g\odot d$; note that
$\zeta$ is a final coalgebra and hence an isomorphism.
% (recall that weakly final coalgebra $J\zeta$ is an isomorphism).
Then 
%\end{myproof}\end{minipage}
%\begin{minipage}{0.32\hsize}
%\vspace*{-6mm}
%{\small
%\begin{xy}
% \xymatrix@R=2.0em@C=1.3em{
% {\overline{F}Y} \ar@{}[drr]|{\sqsupseteq} \kar[rr]_{\overline{F} f } \kar@/^4mm/[rrrr]^{\overline{F}(\trinf(d))} & & {\overline{F} X} \ar@{}[drr]|{=} \kar[rr]_{\overline{F}(\trinf(c))} & & {\overline{F} Z} \\
% {Y} \kar@/_4mm/[rrrr]_(.8){\trinf(d)} \kar[u]_{d} \ar@{}[drr]|{\sqsupseteq} \ar^{f}|-*\dir{|}[rr]  & & {X} \kar[u]_{c} \kar[rr]^{\trinf(c)} & & {Z} \kar[u]_{J\zeta} \\
% {} \ar@{}[ru]|(.10)*\dir{-} & {1} \ar`l[lu][lu]^(0)t \ar`r[ru][ru]_(0)s  & \ar@{}[lu]|(.10)*\dir{-} & & }
% \end{xy}}
%\end{minipage}
\begin{align*}
\trinf(c)\odot f 
%&= J\zeta^{-1}\odot J\zeta \odot \trinf(c)\odot f & (J\zeta\text{ is an isomorphism}) \\
&=J\zeta^{-1}\odot \overline{F}(\trinf(c))\odot c \odot f & (\text{$\trinf(c)$ is a homomorphism}) \\ 
&\sqsubseteq J\zeta^{-1}\odot \overline{F}(\trinf(c))\odot \overline{F}f \odot d & (\text{$f$ is a forward simulation}) \\
&= \Phi_d(\trinf(c)\odot f) & (\text{by definition of $\Phi_{Y}$}).
%&\sqsubseteq (J\zeta)^{-1}\odot \overline{F}(\trinf(c))\odot \overline{F}f \odot f & (\trinf(c)\text{ is a homomorphism}) \\ 
%&= \Phi_Y(\trinf(c)\odot f) & (\text{by definition}).
\end{align*}

\begin{minipageparindent}
\begin{wrapfigure}[8]{r}{0pt}
\vspace*{-\intextsep}
\raisebox{-5mm}[0pt][0cm]{
{\small
\begin{xy}
 \xymatrix@R=2.0em@C=1.3em{
 {\overline{F}Y} \ar@{}[drr]|{\sqsupseteq} \kar[rr]_{\overline{F} f } \kar@/^4mm/[rrrr]^{\overline{F}(\trinf(d))} & & {\overline{F} X} \ar@{}[drr]|{=} \kar[rr]_{\overline{F}(\trinf(c))} & & {\overline{F} Z} \\
 {Y} \kar@/_4mm/[rrrr]_(.8){\trinf(d)} \kar[u]_{d} \ar@{}[drr]|{\sqsupseteq} \ar^{f}|-*\dir{|}[rr]  & & {X} \kar[u]_{c} \kar[rr]^{\trinf(c)} & & {Z} \kar[u]_{J\zeta} \\
 {} \ar@{}[ru]|(.10)*\dir{-} & {1} \ar`l[lu][lu]^(0)t \ar`r[ru][ru]_(0)s  & \ar@{}[lu]|(.10)*\dir{-} & & }
 \end{xy}}}
%\hbox{}
% \vspace{-30pt}
\end{wrapfigure}
%
%This means that $\trinf(c)\odot f$ is a prefixed
By the assumption that $\overline{F}$ and the composition are   monotone,
$\Phi_d$ is also monotone.
%Moreover, it is easy to see that $\Kl(\pow)(Y,Z)$ is a complete lattice whose
%join and meet are given by element-wise union and intersection, respectively.
%
Therefore by repeatedly applying  $\Phi_d$ to the both sides of the above inequality, we obtain an increasing sequence
$\trinf(c)\odot f\sqsubseteq\Phi_d(\trinf(c)\odot f)\sqsubseteq\Phi_d^2(\trinf(c)\odot f)\sqsubseteq\cdots$ in $\Kl(T)(Y,Z)$.

%By the assumption that each homset of $\Kl(T)$ is $\omega$-complete, % in upward manner,
As $\Kl(T)(Y,Z)$ is $\omega$-complete,
the least upper bound $\bigsqcup_{i<\omega}\Phi^i(\trinf(c)\odot f)$ exists.
By the assumption that $\overline{F}$ and $\odot$ are both  $\omega$-continuous,
$\Phi_d$ is also $\omega$-continuous.
Therefore we have
%\begin{displaymath}
$\Phi(\bigsqcup_{i<\omega}\Phi^i(\trinf(c)\odot f))=\bigsqcup_{i<\omega}\Phi^{i+1}(\trinf(c)\odot f)=\bigsqcup_{i<\omega}\Phi^i(\tr^{\infty}_{z}(c)\odot f)$. % and
%\end{displaymath}
%holds.
This means that 
%$\Phi_Y^{\omega}(\trinf(c)\odot f)$ 
$\bigsqcup_{i<\omega}\Phi^i(\trinf(c)\odot f)$ is a fixed point of $\Phi_Y$, hence 
a homomorphism from $d$ to $J\zeta$.
As $\trinf(d)$ is the largest homomorphism from $d$ to $J\zeta$, this implies 
$\trinf(c)\odot f \sqsubseteq \bigsqcup_{i<\omega}\Phi^i(\trinf(c)\odot f)\sqsubseteq\trinf(d)$.
Combining with the assumption that $f$ is a forward simulation (specifically its
condition on initial states), 
we have $\trinf(c)\odot s \sqsubseteq \trinf(c)\odot f\odot t \sqsubseteq \trinf(d)\odot t$.
%Hence
%\begin{align*}
%\trinf(c)\odot s
%&\sqsubseteq \trinf(c)\odot f\odot t &(s \text{ is a fwd. simulation})\\
%&\sqsubseteq \trinf(d)\odot t &(\text{above result})
%\end{align*}
%This concludes the proof.
\qed
%\vspace{3mm}
\end{minipageparindent}
\vspace{1mm}
%\end{myproof}

% This lemma immediately implies the following---soundness of forward Kleisli simulation.
% %
% \begin{thm}\label{thm:soundnessFwd}
% We assume the conditions in Lem.~\ref{lem:soundnessFwd}.
% For  two $(T,F)$-system $\mathcal{X}=(X,s,c)$ and $\mathcal{Y}=(Y,t,d)$,
% $\mathcal{X}\fwd\mathcal{Y}$ implies $\trinf(c)\odot s\sqsubseteq \trinf(d)\odot t$.
% \qed
% \end{thm}
%

\auxproof{
\begin{wrapfigure}[8]{r}{0pt}
\vspace*{-\intextsep}
\begin{xy}
 \xymatrix@R=2.3em@C=2.8em{
{\overline{F}X} \ar@{}[dr]|{\sqsubseteq} \kar[r]_{\overline{F}b} \kar@/^5mm/[rr]^{\overline{F}\top_{X,Z}} & {\overline{F}Y} \ar@{}[dr]|{\sqsubseteq} \kar[r]_{\overline{F}\top_{Y,Z}} & \overline{F}Z \\
 {X} \kar[u]^{c} \kar[r]^{b} \kar@/_5mm/[rr]_{\top_{X,Z}} &  {Y} \kar[u]^{d} \kar[r]^(.47){\top_{Y,Z}}  & {Z} \kar[u]^{J\zeta}_{\cong} }
\end{xy}
%\hbox{}
\end{wrapfigure}
\proofmark%\begin{myproof}
Let $\zeta:Z\to FZ$ be a final $F$-coalgebra in $\mathbb{C}$.
%and $b:X\kto Y$ be a restricted backward simulation from $\mathcal{X}$ to $\mathcal{Y}$.
We define  $\Phi_X:\Kl(T)(X,Z)\to\Kl(T)(X,Z)$ and $\Phi_Y:\Kl(T)(Y,Z)\to\Kl(T)(Y,Z)$ as in the proof of Proposition~\ref{prop:generalconstructiontop}. 
Moreover, 
in the same manner as in
%similarly to
%in the same manner as 
the proof of Proposition~\ref{prop:generalconstructiontop},
for each 
%$g:X\kto Z$ and  
ordinal $\mathfrak{a}$, we define $\Phi_X^{\mathfrak{a}}(\top_{X,Z}):X\kto Z$ and $\Phi_Y^{\mathfrak{a}}(\top_{Y,Z}):Y\kto Z$ by the transfinite induction on $\mathfrak{a}$.
As we have seen in the proof of Proposition~\ref{prop:generalconstructiontop}, 
there exist ordinals $\mathfrak{l}_c$ and $\mathfrak{l}_d$ such that
$\trinf(c)=\Phi_X^{\mathfrak{l}_c}(\top_{X,Z})$ and $\trinf(d)=\Phi_Y^{\mathfrak{l}_d}(\top_{Y,Z})$. Let $\mathfrak{l}=\max(\mathfrak{l}_c,\mathfrak{l}_d)$.
%\end{myproof}
%\end{minipage}
%\begin{minipage}{0.3\hsize}{\small
%\begin{xy}
% \xymatrix@R=2.3em@C=2.8em{
%{\overline{F}X} \ar@{}[dr]|{\sqsubseteq} \kar[r]_{\overline{F}b} \kar@/^5mm/[rr]^{\overline{F}\top_{X,Z}} & {\overline{F}Y} \ar@{}[dr]|{\rotatebox{270}{\sqsubseteq}} \kar[r]_{\overline{F}\top_{Y,Z}} & \overline{F}Z \\
% {X} \kar[u]^{c} \kar[r]^{b} \kar@/_5mm/[rr]_{\top_{X,Z}} &  {Y} \kar[u]^{d} \kar[r]^(.47){\top_{Y,Z}}  & {Z} \kar[u]^{J\zeta}_{\rotatebox{270}{$\cong$}} }
%\end{xy}}
%\end{minipage}\vspace{1mm}
%
%From assumption~\ref{item:restrictedbwdsim2}, we have $\top_{X,Z}=\top_{Y,Z}\odot b$.
%Then by transfinite induction, we can prove $\trinf(c)=\Phi_X^{\mathfrak{l}}(\top_{X,Z})=\Phi_X^{\mathfrak{l}}(\top_{Y,Z}\odot b)$.
%
We shall now prove by 
transfinite induction that, 
for each 
%ordinal 
$\mathfrak{a}$, we have
$\Phi_X^{\mathfrak{a}}(\top_{X,Z})\sqsubseteq\Phi_Y^{\mathfrak{a}}(\top_{Y,Z})\odot
b$; this will yield our goal by taking $\mathfrak{a}=\mathfrak{l}$.

% by the
% transfinite induction on $\mathfrak{a}$ as follows.

For $\mathfrak{a}=0$, from Assumption~\ref{item:restrictedbwdsim2top} of
Definition~\ref{def:restrictedbwdsimtop}, we have 
$\Phi_X^{\mathfrak{a}}(\top_{X,Z})=\top_{X,Z} = \top_{Y,Z}\odot b=\Phi_Y^{\mathfrak{a}}(\top_{Y,Z})\odot b$.

Assume that $\mathfrak{a}$ is a successor ordinal and $\Phi_X^{\mathfrak{a-1}}(\top_{X,Z})\sqsubseteq\Phi_Y^{\mathfrak{a-1}}(\top_{Y,Z})\odot b$.
Then %we have
\begin{align*}
\Phi^{\mathfrak{a}}_c(\top_{X,Z})
%&= J\zeta^{-1}\odot \overline{F}(\Phi^{\mathfrak{a}-1}_X(\top_{X,Z})) \odot d & (\text{by definition of $\Phi_X$ }) \\
&\sqsubseteq J\zeta^{-1}\odot \overline{F}(\Phi^{\mathfrak{a}-1}_d(\top_{Y,Z}))\odot \overline{F}b \odot c  & (\text{by induction hypothesis}) \\
&\sqsubseteq  \Phi^{\mathfrak{a}}_d(\top_{Y,Z})\odot b & (\text{$b$ is a bwd.\ simulation})\,.
%&\sqsubseteq J\zeta^{-1}\odot \overline{F}(\Psi^{\mathfrak{a}-1}_Y(\top_{Y,Z}))\odot d\odot b & (\text{$b$ is a bwd.\ simulation}) \\
%&=  \Phi^{\mathfrak{a}}_Y(\top_{Y,Z})\odot b & (\text{by definition of $\Phi_Y$})\,.
\end{align*}

Let $\mathfrak{a}$ be a limit ordinal and assume that $\Phi^{\mathfrak{i}}_c(\top_{X,Z})\sqsubseteq \Psi^{\mathfrak{i}}_d(\top_{Y,Z})\odot b$ for all $\mathfrak{i}<\mathfrak{a}$.
Then 
\begin{align*}
\Phi_X^{\mathfrak{a}}(\top_{X,Z})
%&= \bigsqcap_{\mathfrak{i}<\mathfrak{a}}\Phi_X^{\mathfrak{i}}(\top_{X,Z}) & (\text{by definition}) \\
&\sqsubseteq\textstyle \bigsqcap_{\mathfrak{i}<\mathfrak{a}}\left(\Phi_Y^{\mathfrak{i}}(\top_{Y,Z}) \odot b\right) & (\text{by induction hypothesis}) \\
&= \Phi^{\mathfrak{a}}_d(\top_{Y,Z})\odot b & (\text{by
 Assumption~\ref{item:restrictedbwdsim3top} of Definition~\ref{def:restrictedbwdsimtop}})\,.
%&=\left(\bigsqcap_{\mathfrak{i}<\mathfrak{a}}\Phi_Y^{\mathfrak{i}}(\top_{Y,Z})\right) \odot b & (\text{by assumption~\ref{item:restrictedbwdsim3top}}) \\
%&= \Phi^{\mathfrak{a}}_Y(\top_{Y,Z})\odot b & (\text{by definition})\,.
\end{align*} 
%Therefore we have $\Phi^{\mathfrak{a}}_X(\top_{X,Z})\sqsubseteq \Phi^{\mathfrak{a}}_Y(\top_{Y,Z})\odot b$ for each ordinal $\mathfrak{a}$.
%Letting $\mathfrak{a}=\mathfrak{l}$, 
Thus $\trinf(c)\sqsubseteq
\trinf(d)\odot b$. The last claim follows from $b$'s condition on
initial states.
\qed\vspace{3mm}
}

%{\proof[Proof of Thm.\ \ref{thm:soundnessFwdpow}]
%Immediate from Lem.~\ref{lem:soundnessFwd}
%% and 
%%the assumption that $f$ is a forward simulation (its
%%condition on initial states).
%\qed
%\vspace{3mm}

It is known from~\cite{hasuo07generictrace} that the combination of polynomial $F$ and
$T=\pow$ satisfy the conditions of Lemma~\ref{lem:soundnessFwd}. Hence we
obtain Theorem~\ref{thm:soundnessFwdpow}, i.e.\ soundness of forward
simulation in the nondeterministic setting.

\subsubsection{Backward Simulations}\label{subsubsec:powbwd}
Next we 
%focus on
%want to show the soundness of 
wish to establish soundness of
\emph{backward} Kleisli simulations
with respect to \emph{infinitary} traces (for finite traces it is shown in~\cite{hasuo06genericforward}). 
In fact, the desired soundness fails in general:
here are counterexamples.
%---a counterexample is in
%Example~\ref{example:bwdsimnotsound}. 

\begin{exa}\label{example:bwdsimnotsound}
There exists a (not total) backward simulation from the nondeterministic automaton 
$\mathcal{X}$ to $\mathcal{Y}$ below.
Concretely, the backward simulation $b:\{x_0,x_1,x_2\}\to\pow(\{y_0,y_1\})$ is given by
$b(x_0)=\{y_0\}$, $b(x_1)=\{y_1\}$ and $b(x_2)=\emptyset$.
However, the simulated automaton $\mathcal{X}$ outputs an infinite word $aaa\ldots$ while $\mathcal{Y}$ does not.
Therefore the infinitary traces of $\mathcal{X}$ are not included in those of $\mathcal{Y}$.
%We can see that the backward simulation between them does not satisfy assumption~\ref{item:restrictedbwdsim2top} 
%in the definition of restricted backward simulation (
%of Def.~\ref{def:restrictedbwdsimtop}.

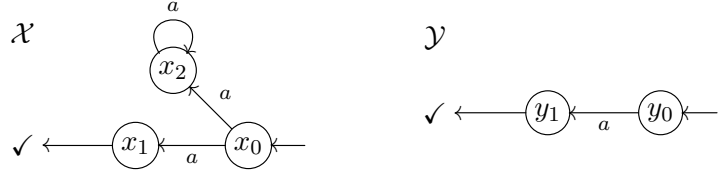
\begin{minipage}{0.5\hsize}
\begin{center}
\begin{xy}
(-25,0)*{} = "",
(0,15)*{\mathcal{X}} = "",
(0,0)*+{\checkmark} = "xc",
(15,0)*+[Fo]{x_1} = "x0",
(30,0)*+[Fo]{x_0} = "x1",
(20,10)*+[Fo]{x_2} = "x2",
\ar (37.5,0);"x1"
\ar @(ul,ur)^{a} "x2";"x2"
\ar ^{a} "x1";"x0"
\ar _{a} "x1";"x2"
\ar ^{} "x0";"xc"
\end{xy}
%
%\begin{xy}
%(-25,0)*{} = "",
%(0,10)*{\mathcal{X}} = "",
%(0,0)*+{\checkmark} = "xc",
%(10,0)*+{\circ} = "x0",
%(20,0)*+{\circ} = "x1",
%(13,7)*+{\circ} = "x2",
%\ar (25,0);"x1"
%\ar @(ul,ur)^{a} "x2";"x2"
%\ar ^{a} "x1";"x0"
%\ar _{a} "x1";"x2"
%\ar ^{} "x0";"xc"
%\end{xy}
\end{center}
\end{minipage}
\begin{minipage}{0.5\hsize}
\begin{center}
\begin{xy}
%(-25,0)*{} = "",
(0,10)*{\mathcal{Y}} = "",
(0,0)*+{\checkmark} = "xc",
(15,0)*+[Fo]{y_1} = "x0",
(30,0)*+[Fo]{y_0} = "x1",
%(13,7)*+{\circ} = "x2",
\ar (37.5,0);"x1"
%\ar @(ul,ur)^{a} "x2";"x2"
\ar ^{a} "x1";"x0"
%\ar _{a} "x1";"x2"
\ar ^{} "x0";"xc"
\end{xy}
%
%\begin{xy}
%%(-25,0)*{} = "",
%(0,10)*{\mathcal{Y}} = "",
%(0,0)*+{\checkmark} = "xc",
%(10,0)*+{\circ} = "x0",
%(20,0)*+{\circ} = "x1",
%%(13,7)*+{\circ} = "x2",
%\ar (25,0);"x1"
%%\ar @(ul,ur)^{a} "x2";"x2"
%\ar ^{a} "x1";"x0"
%%\ar _{a} "x1";"x2"
%\ar ^{} "x0";"xc"
%\end{xy}
\end{center}
\end{minipage}
%
%The following pair of nondeterministic automata $\mathcal{Z}$ and $\mathcal{W}$ 
There exists a (not image-finite) backward simulation from $\mathcal{Z}$ to $\mathcal{W}$ below.
Concretely, the backward simulation $b:\{z_0,z_1\}\to\pow(\{w_{\omega},w_0,w_1,\ldots\})$ is given by
$b(z_0)=\{w_{\omega}\}$ and $b(z_1)=\{w_0,w_1,\ldots\}$.
However, trace inclusion from $\mathcal{Z}$ to $\mathcal{W}$  does not hold.
%The automata $\mathcal{Z}$ and $\mathcal{W}$ below
%constitute another pair of such an example.
%The backward simulation from $\mathcal{Z}$ to $\mathcal{W}$ 
%does not satisfy assumption~\ref{item:restrictedbwdsim3top} 
%in the definition of restricted backward simulation (
%of Def.~\ref{def:restrictedbwdsimtop}.

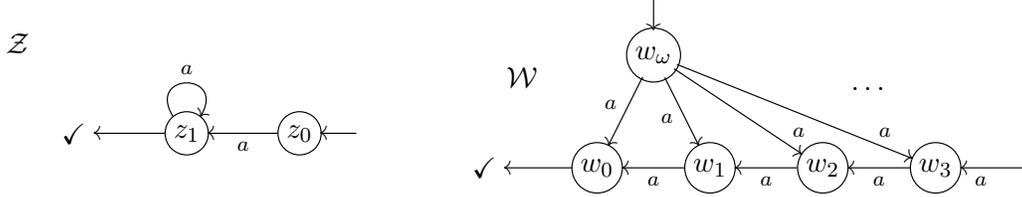
\begin{minipage}{0.4\hsize}
\begin{center}
\begin{xy}
(-10,0)*{} = "",
(-7.5,12)*{\mathcal{Z}} = "",
(0,0)*+{\checkmark} = "xc",
(15,0)*+[Fo]{z_1} = "xo",
(30,0)*+[Fo]{z_0} = "xo2",
\ar (37.5,0);"xo2"
\ar ^{a} "xo2";"xo"
\ar @(ul,ur)^{a} "xo";"xo"
\ar ^{} "xo";"xc"
\end{xy}
%
%\begin{xy}
%(-25,0)*{} = "",
%(-5,8)*{\mathcal{Z}} = "",
%(0,0)*+{\checkmark} = "xc",
%(10,0)*+{\circ} = "xo",
%(20,0)*+{\circ} = "xo2",
%\ar (25,0);"xo2"
%\ar ^{a} "xo2";"xo"
%\ar @(ul,ur)^{a} "xo";"xo"
%\ar ^{} "xo";"xc"
%\end{xy}
\end{center}
\end{minipage}
\begin{minipage}{0.6\hsize}
\begin{center}
\begin{xy}
(5,12)*{\mathcal{W}} = "",
(0,0)*+{\checkmark} = "xc",
(15,0)*+[Fo]{w_0}*+++[o]{} = "x0",
(30,0)*+[Fo]{w_1}*+++[o]{} = "x1",
(45,0)*+[Fo]{w_2}*+++[o]{} = "x2",
(60,0)*+[Fo]{w_3}*+++[o]{} = "x3",
(72,0)*{} = "x4",
(84,0)*{} = "x5",
(22.5,15)*+[Fo]{w_{\omega}} *+++[o]{}= "xo",
(51,10.5)*{\dots}="",
(117,0)*{\cdots}="",
\ar ^{} "x0";"xc"
\ar ^{a} "x1";"x0"
\ar ^{a} "x2";"x1"
\ar ^{a} "x3";"x2"
\ar ^{a} "x4";"x3"
%\ar @{.}^{} "x5";"x4"
\ar _{a} "xo";"x0"
\ar _{a} "xo";"x1"
\ar ^(.8){a} "xo";"x2"
\ar ^(.8){a} "xo";"x3"
\ar (22.5,22.5);"xo"
\end{xy}
%
%\begin{xy}
%(5,12)*{\mathcal{W}} = "",
%(0,0)*+{\checkmark} = "xc",
%(10,0)*+{\circ} = "x0",
%(20,0)*+{\circ} = "x1",
%(30,0)*+{\circ} = "x2",
%(40,0)*+{\circ} = "x3",
%(48,0)*{} = "x4",
%(56,0)*{} = "x5",
%(15,10)*+{\circ} = "xo",
%(34,7)*{\dots}="",
%(52,0)*{\cdots}="",
%\ar ^{} "x0";"xc"
%\ar ^{a} "x1";"x0"
%\ar ^{a} "x2";"x1"
%\ar ^{a} "x3";"x2"
%\ar ^{a} "x4";"x3"
%%\ar @{.}^{} "x5";"x4"
%\ar _{a} "xo";"x0"
%\ar _{a} "xo";"x1"
%\ar ^(.8){a} "xo";"x2"
%\ar ^(.8){a} "xo";"x3"
%\ar (15,15);"xo"
%\end{xy}
\end{center}
\end{minipage}
\end{exa}

Nevertheless, for nondeterministic word automata,
it is known that 
imposing certain restrictions (\emph{totality} and \emph{image-finiteness}) 
leads to
%on the simulation, we can ensure 
soundness of 
backward simulation~\cite{lynch95forwardand}.
In this section, for $(\pow,F)$-system with polynomial $F$, we prove a similar result in general coalgebraic terms.
This allows us to use (restricted) backward simulation to check language inclusion between 
not only nondeterministic \emph{word} automata but also nondeterministic \emph{tree} automata.
Moreover, this also gives us an idea about how we should impose restriction to make backward simulation sound in the probabilistic setting 
(Section~\ref{subsec:bwdsimprob}).
%impose certain restriction (\emph{totality} and \emph{image-finiteness})
%on backward simulation for nondeterministic word automata by Lynch and Vaandrager is sound with respect to infinite-length words if we 
%i ~\cite{lynch95forwardand}.

%Nevertheless, it turns out that we can impose
%certain restrictions on backward Kleisli simulations and ensure soundness.
%\marginpar{cite \cite{lynch95forwardand}}

% counterexamples show that soundness of backward Kleisli simulation with respect to infinite trace 
% cannot be proved as long as we put an assumption on a functor and a monad so that $\FSigma$ and $\pow$ are allowed.
% More concretely, there exist pairs of nondeterministic automata where a backward simulation exists but infinite language inclusion does not hold (see Example~\ref{example:bwdsimnotsound}).
% To recover this situation, we put some restriction on the definition of backward simulation.

\begin{defi}[totality, image-finiteness, TIF-backward simulation]\label{def:restrictedbwdsimpow}
%Let $F$ be an endofunctor and $T$ be a monad on $\mathbb{C}$ that 
%satisfy the conditions in Prop.~\ref{prop:generalconstructiontop} wrt.\ $\sqsubseteq$. 
%constitute an infinite trace situation with respect to an order $\sqsubseteq$.
%
Let $\mathcal{X}=(X,s,c)$ and $\mathcal{Y}=(Y,t,d)$ be
 $(\pow,F)$-systems. 
A backward simulation $b:X\kto Y$ 
from $\mathcal{X}$ to $\mathcal{Y}$ is \emph{total} if
$b(x)\neq\emptyset$ for all $x\in X$; it is \emph{image-finite} if
 $b(x)\subseteq Y$ is finite for all $x\in X$.
If $b$ satisfies both of the two conditions, it is called
a \emph{TIF-backward simulation}. 
We write $\mathcal{X}\tifbwd\mathcal{Y}$ if there exists a TIF-backward simulation from $\mathcal{X}$ to $\mathcal{Y}$.
\end{defi}

\begin{thm}[soundness of $\tifbwd$]\label{thm:soundnesstifbwdpow}
For  two $(\pow,F)$-systems $\mathcal{X}=(X,s,c)$ and $\mathcal{Y}=(Y,t,d)$,
$\mathcal{X}\tifbwd\mathcal{Y}$ implies $\trinf(c)\odot s\sqsubseteq \trinf(d)\odot t$.
%\qed
\end{thm}

The proof of Theorem~\ref{thm:soundnesstifbwdpow} is, yet again, via the following axiomatic development.
We first characterize \emph{totality} and \emph{image-finiteness} using categorical terms.
%Assumption~(\ref{item:restrictedbwdsim3top}) of
%%Def.~\ref{def:restrictedbwdsimtop} 
%the following definition
%resembles how ``finiteness'' is
%formulated in category theory, e.g.\ in the definition of \emph{finitary} objects.
%
\begin{defi}[TIF-backward simulation, generally]\label{def:restrictedbwdsimtop}$\!\!\!{}^\dagger$
Let $F$ be an endofunctor and $T$ be a monad on $\mathbb{C}$ that 
satisfy the conditions in Proposition~\ref{prop:generalconstructiontop} with respect to $\sqsubseteq$. 
%constitute an infinite trace situation with respect to an order $\sqsubseteq$.
%
For  two $(T,F)$-systems $\mathcal{X}=(X,s,c)$ and
 $\mathcal{Y}=(Y,t,d)$,
a \emph{TIF-backward simulation} 
from $\mathcal{X}$ to $\mathcal{Y}$ 
% a \emph{total and image finite backward Kleisli simulation} (\emph{TIF-backward simulation} for short) is 
is a backward simulation $b:X\kto Y$ 
that satisfies the following conditions.
\begin{enumerate}
\item The arrow $b:X\kto Y$ satisfies $\top_{Y,Z}\odot b = \top_{X,Z}$ for any $Z\in\Kl(T)$.
 \label{item:restrictedbwdsim2top}
%\item One of the following is satisfied.
%\begin{enumerate}
%\item Assumption~\ref{asm:generalconstructiontopGFPcont} in Prop.~\ref{prop:generalconstructiontop} is satisfied. \label{item:restrictedbwdsim3cont}
\item
\label{item:restrictedbwdsim3top}  
Precomposing $b:X\kto Y$ preserves the greatest lower bound of any decreasing transfinite sequence.
That is, let   $A\in\Kl(T)$, $\mathfrak{a}$ be an ordinal, and
       $(g_{\mathfrak{i}}:Y\kto A)_{\mathfrak{i}<\mathfrak{a}}$ be 
a family of Kleisli arrows such that $\mathfrak{i}\leq \mathfrak{j}$
       implies $g_{\mathfrak{i}}\sqsupseteq g_{\mathfrak{j}}$. Then
%$g_0\sqsupseteq g_1 \sqsupseteq g_2\ldots :Y\kto A$, 
we have $\bigsqcap_{\mathfrak{i}\in\mathfrak{a}}(g_\mathfrak{i}\odot
       b)=(\bigsqcap_{\mathfrak{i}\in\mathfrak{a}}g_\mathfrak{i})\odot
       b$.
\end{enumerate}
We write $\mathcal{X}\tifbwd\mathcal{Y}$ if there exists a TIF-backward simulation from $\mathcal{X}$ to $\mathcal{Y}$.
\end{defi}

Assumption~(\ref{item:restrictedbwdsim3top}) of
Definition~\ref{def:restrictedbwdsimtop} resembles how ``finiteness'' is
formulated in category theory, e.g.\ in the definition of \emph{finitely presented} objects.

This general TIF-backward simulation satisfies soundness. For its proof
we have to look into the inductive 
% It can be proved along the
 construction of the largest homomorphism in Proposition~\ref{prop:generalconstructiontop}.
% Section~\ref{subsec:constructlargesthompow}.

\begin{lem}\label{lem:soundnessRBwdtop}$\!\!\!{}^\dagger$
Let $F$ and $T$ be
%We assume that $F$ and $T$ 
as in Proposition~\ref{prop:generalconstructiontop}.
% wrt.\ $\sqsubseteq$. 
% that constitute an infinite trace situation with respect to an order $\sqsubseteq$.
%We assume that the weakly final coalgebra is obtained by using Prop.~\ref{prop:generalconstructiontop}. 
For two $(T,F)$-systems $\mathcal{X}=(X,s,c)$ and $\mathcal{Y}=(Y,t,d)$, 
$\mathcal{X}\tifbwd\mathcal{Y}$ (in the sense of Definition~\ref{def:restrictedbwdsimtop}) implies $\trinf(c)\sqsubseteq
 \trinf(d)\odot b$.
 Furthermore it follows that  $\trinf(c)\odot s\sqsubseteq \trinf(d)\odot t$.
\end{lem}
%
%\begin{minipage}{0.7\hsize}

\noindent\begin{minipageparindent}
\begin{wrapfigure}[8]{r}{0pt}
%\vspace*{-\intextsep}
\raisebox{-5mm}[0pt][0cm]{
\begin{xy}
 \xymatrix@R=2.3em@C=2.8em{
{\overline{F}X} \ar@{}[dr]|{\sqsubseteq} \kar[r]_{\overline{F}b} \kar@/^5mm/[rr]^{\overline{F}\top_{X,Z}} & {\overline{F}Y} \ar@{}[dr]|{\sqsubseteq} \kar[r]_{\overline{F}\top_{Y,Z}} & \overline{F}Z \\
 {X} \kar[u]^{c} \kar[r]^{b} \kar@/_5mm/[rr]_{\top_{X,Z}} &  {Y} \kar[u]^{d} \kar[r]^(.47){\top_{Y,Z}}  & {Z} \kar[u]^{J\zeta}_{\cong} }
\end{xy}}
%\vspace{1cm}
%\hbox{}
\end{wrapfigure}
\proofmark%\begin{myproof}
%\proof%
Let $\zeta:Z\to FZ$ be a final $F$-coalgebra in $\mathbb{C}$.
%and $b:X\kto Y$ be a restricted backward simulation from $\mathcal{X}$ to $\mathcal{Y}$.
We define  $\Phi_c:\Kl(T)(X,Z)\to\Kl(T)(X,Z)$ and $\Phi_d:\Kl(T)(Y,Z)\to\Kl(T)(Y,Z)$ as in the proof of Proposition~\ref{prop:generalconstructiontop}. 
Moreover, in the same manner as in
%similarly to
%in the same manner as 
the proof of Proposition~\ref{prop:generalconstructiontop},
for each 
%$g:X\kto Z$ and  
ordinal $\mathfrak{a}$, we define $\Phi_c^{\mathfrak{a}}(\top_{X,Z}):X\kto Z$ and $\Phi_d^{\mathfrak{a}}(\top_{Y,Z}):Y\kto Z$ by  transfinite induction on $\mathfrak{a}$.
As we have seen in the proof of Proposition~\ref{prop:generalconstructiontop}, 
there exist ordinals $\mathfrak{l}_c$ and $\mathfrak{l}_d$ such that
$\trinf(c)=\Phi_c^{\mathfrak{l}_c}(\top_{X,Z})$ and $\trinf(d)=\Phi_d^{\mathfrak{l}_d}(\top_{Y,Z})$. Let $\mathfrak{l}=\max(\mathfrak{l}_c,\mathfrak{l}_d)$.
%\end{myproof}
%\end{minipage}
%\begin{minipage}{0.3\hsize}{\small
%\begin{xy}
% \xymatrix@R=2.3em@C=2.8em{
%{\overline{F}X} \ar@{}[dr]|{\sqsubseteq} \kar[r]_{\overline{F}b} \kar@/^5mm/[rr]^{\overline{F}\top_{X,Z}} & {\overline{F}Y} \ar@{}[dr]|{\rotatebox{270}{\sqsubseteq}} \kar[r]_{\overline{F}\top_{Y,Z}} & \overline{F}Z \\
% {X} \kar[u]^{c} \kar[r]^{b} \kar@/_5mm/[rr]_{\top_{X,Z}} &  {Y} \kar[u]^{d} \kar[r]^(.47){\top_{Y,Z}}  & {Z} \kar[u]^{J\zeta}_{\rotatebox{270}{$\cong$}} }
%\end{xy}}
%\end{minipage}\vspace{1mm}
%
%From assumption~\ref{item:restrictedbwdsim2}, we have $\top_{X,Z}=\top_{Y,Z}\odot b$.
%Then by transfinite induction, we can prove $\trinf(c)=\Phi_X^{\mathfrak{l}}(\top_{X,Z})=\Phi_X^{\mathfrak{l}}(\top_{Y,Z}\odot b)$.
%
We shall now prove by 
transfinite induction that, 
for each 
%ordinal 
$\mathfrak{a}$, we have
$\Phi_c^{\mathfrak{a}}(\top_{X,Z})\sqsubseteq\Phi_d^{\mathfrak{a}}(\top_{Y,Z})\odot
b$; this will yield our goal by taking $\mathfrak{a}=\mathfrak{l}$.
\end{minipageparindent}
% by the
% transfinite induction on $\mathfrak{a}$ as follows.

For $\mathfrak{a}=0$, from Assumption~(\ref{item:restrictedbwdsim2top}) of
Definition~\ref{def:restrictedbwdsimtop}, we have 
$\Phi_c^{\mathfrak{a}}(\top_{X,Z})=\top_{X,Z} = \top_{Y,Z}\odot b=\Phi_d^{\mathfrak{a}}(\top_{Y,Z})\odot b$.

Assume that $\mathfrak{a}$ is a successor ordinal and $\Phi_c^{\mathfrak{a-1}}(\top_{X,Z})\sqsubseteq\Phi_d^{\mathfrak{a-1}}(\top_{Y,Z})\odot b$.
Then %we have
\begin{align*}
\Phi^{\mathfrak{a}}_c(\top_{X,Z})
%&= J\zeta^{-1}\odot \overline{F}(\Phi^{\mathfrak{a}-1}_X(\top_{X,Z})) \odot d & (\text{by definition of $\Phi_X$ }) \\
&\sqsubseteq J\zeta^{-1}\odot \overline{F}(\Phi^{\mathfrak{a}-1}_d(\top_{Y,Z}))\odot \overline{F}b \odot c  & (\text{by induction hypothesis}) \\
&\sqsubseteq 
 J\zeta^{-1}\odot \overline{F}(\Phi^{\mathfrak{a}-1}_d(\top_{Y,Z}))\odot d\odot b & (\text{$b$ is a backward simulation})\\
&=  \Phi^{\mathfrak{a}}_d(\top_{Y,Z})\odot b & (\text{by definition})\,.
%&\sqsubseteq J\zeta^{-1}\odot \overline{F}(\Psi^{\mathfrak{a}-1}_d(\top_{Y,Z}))\odot d\odot b & (\text{$b$ is a bwd.\ simulation}) \\
%&=  \Phi^{\mathfrak{a}}_Y(\top_{Y,Z})\odot b & (\text{by definition of $\Phi_Y$})\,.
\end{align*}

Let $\mathfrak{a}$ be a limit ordinal and assume that $\Phi^{\mathfrak{i}}_c(\top_{X,Z})\sqsubseteq \Phi^{\mathfrak{i}}_d(\top_{Y,Z})\odot b$ for all $\mathfrak{i}<\mathfrak{a}$.
Then 
\begin{align*}
\Phi_c^{\mathfrak{a}}(\top_{X,Z})
%&= \bigsqcap_{\mathfrak{i}<\mathfrak{a}}\Phi_X^{\mathfrak{i}}(\top_{X,Z}) & (\text{by definition}) \\
&\sqsubseteq\textstyle \bigsqcap_{\mathfrak{i}<\mathfrak{a}}\left(\Phi_Y^{\mathfrak{i}}(\top_{Y,Z}) \odot b\right) & (\text{by induction hypothesis}) \\
&= \Phi^{\mathfrak{a}}_d(\top_{Y,Z})\odot b & (\text{by
 Assumption~(\ref{item:restrictedbwdsim3top}) of Definition~\ref{def:restrictedbwdsimtop}})\,.
%&=\left(\bigsqcap_{\mathfrak{i}<\mathfrak{a}}\Phi_Y^{\mathfrak{i}}(\top_{Y,Z})\right) \odot b & (\text{by assumption~\ref{item:restrictedbwdsim3top}}) \\
%&= \Phi^{\mathfrak{a}}_Y(\top_{Y,Z})\odot b & (\text{by definition})\,.
\end{align*} 
%Therefore we have $\Phi^{\mathfrak{a}}_X(\top_{X,Z})\sqsubseteq \Phi^{\mathfrak{a}}_Y(\top_{Y,Z})\odot b$ for each ordinal $\mathfrak{a}$.
%Letting $\mathfrak{a}=\mathfrak{l}$, 
Thus $\trinf(c)\sqsubseteq
\trinf(d)\odot b$. The last claim follows from $b$'s condition on
initial states.
\qed\vspace{3mm}

%The following lemma states
Next we show that a TIF-backward
 simulation
in the specific sense of Definition~\ref{def:restrictedbwdsimpow} is also 
a TIF-backward
 simulation in the general sense of Definition~\ref{def:restrictedbwdsimtop}.
 To this end, we first prove the following ``pigeon-hole'' sublemma.
 
\begin{sublem}\label{sublem:ordinalramsey}
Let $\mathfrak{a}$ be a limit ordinal, $C$ be a finite set and $f:\mathfrak{a}\to C$.
Then 
\begin{displaymath}
\exists c\in C.\;\; \forall \mathfrak{i}<\mathfrak{a}.\;\; \exists \mathfrak{b}\geq\mathfrak{i}.\;\; f(\mathfrak{b})=c\,.
\end{displaymath}
%there exists $c\in C$ such that for all ordinals $\mathfrak{i}<\mathfrak{a}$, 
%there exists an ordinal $\mathfrak{b}\geq\mathfrak{i}$ such that $f(\mathfrak{b})=c$.
\end{sublem}

%\begin{myproof}
\proof
For each $c\in C$, we define $A_c\subseteq \mathfrak{a}$ by $A_c=\{\mathfrak{d}\in\mathfrak{a}\mid f(\mathfrak{d})=c\}$.
We prove the statement by contradiction.
%Assume the following.
Assume the negation of the claim, that is,
\begin{displaymath}
\forall c\in C.\;\; \exists \mathfrak{i}_c<\mathfrak{a}.\;\; \forall \mathfrak{b}\geq\mathfrak{i}_c.\;\; f(\mathfrak{b})\neq c\,.
\end{displaymath}
This is equivalent to assuming
that for all $c\in C$, there exists $\mathfrak{i}_c<\mathfrak{a}$ such that for all $\mathfrak{j}\in A_c$, $\mathfrak{j}<\mathfrak{i}_c$ holds.
Then $\bigsqcup_{\mathfrak{j}\in A_c}\mathfrak{j}\leq \mathfrak{i}_c<\mathfrak{a}$ for all $c\in C$.
As $C$ is finite and $\bigcup_{c\in C} A_c=\mathfrak{a}$, this implies 
$\mathfrak{a}=\bigsqcup_{\mathfrak{j}<\mathfrak{a}}\mathfrak{j}=
\bigsqcup_{c\in C}\bigsqcup_{\mathfrak{j}\in A_c}\mathfrak{j}\leq\bigsqcup_{c\in C}\mathfrak{i}_c<\mathfrak{a}$.
This contradicts and the statement is proved.
\qed
%\end{myproof}

\begin{lem}\label{lem:whenrestrictedpow}
In Definition~\ref{def:restrictedbwdsimtop}, let $T=\pow$ and $F$ be a polynomial functor.
%then
%\begin{enumerate}
%\item 
Assumption~(\ref{item:restrictedbwdsim2top}) is satisfied if $b(x)\neq\emptyset$ for each $x\in X$;
%\label{item:whenrestrictedpow2}
%\item 
%while 
Assumption~(\ref{item:restrictedbwdsim3top}) is satisfied if $b(x)$ is finite for each $x\in X$. 
%\label{item:whenrestrictedpow3}
%\qed
%\end{enumerate}
\end{lem}

%\newtheorem*{proof:whenrestrictedpow}{Proof of Prop.\ \ref{prop:whenrestrictedpow}}{\itshape}{\rmfamily}
%\begin{proof:whenrestrictedpow}
%{\proof[Proof of Lem.\ \ref{lem:whenrestrictedpow}]
%\begin{myproof}
\proof
%\noindent{\bf \ref{item:whenrestrictedpow2}.} 
Assume that $b(x)\neq\emptyset$ for all $x\in X$.
To show that Assumption~(\ref{item:restrictedbwdsim2top}) in Definition~\ref{def:restrictedbwdsimtop} is satisfied,
it suffices to prove $z\in \top_{Y,Z}\odot b(x)$ for all $z\in Z$ and $x\in X$.
By the assumption, there exists $y\in Y$ such that $y\in b(x)$.
Therefore for all $z\in Z$, $z \in \top_{Y,Z}(y)\subseteq \top_{Y,Z}\odot b(x)$.

%\noindent{\bf \ref{item:whenrestrictedpow3}.}
%Next we prove the latter half of the lemma.
Assume that $b(x)$ is finite for all $x\in X$.
%Next we assume that $b(x)$ is finite for all $x\in X$ and 
%prove that Assumption~(\ref{item:restrictedbwdsim3top}) in Def.~\ref{def:restrictedbwdsimtop} is satisfied.
%Here,
Note that $\bigsqcap_{\mathfrak{i}<\mathfrak{a}}(g_{\mathfrak{a}}\odot b)(x)=\bigcap_{\mathfrak{i}<\mathfrak{a}}\bigcup_{y\in b(x)} g_{\mathfrak{i}}(y)$
while $\bigl(\bigsqcap_{\mathfrak{i}<\mathfrak{a}}g_{\mathfrak{i}}\bigr)\odot b(x)=\bigcup_{y\in b(x)}\bigcap_{\mathfrak{i}<\mathfrak{a}} g_{\mathfrak{i}}(y)$.
As it is easily shown that the latter is always included in the former,
%Therefore 
it suffices to prove that $z\in \bigcap_{\mathfrak{i}<\mathfrak{a}}\bigcup_{y\in b(x)} g_{\mathfrak{i}}(y)$ implies 
 $z\in\bigcup_{y\in b(x)}\bigcap_{\mathfrak{i}<\mathfrak{a}} g_{\mathfrak{i}}(y)$.

%Assume that 
If $z\in \bigcap_{\mathfrak{i}<\mathfrak{a}}\bigcup_{y\in b(x)} g_{\mathfrak{i}}(y)$,
then for all $\mathfrak{i}<\mathfrak{a}$, there exists $y_{\mathfrak{i}}\in b(x)$ such that $z\in g_{\mathfrak{i}}(y_{\mathfrak{i}})$.
As $b(x)$ is assumed to be finite, from Sublemma~\ref{sublem:ordinalramsey}, we have
\begin{displaymath}
\exists y\in b(x).\;\; \forall \mathfrak{i}<\mathfrak{a}.\;\; \exists \mathfrak{j}\geq\mathfrak{i}.\;\; 
z\in g_{\mathfrak{j}}(y)\,.
\end{displaymath}
%for each $\mathfrak{i}<\mathfrak{a}$, 
%there exists an ordinal $\mathfrak{j}$ such that $\mathfrak{i}\leq \mathfrak{j}<\mathfrak{a}$ and 
%$z\in g_{\mathfrak{j}}(y)$.
As $\mathfrak{i}\leq \mathfrak{j}$ implies $g_{\mathfrak{i}} \sqsupseteq g_{\mathfrak{j}}$, 
%we have 
$z\in g_{\mathfrak{j}}(y)$ implies $z\in g_{\mathfrak{i}}(y)$.
Therefore $z\in\bigcup_{y\in b(x)}\bigcap_{\mathfrak{i}<\mathfrak{a}} g_{\mathfrak{i}}(y)$ holds.
\qed
%}
%\end{myproof}
%\end{proof:whenrestrictedpow}

%\newtheorem*{proof:soundnesstifbwdpow}{Proof of Thm.\ \ref{thm:soundnesstifbwdpow}}{\itshape}{\rmfamily}
%\begin{proof:soundnesstifbwdpow}
{\proof[Proof of Theorem \ref{thm:soundnesstifbwdpow}]
Immediate from Lemma~\ref{lem:soundnessRBwdtop} and Lemma~\ref{lem:whenrestrictedpow}.
%
%In Lem.~\ref{lem:whenrestrictedpow} we prove that a TIF-backward
% simulation
%in the specific sense of Def.~\ref{def:restrictedbwdsimpow} is also 
%a TIF-backward
% simulation in the general sense of Def.~\ref{def:restrictedbwdsimtop}.
% Therefore Lem.~\ref{lem:soundnessRBwdtop} yields trace inclusion.
% It is immediate from  Lem.~\ref{lem:soundnessRBwdtop} and that
% if $T=\pow$ and $F=\FSigma$, for a backward simulatio $b:X\kto Y$,
% Assumption~\ref{item:restrictedbwdsim2} In Def.~\ref{def:restrictedbwdsimtop} is satisfied if $b(x)\neq\emptyset$ for each $x\in X$, 
% while Assumption~\ref{item:restrictedbwdsim3} is satisfied if $b(x)$ is finite for each $x\in X$ (Lem.~\ref{lem:whenrestrictedpow}).
\qed}
%\end{proof:soundnesstifbwdpow}

Even with the additional constraints of totality and image-finiteness,
backward Kleisli simulations seems to be a viable method for establishing
infinitary trace inclusion,
because there exists a pair of nondeterministic automata such that a TIF-backward simulation can prove trace inclusion between them but
 a forward simulation cannot.
An example of such a pair of automata
%An example where a TIF-backward simulation can prove infinitary trace inclusion
is shown below.
%in Example~\ref{example:bwdsimnecessary}, where a fwd.\ simulation does not
%exist but a TIF-bwd.\ simulation does.

\begin{exa}\label{example:bwdsimnecessary}
The infinitary traces of the nondeterministic automata $\mathcal{X}$ below are included in the infinitary traces of $\mathcal{Y}$.
%The following pair of nondeterministic automata $\mathcal{X}$ and $\mathcal{Y}$ exhibits infinite trace inclusion.
There exists no forward simulation from $\mathcal{X}$ to $\mathcal{Y}$
% but
 while a TIF-backward simulation does exist.

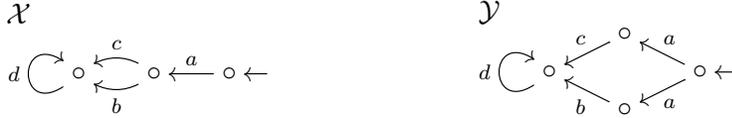
\begin{minipage}{0.5\hsize}
\begin{center}
\begin{xy}
(-25,0)*{} = "",
(-8,8)*{\mathcal{X}} = "",
(20,0)*+{\circ} = "x0",
(10,0)*+{\circ} = "x1",
(0,0)*+{\circ} = "x2",
\ar (25,0);"x0"
\ar _{a} "x0";"x1"
\ar @/^/ ^{b} "x1";"x2"
\ar @/_/ _{c} "x1";"x2"
\ar @(dl,ul)^{d} "x2";"x2"
\end{xy}
\end{center}
\end{minipage}
\begin{minipage}{0.5\hsize}
\begin{center}
\begin{xy}
%(-25,0)*{} = "",
(-8,8)*{\mathcal{Y}} = "",
(20,0)*+{\circ} = "x0",
(10,-5)*+{\circ} = "x11",
(10,5)*+{\circ} = "x12",
(0,0)*+{\circ} = "x2",
\ar (25,0);"x0"
\ar ^{a} "x0";"x11"
\ar _{a} "x0";"x12"
\ar ^{b} "x11";"x2"
\ar _{c} "x12";"x2"
\ar @(dl,ul)^{d} "x2";"x2"
\end{xy}
\end{center}
\end{minipage}
\end{exa}

% The TIF-backward simulations increases the number of pairs of automata where we can check infinite trace inclusion.
% Namely, there exists a pair of %nondeterministic 
% automata where forward  simulation does not exist but TIF-backward  simulation exists 
% (e.g.\ Example~\ref{example:bwdsimnecessary}).

\subsection{Forward Partial Execution for Nondeterministic
Systems}\label{subsec:FPEpow}
%We now apply \emph{forward partial execution} (FPE)~\cite{urabeH14genericforward}---a transformation of coalgebraic
%systems that potentially
%increases the likelihood of existence of simulations---in the current
%setting of nondeterminism and infinite traces. We follow the setting
%in~\cite{urabeH14genericforward} for the \emph{finite} traces, and
%formulate FPE's ``correctness'' in the following theorem.
%
Recall that in Definition~\ref{def:fpe}, we have reviewed \emph{forward partial execution} 
(FPE)~\cite{urabeH14genericforward,urabeH17matSim}---a transformation of coalgebraic
systems that potentially
increases the likelihood of existence of simulations.
We now apply FPE in the current
setting of nondeterminism and infinitary traces. We follow the setting
in~\cite{urabeH17matSim} for the \emph{finite} traces, and
formulate FPE's ``correctness'' in the following theorem.

\begin{thm}\label{thm:propertiesFPEpow}
Let $F$ be a polynomial functor on $\Sets$.
For $(\pow,F)$-systems $\mathcal{X}=(X,s,c)$ and $\mathcal{Y}=(Y,t,d)$, the following  hold.
\begin{enumerate}[itemsep=1ex]
\item
\begin{enumerate}
\item (soundness of FPE for forward simulation)  
$\mathcal{X}_{\FPE}\fwd\mathcal{Y}$ implies $\trinf(c)\odot s\sqsubseteq \trinf(d)\odot t$.
\label{item:soundnessfpefwdpow}
\item (adequacy of FPE for forward simulation) $\mathcal{X}\fwd\mathcal{Y}$ implies $\mathcal{X}_{\FPE}\fwd\mathcal{Y}$.
\label{item:adequacyfpefwdpow}
\end{enumerate}
\label{item:safpefwdpow}
\item
\begin{enumerate}
\item (soundness of FPE for backward simulation)  $\mathcal{X}\tifbwd\mathcal{Y}_{\FPE}$ implies 
$\trinf(c)\odot s\sqsubseteq \trinf(d)\odot t$.%
\label{item:soundnessfpebwdtoppow}
\item (adequacy of FPE for backward simulation) 
$\mathcal{X}\tifbwd\mathcal{Y}$ implies
      $\mathcal{X}\tifbwd\mathcal{Y}_{\FPE}$, assuming that the following hold.
\begin{enumerate}
\item 
$d(y)\neq\emptyset$ for all $y\in Y$.
\label{item:restrictedfpe2toppow}
\item 
$d(y)$ is finite for all $y\in Y$.
\label{item:restrictedfpe3toppow}
%\qed
\end{enumerate}
\label{item:adequacyfpebwdtoppow}
\end{enumerate}
\label{item:safpebwdpow}
\end{enumerate}
\end{thm}
Informally: \emph{soundness} means that discovery of a simulation after applying FPE
still witnesses the trace inclusion between the original systems; and
\emph{adequacy} means that the relationship $\fwd$ (or $\tifbwd$) is not
destroyed by application of FPE. The theorem also implies that FPE must
be applied to the ``correct side'' of the desired trace inclusion $\langinf(\mathcal{X})\sqsubseteq\langinf(\mathcal{Y})$: 
$\mathcal{X}$ in the search for a forward simulation; and 
$\mathcal{Y}$ in the search for a backward one.

Note that the adequacy property is independent from the choice of trace
semantics (finite or infinitary). Therefore the
statement~(\ref{item:adequacyfpefwdpow}) of
Theorem~\ref{thm:propertiesFPEpow} is the same as its counterpart
in~\cite{urabeH17matSim}. For the
statement~(\ref{item:adequacyfpebwdtoppow}), however, we have to 
check that the TIF restriction (that is absent in~\cite{urabeH17matSim}) is indeed carried over.

In~\cite{urabeH17matSim} it is shown that FPE can indeed create
a simulation that does not exist between the original systems. Its
practical
use is witnessed by experimental results
in~\cite{urabeH14genericforward,urabeH17matSim}, 
where FPE was used in verifications of security protocols.
%too. 
It would not be hard to observe the same in the current setting for \emph{infinitary} traces.

For the proof of Theorem~\ref{thm:propertiesFPEpow}, once again, we turn 
to an axiomatic development.
\begin{thm}[FPE and forward simulation]%[correctness of FPE for fwd.\ sim.]
\label{thm:soundnessFPEfwd}$\!\!\!{}^\dagger$
Let $F$ be an endofunctor and $T$ be a monad on $\mathbb{C}$, as in
Lemma~\ref{lem:soundnessFwd} (that is, they constitute an infinitary trace
 situation and satisfy the two additional assumptions.)
% that
%  constitute an infinite trace situation with respect to $\sqsubseteq$ (Def.~\ref{def:infiniteTraceSituation}).
Let $\mathcal{X}=(X,s,c)$ and $\mathcal{Y}=(Y,t,d)$ be $(T,F)$-systems. % and $X=X_1+X_2$.
%Assume that $F$ and $T$ satisfy the assumptions in  Lem.~\ref{lem:soundnessFwd}.
Then  
\begin{enumerate}
%\item (monotonicity) If $X_1=X_{1,1}+X_{1,2}$, then $\mathcal{X}_{\FPE,X_{1,1}}\fwd\mathcal{Y}$ implies $\mathcal{X}_{\FPE,X_1}\fwd\mathcal{Y}$.
%\label{item:monotonicityfpefwd}
\item (soundness for forward simulation)  $\mathcal{X}_{\FPE}\fwd\mathcal{Y}$ implies $\trinf(c)\odot s\sqsubseteq \trinf(d)\odot t$.
\label{item:soundnessfpefwd}
\item (adequacy for forward simulation) $\mathcal{X}\fwd\mathcal{Y}$ implies $\mathcal{X}_{\FPE}\fwd\mathcal{Y}$.
%\qed
\label{item:adequacyfpefwd}
\end{enumerate}
\end{thm}

%\begin{myproof}
\proof
%{\proof[Proof of Thm.\ \ref{thm:soundnessFPEfwd}]
%\paragraph*{\ref{item:soundnessfpefwd}~(soundness).}
\noindent{\bf (\ref{item:soundnessfpefwd})(soundness).}
\begin{equation*}
\begin{xy}
 \xymatrix@R=2.4em@C=1.7em{
 {\overline{F}X} \ar@{}[drr]|{=} \kar[rr]^{\overline{F} c } \kar@/^6mm/[rrrr]^{\overline{F}(\trinf(c))} & & {\overline{F}^2 X} \ar@{}[drr]|{=} \kar[rr]_{\overline{F}(\trinf(\overline{F}c))} & & {\overline{F} Z} \ar@{}[drr]|{=} & & {\overline{F}Y} \kar[ll]_{ \overline{F}(\trinf(d))}\\  
 {X} \kar@/_4mm/[rrrr]_(.3){\trinf(c)} \kar[u]_{c} \ar@{}[drrr]|(.65){=} \kar[rr]^{c}  & & {\overline{F}X} \kar[u]_{\overline{F}c} \kar[rr]^{\trinf(\overline{F}c)} & & {Z} \kar[u]_{J\zeta} & &  {Y} \kar[ll]_{\trinf(d)} \kar[u]^{d}\\
 {} \ar@{}[ru]|(.10)*\dir{-} &   & & & {1} \ar@{}[ur]|(.5){\sqsubseteq} \ar`l[llllu][llllu]^(0){s} \kar[llu]^{c\odot s} \ar`r[rru][rru]_(0){t} & &  \ar@{}[lu]|(.10)*\dir{-} }
\end{xy}
\end{equation*}
%As $X\in\sigalg_X$, by letting $\mathcal{Y}=\mathcal{X}_{\FPE,X}$ in Prop.~\ref{prop:catmonotonicityOfFPEInf},
%from reflexivity of FPE, we obtain $\mathcal{X}_{\FPE,X}\fwd\mathcal{X}_{\FPE,X_1}$.
%Therefore by transitivity of FPE, we get $\mathcal{X}_{\FPE,X}\fwd\mathcal{Y}$.
%Hence it suffices to prove that  $\mathcal{X}_{\FPE,X}\fwd\mathcal{Y}$ implies $\tr(\mathcal{X})\sqsubseteq \tr(\mathcal{Y})$.
%
Let 
%$\mathcal{X}=((X,\sigalg_X),s,c)$, $\mathcal{Y}=((Y,\sigalg_Y),t,d)$ and 
$\zeta:Z\to FZ$ be a final $F$-coalgebra.
By definition, $\mathcal{X}_{\FPE}=(\overline{F}X,c\odot s, \overline{F}c)$.
Assume $\mathcal{X}_{\FPE}\fwd\mathcal{Y}$.
%By the assumption and 
Then by
 soundness of a forward Kleisli simulation, we have:
\begin{equation}
\trinf(\overline{F}c)\odot (c\odot s)\sqsubseteq\trinf(d)\odot t \;. \label{eq:catsoundnessFPEinf1}
\end{equation}
%Note that $\trinf(\overline{F}c)$ is the largest homomorphism from 
%an $\overline{F}$-coalgebra $\overline{F}c:\overline{F}X\kto\overline{F}^2X$ to $J\zeta$.
As $\trinf(c)$ is a homomorphism from $c$ to $J\zeta$, 
%and as $J\zeta$ is an isomorphism, 
we have:
\begin{equation}
\trinf(c)=(J\zeta)^{-1}\odot \overline{F}(\trinf(c))\odot c \;. \label{eq:catsoundnessFPEinf2}
\end{equation}
Here $(J\zeta)^{-1}\odot \overline{F}(\trinf(c))$ is a homomorphism from an $\overline{F}$-coalgebra $\overline{F}c:\overline{F}X\kto\overline{F}^2X$ to $J\zeta$
because of the following equation.
\begin{align*}
J\zeta\odot \bigl((J\zeta)^{-1}\odot \overline{F}(\trinf(c))\bigr) &= \overline{F}(\trinf(c)) & \\
&= \overline{F}\bigl((J\zeta)^{-1}\odot \overline{F}(\trinf(c))\odot c\bigr) & (\text{by  (\ref{eq:catsoundnessFPEinf2})}) \\
&= \overline{F}\bigl((J\zeta)^{-1}\odot \overline{F}(\trinf(c))\bigr)\odot \overline{F}(c) &
\end{align*}
%$\zeta^{-1}\odot \overline{F}(\tr^{\infty}_{\zeta}(c))$ is a homomorphism from $\overline{F}c$ to $\zeta$.
%Therefore, 
%This implies the following equation because 
%Note that 
As 
$\trinf(\overline{F}c)$ is the largest homomorphism from 
 $\overline{F}c$ to $J\zeta$, we have:
%$\trinf(\overline{F}c)$ is the largest homomorphism. % from  $\overline{F}c$ to $J\zeta$.
%As $\trinf(\overline{F}c)$ is the largest homomorphism from  $\overline{F}c$ to $J\zeta$, this implies
\begin{equation}
(J\zeta)^{-1}\odot \overline{F}(\trinf(c))\sqsubseteq\trinf(\overline{F}c)\,. \label{eq:catsoundnessFPEinf3}
\end{equation}
From the equations (\ref{eq:catsoundnessFPEinf1}--\ref{eq:catsoundnessFPEinf3}), % and the monotonicity of composition, 
$\trinf(c)\odot s\sqsubseteq\trinf(d)\odot t$ follows.

\vspace*{2mm}\noindent\begin{minipage}{0.70\hsize}
\noindent{\bf (\ref{item:adequacyfpefwd})(adequacy).} 
Let $f:Y\kto X$ be a forward Kleisli simulation from $\mathcal{X}$ to $\mathcal{Y}$.
Then %as $f$ is a forward simulation, 
we have:
\begin{equation*}
\overline{F}c\odot (c\odot f)
\sqsubseteq \overline{F}c\odot (\overline{F}f\odot d) 
= \overline{F}(c\odot f)\odot d 
\end{equation*}
and
\begin{equation*}
c\odot s
\sqsubseteq  c\odot (f\odot t)
=(c\odot f)\odot t \,. %& (\text{$f$ is a fwd.\ simulation})\,.
\end{equation*}
Hence $c\odot f:Y\kto \overline{F}X$ is a forward simulation from $\mathcal{X}_{\FPE}$ to $\mathcal{Y}$.
\qed
\end{minipage}
\begin{minipage}{0.30\hsize}
\begin{xy}
 \xymatrix@R=2.4em@C=2.6em{
 {\overline{F}^2X} \ar@{}[dr]|{=} & {\overline{F}X} \ar@{}[dr]|{\sqsubseteq} \kar[l]_{\overline{F}c} & {\overline{F}Y} \kar[l]_{\overline{F}f} \\
  {\overline{F}X}  \kar[u]^{\overline{F}c} & {X} \ar@{}[dl]|(.45){=} \ar@{}[dr]|(.45){\sqsubseteq} \kar[l]_{c} \kar[u]^{c} & {Y} \kar[l]_{f} \kar[u]_{d} \\
  & {1} \kar@/^5mm/[ul]^{c\odot s}  \kar[u]^{s} \kar@/_5mm/[ur]_{t} & }
\end{xy}
\end{minipage}
%\qed
%\end{myproof}
%\end{proof:soundnessFPEfwd}

%The next aim is %we want 
%to show that FPE satisfy soundness and adequacy with respect to  infinite trace when it is used to increase restricted backward simulation.

\begin{thm}[FPE and backward simulation]%[correctness of FPE for restricted bwd.\ sim.]
\label{thm:soundnessFPEbwdtop}$\!\!\!{}^\dagger$
Let $F$ be an endofunctor and $T$ be a monad on $\mathbb{C}$
that satisfy the conditions in Proposition~\ref{prop:generalconstructiontop}
 (hence those in Lemma~\ref{lem:soundnessRBwdtop}). 
%Let $F$ be an endofunctor and $T$ be a monad on $\mathbb{C}$ that constitute an infinite trace situation with respect to 
%%an order 
%$\sqsubseteq$.
%We assume that the largest homomorphism is obtained by using Prop.~\ref{prop:generalconstructiontop}. 
%satisfies the assumptions \ref{item:restrictedbwdsima} and \ref{item:restrictedbwdsimb} in Def.~\ref{def:restrictedbwdsim}.
Let $\mathcal{X}=(X,s,c)$ and $\mathcal{Y}=(Y,t,d)$ be $(T,F)$-systems. % and $Y=Y_1+Y_2$.
%Moreover, we define $d_i:Y_i\kto \overline{F}Y$ ($i\in\{1,2\}$) by 
%$d_i=d\odot \kappa_i$.
\begin{enumerate}
%\item (monotonicity) If $Y_1=Y_{1,1}+Y_{1,2}$, then $\mathcal{X}\tifbwd\mathcal{Y}_{\FPE,Y_{1,1}}$ implies $\mathcal{X}\bwd\mathcal{Y}_{\FPE,Y_1}$.
%\label{item:monotonicityfpebwd}
\item (soundness for backward simulation)  $\mathcal{X}\tifbwd\mathcal{Y}_{\FPE}$ implies $\trinf(c)\odot s\sqsubseteq \trinf(d)\odot t$.
\label{item:soundnessfpebwdtop}
\item (adequacy for backward simulation) 
$\mathcal{X}\tifbwd\mathcal{Y}$ implies $\mathcal{X}\tifbwd\mathcal{Y}_{\FPE}$ if the following conditions are satisfied.
%Assume the following conditions.
\begin{enumerate}
\item 
The coalgebra $d:Y\kto \overline{F}Y$ satisfies 
$\top_{\overline{F}Y,Z}\odot d = \top_{Y,Z}$. 
%$J!_{FY}\odot d =      J!_Y$. 
%The arrow $(d_1+\id_{X_2}):Y\kto \overline{F}Y+Y_2$ satisfies $J!_{\overline{F}Y+Y_2}\odot (c_1+\id_{X_2}) = J!_Y$. 
\label{item:restrictedfpe2top}
%\item One of the following is satisfied.
%\begin{enumerate}
%\item Assumption~\ref{asm:existenceWeaklyFinalCoalgGFPcont} in Prop.~\ref{prop:generalconstructionnotop} is satisfied. \label{item:restrictedfpe3cont}
\item Precomposing $d$ preserves the greatest lower bound of a (possibly transfinite) decreasing  sequence.
%Namely, for $A\in\Kl(T)$ and a family of Kleisli arrows $(g_{\mathfrak{i}}:\overline{F}Y\kto A)_{\mathfrak{i}\in\mathfrak{a}}$ where 
%$\mathfrak{a}$ is a limit ordinal and $\mathfrak{i}\leq \mathfrak{j}$ implies $g_{\mathfrak{i}}\sqsupseteq g_{\mathfrak{j}}$,
%%$g_0\sqsupseteq g_1 \sqsupseteq g_2\ldots :Y\kto A$,
%$\bigsqcap_{\mathfrak{i}<\mathfrak{a}}(g_{\mathfrak{i}}\odot d)=(\bigsqcap_{\mathfrak{i}<\mathfrak{a}}g_{\mathfrak{i}})\odot d$ holds.
%\item Precomposing $b:X\kto Y$ preserves the greatest fixed point of $\Psi$. 
%Namely, if $f:Y\kto 1$ is the greatest fixed point of $\Psi_Y$, then $f\odot b:X\kto 1$ is the greatest fixed point of $\Psi_X$.
\label{item:restrictedfpe3top}
\end{enumerate}
%Then $\mathcal{X}\tifbwd\mathcal{Y}$ implies $\mathcal{X}\tifbwd\mathcal{Y}_{\FPE}$.
\label{item:adequacyfpebwdtop}
\end{enumerate}
\end{thm}

%\newtheorem*{proof:soundnessFPEbwdtop}{Proof of Thm.\ \ref{thm:soundnessFPEbwdtop}}{\itshape}{\rmfamily}
%\begin{proof:soundnessFPEbwdtop}
%\begin{myproof}
\proof
%{\proof[Proof of Thm.\ \ref{thm:soundnessFPEbwdtop}]
\noindent{\bf (\ref{item:soundnessfpebwdtop})~(soundness).} 
\begin{equation*}
\begin{xy}
 \xymatrix@R=2.4em@C=1.7em{
 {\overline{F}X} \ar@{}[drr]|{=} \kar[rr]^{\overline{F}(\trinf(c))} & & {\overline{F}Z} \ar@{}[drr]|{=}  & & {\overline{F}^2Y} \ar@{}[drr]|{=} \kar[ll]^{\overline{F}(\trinf(\overline{F}d))} & & {\overline{F}Y} \kar[ll]^{\overline{F}d} \kar@/_6mm/[llll]_{\overline{F}(\trinf(d))} \\  
 {X}  \kar[u]_{c}  \kar[rr]^{\trinf(c)}  & & {Z}\ar@{}[drrr]|(.65){=} \kar[u]_{J\zeta} & & {\overline{F}Y} \kar[u]_{\overline{F}d}  \kar[ll]_{\trinf(\overline{F}d)} & &  {Y} \kar[ll]_{d} \kar[u]^{d} \kar@/^4mm/[llll]^(.3){\trinf(d)}\\
 {} \ar@{}[ru]|(.10)*\dir{-} &   & {1} \ar@{}[ul]|(.5){\sqsubseteq} \ar`l[llu][llu]_(0){s} \ar`r[rrrru][rrrru]^(0){t} \kar[rru]^{d\odot t}  & &  & &  \ar@{}[lu]|(.10)*\dir{-} }
\end{xy}
\end{equation*}
Let $b:X\kto \overline{F}Y$ be a TIF-backward simulation from $\mathcal{X}$ to $\mathcal{Y}_{\FPE}$.
Then by  soundness of TIF-backward simulation, % (Lemma~\ref{lem:soundnessRBwdnotop}), 
we have:
\begin{equation}
\trinf(c)\odot s\sqsubseteq \trinf(\overline{F}d)\odot (d\odot t)\,.
\label{eq:soundnessfpebwd1}
\end{equation}
It is easy to see that $d:Y\kto \overline{F}Y$ is a forward simulation from $\mathcal{Y}_{\FPE}$ to $\mathcal{Y}$.
Therefore by soundness of forward simulation, % (Lemma~\ref{lem:soundnessFwd}), 
we have:
\begin{equation}
\trinf(\overline{F}d)\odot (d\odot t)\sqsubseteq \trinf(d)\odot t\,.
\label{eq:soundnessfpebwd2}
\end{equation}
From the inequalities (\ref{eq:soundnessfpebwd1}) and (\ref{eq:soundnessfpebwd2}), we have 
$\trinf(c)\odot s\sqsubseteq \trinf(d)\odot t$.

\vspace{2mm}\noindent\begin{minipage}{0.65\hsize}
\noindent{\bf (\ref{item:adequacyfpebwdtop})~(adequacy).} 
Let $b:X\kto Y$ be a TIF-backward simulation from $\mathcal{X}$ to $\mathcal{Y}$.
In a similar manner to the proof of Theorem~\ref{thm:soundnessFPEfwd} (\ref{item:soundnessfpefwd}),
we can prove that $d\odot b:X\kto \overline{F}Y$ is a backward simulation from $\mathcal{X}$ to $\mathcal{Y}_{\FPE}$.
Moreover, the Assumptions~(\ref{item:restrictedfpe2top}) and (\ref{item:restrictedfpe3top}) imply that 
$d\odot b$ satisfy Assumptions~(\ref{item:restrictedbwdsim2top}) and (\ref{item:restrictedbwdsim3top})  in 
Definition~\ref{def:restrictedbwdsimtop}.
%the definition of a TIF-backward simulation (Definition~\ref{def:restrictedbwdsimtop}).
Therefore $d\odot b$ is a TIF-backward simulation from $\mathcal{X}$ to $\mathcal{Y}_{\FPE}$.
\qed
\end{minipage}
\begin{minipage}{0.35\hsize}
\begin{xy}
 \xymatrix@R=2.0em@C=2.8em{
 {\overline{F}X} \ar@{}[dr]|{\sqsubseteq} \kar[r]^{\overline{F}b} & {\overline{F}Y} \ar@{}[dr]|{=} \kar[r]^{\overline{F}d} & {\overline{F}^2Y}  \\
  {X}  \kar[u]^{c} \kar[r]^{b} & {Y} \ar@{}[dl]|(.45){\sqsubseteq} \ar@{}[dr]|(.45){=}  \kar[r]^{d} \kar[u]^{\overline{F}d} & {\overline{F}Y} \kar[u]_{\overline{F}d} \\
  & {1} \kar@/^5mm/[ul]^{s}  \kar[u]^{t} \kar@/_5mm/[ur]_{d\odot t} & }
\end{xy}
\end{minipage}
%}
%\end{myproof}
%\end{proof:soundnessFPEbwdtop}

%\newtheorem*{proof:propertiesFPEpow}{Proof of Thm.\ \ref{thm:propertiesFPEpow}}{\itshape}{\rmfamily}
%\begin{proof:propertiesFPEpow}
{\proof[Proof of Theorem \ref{thm:propertiesFPEpow}]
%\begin{myproof} (Of Thm.~\ref{thm:propertiesFPEpow})
% NATSUKI: proof environment
(\ref{item:safpefwdpow}) is immediate from Theorem~\ref{thm:soundnessFPEfwd}.
In a similar manner to Lemma~\ref{lem:whenrestrictedpow},
we can prove (\ref{item:safpebwdpow}) using Theorem~\ref{thm:soundnessFPEbwdtop}.
%\end{myproof}
\qed}
%\end{proof:propertiesFPEpow}

%\subsection{Nondeterministic Tree Automaton}
\subsection{Coincidence between Automata-theoretic and Coalgebraic Infinitary Trace Semantics}
\label{subsec:automcharapow}
In this section we give a sanity-check result for the 
coalgebraic infinitary trace semantics that is defined in Section~\ref{subsec:constructlargesthompow}.
%the previous section.
Namely, 
 for nondeterministic systems,
we show a coincidence between:
%that the 
the coalgebraic infinitary trace semantics formalized in the previous section; and
%that is defined by the largest homomorphism 
%defined in the previous section 
%coincides
%with 
the (infinitary) tree language that is defined using automata-theoretic terms.
%the automata-theoretically defined 
%We first 
%Here, the latter is formalized as follows. 
%The former is formalized in the previous section.
%In this paper, we The latter is formalized as follows.
Here we formalize the latter as follows (Recall the notations from Section~\ref{subsec:rankedAlphabetAndInfiniteTrees}).
%the usual definition of (infinite) tree languages for nondeterministic systems.
%we regard a $(\pow,\FSigma)$-system as an automaton that nondeterministically generate an infinite tree. 
%Then we show that its automata-theoretic semantics coincides with coalgebraic infinite trace semantics.
%

\begin{defi}\label{def:nondettreeautomaton}
Let $\Sigma$ be a ranked alphabet.
%A \emph{$\Sigma$-labeled nondeterministic tree automaton} is 
A $(\pow,\FSigma)$-system $\mathcal{X}=(X,s,c)$ is called a \emph{$\Sigma$-labeled nondeterministic tree automaton}.
%
%Let $\mathcal{X}=(X,s,c)$ be a nondeterministic tree automaton. 
For a $\Sigma$-labeled  infinitary tree $t=(D,l)$ and a state $x\in X$, 
a \emph{run tree of $\mathcal{X}$ from $x$ that generates $t$} 
is a $(X)_{n\in\omega}$-labeled\footnote{Note that $(X)_{n\in\omega}$ is the ranked alphabet $\Sigma'=(\Sigma'_n)_{n\in\omega}$
such that $\Sigma'_n=X$ for each $n\in\omega$.} 
infinitary tree $t_r=(D,l_r)$,
with the same domain as $t$,
 %that %has the same domain $D$ as $t$ and
such that:
%satisfies the following conditions.
%For a $\Sigma$-labeled  infinite tree $t=(D,l)$ and a state $x\in X$, 
%a $(X)_{n\in\omega}$-labeled  infinite tree $t_r=(D,l_r)$ that has the same domain $D$ as $t$ is 
%called a \emph{run tree of $\mathcal{X}$ from $x$ that generates $t$} if the following conditions are satisfied:
\begin{itemize}
\item $l_r(\empseq)=x$; and
\item 
for any 
%Let $\alpha\in D$ be an 
element 
$\alpha\in D$
of the common domain, % such that: 
assume that
$l(\alpha)=a\in\Sigma_n$,
$l_r(\alpha)=y$ and $l_r(\alpha i)=y_i$ for each $i\in\{0,\ldots,n-1\}$.
Then $(a,y_0,\ldots,y_{n-1}) \in c(y)$ holds.
%for all $\alpha\in D$ such that $l(\alpha)=a\in\Sigma_n$, 
%$(a,l_r(\alpha0),l_r(\alpha(n-1)) \in c(l_r(\alpha))$ holds.
%for all $\alpha\in D$ such that $l(\alpha)=a\in\Sigma_n$, $l_r(\alpha)=y$ and 
%$l_r(\alpha i)=y_i$ for all $i\in\{0,\ldots,n-1\}$, we have $(a,y_0,\ldots,y_{n-1}) \in c(y)$.
\end{itemize}
For a state $x\in X$, 
the \emph{infinitary language of $\mathcal{X}$ from $x$} is the set 
$\langinf(\mathcal{X},x)\subseteq \Treeinf(\Sigma)$ that is defined by
$%\begin{equation*}
\langinf(\mathcal{X},x)=\{
t\in \Treeinf(\Sigma)
\mid
\text{there is a  run tree of $\mathcal{X}$ from $x$ that generates $t$}
\}
$. %\end{equation*}
The \emph{infinitary language of $\mathcal{X}$} is the set 
%$\langinf(\mathcal{X})\subseteq \Treeinf(\Sigma)$ that is defined by
$\langinf(\mathcal{X})=\bigcup_{x\in s(*)} \langinf(\mathcal{X},x)$,
where $*$ denotes the unique element of a singleton $1$.
\end{defi}
%
%In the rest of this section, 
The following is the main result of this section.
% coincidence between  coalgebraic infinitary trace semantics and automata-theoretic infinitary language.
%Then we can show that the 
%the automata-theoretic semantics of $\Sigma$-labeled 
%nondeterministic tree automaton coincides with coalgebraic infinite trace semantics.
\begin{prop}\label{prop:largestsemanticspow}
Let $\Sigma$ be a ranked alphabet.
%We define an arrow $\zeta:\Treeinf(\Sigma)\to F_{\Sigma}\Treeinf(\Sigma)$ in $\Sets$ by 
%$\zeta(t)=(a,(t_0,\ldots,t_{n-1}))$.
%where $t=(D,l)$, $a=l(\empseq)\in\Sigma_n$, and $t_i$ is the $i$'th subtree of $t$.
%Then $\zeta$ is a final $F_{\Sigma}$-coalgebra.
%There exists a final $\FSigma$-coalgebra $\zeta$ such that: %whose carrier set 
%\begin{itemize}
%\item the carrier set of $\zeta$ is $\Treeinf(\Sigma)$; and
%\item for a $\Sigma$-labeled nondeterministic tree automaton $\mathcal{X}=(X,s,c)$,
%we have $\trinf(c)(x)=\lang(\mathcal{X},x)$ for all $x\in X$.
%\end{itemize}
%Then 
The carrier set of a final $\FSigma$-coalgebra is given by $\Treeinf(\Sigma)$.
%Moreover,
Moreover for a $\Sigma$-labeled nondeterministic tree automaton $\mathcal{X}=(X,s,c)$,
%With respect to the order in Def.~\ref{def:orderenrichment}, 
 we have $\trinf(c)(x)=\langinf(\mathcal{X},x)$ for all $x\in X$,
%Moreover, 
%Furthermore it implies
and hence
$\trinf(c)\odot s(*)=\langinf(\mathcal{X})$.
\end{prop}
%
%

%\newtheorem*{proof:largestsemanticspow}{Proof of Thm.\ \ref{thm:largestsemanticspow}}{\itshape}{\rmfamily}
%\begin{proof:largestsemanticspow}
%\begin{myproof}
\proof
%We define $f:X\kto \Treeinf(\Sigma)$ by $f(x)=\lang(\mathcal{X},x)$. 
We define an arrow $\zeta:\Treeinf(\Sigma)\to F_{\Sigma}\left(\Treeinf(\Sigma)\right)$ in $\Sets$ by 
$\zeta(t)=(a,(t_0,\ldots,t_{n-1}))$
for each $t\in\Treeinf(\Sigma)$ such that
%where 
$t=(D,l)$, $a=l(\empseq)\in\Sigma_n$, and  the $i$-th subtree of $t$  is $t_i$.
It is known that $\zeta$ is a final $F_{\Sigma}$-coalgebra (see e.g.\ \cite{ruttenT93initialalgebra}).
%
%We first show that $\zeta$ is a final $F_{\Sigma}$-coalgebra.
%
%Let $c:X\to F_{\Sigma}X$ be an $F_{\Sigma}$-coalgebra.
%For all $k\in\mathbb{N}$ and $x\in X$, we define a $\Sigma$-labeled $k$-prefix tree 
%$t_{k,x}=(D_{k,x},l_{k,x})$ by the induction on $k$ as follows:
%\begin{itemize}
%\item $D_{0,x}=\emptyset$ and $l_{0,x}$ is an empty function; and
%\item $D_{k+1,x}=\{i\alpha_i\mid 0\leq i\leq n-1, \alpha_i\in D_{k,x_i}\}$,
%$l_{k+1,x}(\empseq)=a$ and $l_{k+1,x}(i\alpha)=l_{k,x_i}(\alpha)$ 
%where $c(x)=\bigl(a,(x_0,\ldots,x_n-1)\bigr)$, $a\in\Sigma_n$ and $0\leq i\leq n-1$.
%\end{itemize}
%It is easy to see that for each $x\in X$, $D_k\subseteq D_{k+1}$ and 
%
We show that $\langinf(\mathcal{X},\place):X\kto\Treeinf(\Sigma)$ is the largest homomorphism from $c$ to $J\zeta$.
% where $\zeta$ is a final coalgebra defined in Lem.~\ref{lem:finalcoalgsets}.

We first show that $\langinf(\mathcal{X},\place)$ is a homomorphism.
For $x\in X$, we have:
\allowdisplaybreaks[3]\begin{align*}
&(\overline{\FSigma}\langinf(\mathcal{X},\place))\odot c (x) \\
&= (\overline{\FSigma}\langinf(\mathcal{X},\place))\left(\left\{(a,x_0,\ldots,x_{n-1})\relmiddle|
\begin{array}{l}
 n\in\omega, a\in\Sigma_n, x_0,\ldots,x_{n-1}\in X, \\
 (a,x_0,\ldots,x_{n-1})\in c(x)
 \end{array}
 \right\}\right) \\
&= \left\{(a,t_0,\ldots,t_{n-1})\relmiddle|
\begin{array}{l}
n\in\omega, \;a\in\Sigma_n,\; \\
\exists x_0,\ldots,x_{n-1}\in X.\, 
\left(\begin{array}{l}
(a,x_0,\ldots,x_{n-1})\in c(x), \\
t_i\in \langinf(\mathcal{X},x_i)\text{ for each $i$}%$in\{0,\ldots,n-1\}$} 
\end{array}\right)
\end{array}
\right\}\\
&= J\zeta\left(\left\{(D,l)\in\Treeinf(\Sigma) \relmiddle|
\begin{array}{l}
n\in\omega, l(\empseq)\in\Sigma_n, \\
\exists x_0,\ldots,x_{n-1}\in X. 
\left(\begin{array}{l}(l(\empseq),x_0,\ldots,x_{n-1})\in c(x), \\
l(i)\in \langinf(\mathcal{X},x_i) \text{ for each $i$}%\in\{0,\ldots,n-1\}$}
\end{array}\right)
\end{array}
\right\}\right) \\
&= J\zeta\left(\left\{(D,l)\in\Treeinf(\Sigma) \relmiddle|
\!\begin{array}{l}
n\in\omega, l(\empseq)\in\Sigma_n, \\
\exists x_0,\ldots,x_{n-1}\in X. \left(\begin{array}{l}
(l(\empseq),x_0,\ldots,x_{n-1})\in c(x), \\
\text{for each $i$, there is a run tree}\\
\;\;\text{$t_{r,i}$ of $\mathcal{X}$ from $x_i$ that }\\
\;\;\;\text{generates $i$-th subtree $t_i$ of $t$}%\in\{0,\ldots,n-1\}$}\\
\end{array}\right)
\end{array}\!
\right\}\right) \\
&= J\zeta\bigl(\{t\in\Treeinf(\Sigma) \mid \text{there exists a run tree $t_{r}$ of $\mathcal{X}$ from $x$ that generates $t$}\}\bigr) \\
&= \bigl(J\zeta\odot \langinf(\mathcal{X},\place)\bigr)(x)\,.
\end{align*}
Therefore $\langinf(\mathcal{X},\place)$ is a homomorphism from $c$ to $J\zeta$.

It remains to prove that $\langinf(\mathcal{X},\place)$ is the largest homomorphism.
Let $g:X\kto\Treeinf(\Sigma)$ be a homomorphism from $c$ to $J\zeta$.
%we assume that 
We fix $x\in X$ and $t=(D,l)\in g(x)$,
%For $x\in X$ and $t=(D,l)\in g(x)$, %  $x\in X$. 
%we 
and show that $t\in\langinf(\mathcal{X},x)$.
To this end, it suffices to construct a run tree $t_r=(D,l_r)$ of $\mathcal{X}$ from $x$ that generates $t$.
%For $\alpha\in D$, we denote $t_{\alpha}$ for the $\alpha$'th subtree of $t$.
For each $\alpha\in D$, we define a state $l_r(\alpha)\in X$ such that $t_{\alpha}\in g(l_r(\alpha))$
%such that $t_{\alpha}\in g(l_r(\alpha))$
by induction on the length of $\alpha$ as follows.
\begin{itemize}
\item
If $\alpha=\empseq$, we define it by $l_r(\alpha)=x$.
By the assumption, $t_{\alpha}=t\in g(x)=g(l_r(\alpha))$.
%Then $t_{\alpha}=t\in g(x)=g(l_r(\empseq))$.

\item
%Assume that $l(\alpha)\in\Sigma_n$ and $t_{\alpha}\in g(l_r(\alpha))$. 
Let $l(\alpha)\in\Sigma_n$ and assume that $t_{\alpha}\in g(l_r(\alpha))$.
%We assume that $l(\alpha)\in\Sigma_n$.
As $g$ is a homomorphism from $c$ to $J\zeta$, %we have
$t_{\alpha}\in g(l_r(\alpha))=J\zeta^{-1}\odot\overline{\FSigma}g\odot c(l_r(\alpha))$.
By the definition of $\zeta$, this means that there exist a family of states $x_0,\ldots,x_{n-1}$ such that 
$(l(\alpha),x_0,\ldots,x_{n-1})\in c(l_r(\alpha))$ and 
%$\forall i\in\{0,\ldots, n-1\}.\;t_{\alpha i}\in g(x_i)$.
$t_{\alpha i}\in g(x_i)$ for each $i\in\{0,\ldots, n-1\}$.
%  for each $i\in\{0,\ldots, n-1\}$.
We define $l_r(\alpha i)$ by $l_r(\alpha i)=x_i$.
%Then $t_{\alpha i}\in l_r(\alpha i)$.
%
\end{itemize}
By the axiom of dependent choice, this $l_r$ is well-defined.
Moreover, 
by %induction on 
its construction,
% by its construction, 
%we can see that 
$(D,l_r)$ is a run tree of $\mathcal{X}$ from $x$ that generates $t$.
Therefore $t\in \langinf(\mathcal{X},x)$.

%for each $x\in X$ and $t\in g(x)$,
% and this implies 
Therefore $g\sqsubseteq \langinf(\mathcal{X},\place)$ holds, and $\langinf(\mathcal{X},\place)$ is the largest homomorphism from $c$ to $J\zeta$.
Hence we have $\trinf(c)=\langinf(\mathcal{X},\place)$.
This immediately implies  $\trinf(c)\odot s(*)=\langinf(\mathcal{X})$.
\qed
%\end{myproof}
%\end{proof:largestsemanticspow}

Hence the coalgebraic definition of infinitary trace semantics in Definition~\ref{def:coalginftr} indeed characterizes 
the languages of $\Sigma$-labeled nondeterministic tree automata
in Definition~\ref{def:nondettreeautomaton}.

\section{Systems with  Probabilistic Branching}\label{sec:girymonad}
We now turn to probabilistic systems. They are modeled as $(\giry,
F)$-systems in the category $\Meas$. 
Here we establish largely the same statements as
in Section~\ref{sec:powersetmonad}, but many constructions and proofs are different.
Throughout this section $F$ is
assumed to be a (standard Borel) polynomial functor on $\Meas$
(Definition~\ref{def:polyfuncmeas}).

\subsection{Construction of Infinitary Traces}\label{subsec:constructlargesthomgiry}
In this section, like in Section~\ref{subsec:constructlargesthompow}, we 
%prove the constitution of 
establish
an infinitary trace situation.

%Our basic idea of the construction is similar to that for $\pow$
%(Section~\ref{subsec:constructlargesthompow}). 
Our goal is to construct
the largest homomorphism from an $\overline{F}$-coalgebra $c$ in to 
a lifted final coalgebra $J\zeta:Z\kto \overline{F}Z$ in $\Kl(\giry)$; we do so
inductively, 
much like in the nondeterministic setting (Section~\ref{subsec:constructlargesthompow}),
starting from the top element and going down along a
decreasing 
sequence. 
\begin{rem}\label{rem:differencePandG}
Compared to the nondeterministic case ($T=\pow$), major
differences are as follows.
\begin{itemize}
 \item  Composition of Kleisli arrows is $\omega^{\text{op}}$-continuous
	in $\Kl(\giry)$ (see Theorem~\ref{thm:inftrsitgiry} later). This is an advantage, because we can appeal to
	the Kleene fixed point theorem and we only need inductive
	construction 
	up-to $\omega$ steps (while, for $\pow$, we needed transfinite
	induction).
 \item  A big disadvantage, however, is the absence of the top
 element $\top_{X,Z}$ in $\Kl(T)(X,Z)$. One can imagine a top element
	$\top_{X,Z}$ to 
	assign $1$ to every event---this is however not a (probability)
	measure. 
\end{itemize}
\end{rem}

\hspace*{-\parindent}\begin{minipageparindent}
\begin{wrapfigure}[6]{r}{0pt}
\vspace{-\intextsep}
\vspace{-5pt}
\raisebox{-8mm}[0pt][0cm]%
{\tiny
\begin{xy}
(0,0)*+{1} = "F01",
(10,0)*+{\overline{F}1} = "F11",
(22,0)*+{\overline{F}^21} = "F21",
%(30,0)*+{\overline{F}^31} = "F31",
%
(-6,10)*+{X} = "F0X",
(-6,-10)*+{Z} = "F0Z",
(24,6)*+{\dots}= "",
(24,-6)*+{\dots}= "",
(31,0)*+{\dots}= "",
(-16,10)*+{\overline{F}X} ="F1X",
(-16,-10)*+{\overline{F}Z} ="F1Z",
(-11.5,0)*+{=} ="",
\kar _{J!}"F11";"F01"
\kar _{JF!}"F21";"F11"
%\kar _{JF^2!_{F1}}"F31";"F21"
\kar _(.1){JF^2!} (29,0);"F21"
%\ar @{.} (31,0);(34,0)
%
\kar ^{\pi_0} "F0X";"F01"
\kar @/^2mm/ ^(.7){\pi_1} "F0X";"F11"
\kar @/^4mm/ ^(.7){\pi_2} "F0X";"F21"
\kar _(.7){J\gamma_0} "F0Z";"F01"
\kar @/_2mm/ _(.7){J\gamma_1} "F0Z";"F11"
\kar @/_4mm/ _(.7){J\gamma_2} "F0Z";"F21"
\kar @{-->} _{\rotatebox{90}{\raisebox{1ex}{$\trinf(c)$}}} "F0X";"F0Z"
\kar _{c} "F0X";"F1X"
\kar _{\rotatebox{90}{\raisebox{1ex}{$\overline{F}(\trinf(c))$}}} "F1X";"F1Z"
\kar _{J\zeta}^{\cong} "F0Z";"F1Z"
\end{xy}}
\end{wrapfigure}
%Our alternative idea is as follows: the greatest element might not exists in $\Kl(T)(X,Z)$, 
%however, it exists in $\Kl(T)(X,1)$ in many cases. 
%The alternative idea is as follows: 
%\hspace{1cm}
To cope with the latter challenge, 
we turn to the \emph{final $F$-sequence} 
in $\Meas$ that yields a final $F$-coalgebra as its limit. 
Instead of using a sequence like $\top\sqsupseteq\Phi(\top)\sqsupseteq\cdots$ in
$\Kl(T)(X,Z)$ (where the largest element $\top$ does not exist anyway), 
we use a decreasing sequence that goes along the final sequence,
which is known to yield a final $F$-coalgebra~\cite{schubert09terminalcoalgebras}.
%
%Then we construct a cone over the sequence 
Once the largest element along the sequence is obtained,
we can construct the largest homomorphism from it in a similar manner to~\cite{cirstea10genericinfinite}.
%and construct a cone over the sequence.
%
%
%The construction of the largest homomorphism given in
%Proposition~\ref{prop:generalconstructionnotop} is based on the one
%in~\cite{cirstea10genericinfinite}.
\end{minipageparindent}
\vspace{1mm}

% is applicable to both $\giry$ and $\pow$ (c.f.\ Prop.~\ref{prop:satassumptionnondet2}).

The precise construction is found in the proof of the following
proposition. 
%The construction makes use of a well-known construction of a final coalgebra via a \emph{final sequence}.
%(the proof is in Appendix~\ref{subsec:omittedProofsForGiryMonad}). 

\begin{prop}\label{prop:generalconstructionnotop}$\!\!\!{}^\dagger$
Let $\mathbb{C}$ be a category, $F$ be an endofunctor on $\mathbb{C}$, 
and $T$ be a monad on $\mathbb{C}$ where each homset of $\Kl(T)$ carries an order $\sqsubseteq$. 
%Let $J:\mathbb{C}\to\Kl(T)$ be a Kleisli inclusion functor.
We assume the following conditions.
\begin{enumerate}
\item The category $\mathbb{C}$ has a final object $1$;
the final $\omega^{\op}$-sequence $1\overset{!_{F1}}{\leftarrow}F1\overset{F!_{F1}}{\leftarrow}F^2 1\overset{F^2!_{F1}}{\leftarrow} \cdots$ 
has a limit 
$(Z,(\gamma_i:Z\to F^i1)_{i\in\omega})$; and moreover,
$F$ preserves this limit. Hence the limit carries a final $F$-coalgebra~\cite{adamekK79leastfixed}.
%(Then this limit yields a final $F$-coalgebra $\zeta:Z\to FZ$.) 
\label{asm:existenceWeaklyFinalCoalgFinalSeq}
\item There exists a distributive law $\lambda:FT\Rightarrow TF$,
      yielding a lifting $\overline{F}$ on $\Kl(T)$ of $F$.
%and hence $F$ can be lifted to an endofunctor $\overline{F}$ on $\Kl(T)$. 
\label{asm:existenceWeaklyFinalCoalgDistLaw}
%
%\item For each 
%%pair of objects $X$ and $Y$ in $\Kl(T)$, 
%$X,Y\in\Kl(T)$,
%the homset $\Kl(T)(X,Y)$ carries a partial order $\sqsubseteq$. 
%Moreover, $\overline{F}$ and the composition of arrows in $\Kl(T)$ 
%%and the functor  $\overline{F}$ 
%are  monotone  wrt.\ this order.
%%composition is locally monotone (i.e.\ preserves order) with respect to this order.
%\label{asm:existenceWeaklyFinalCoalgFMonotone}
\item 
\label{asm:existenceWeaklyFinalCoalgGFPcont}
For 
%each 
$X,Y\in\Kl(T)$,
%pair of objects $X$ and $Y$ in $\Kl(T)$,  
every decreasing $\omega^{\op}$-sequence $f_0\sqsupseteq f_1\sqsupseteq
      \ldots$ in $\Kl(T)(X,Y)$ has the greatest lower bound $\bigsqcap_{i\in\omega}f_i$.
Moreover,  composition of arrows in $\Kl(T)$ and $\overline{F}$'s action on arrows  are both $\omega^{\text{op}}$-continuous. %, respectively.
That is, for each $g:Z\kto X$ and $h:Y\kto W$, we have 
$g\odot (\bigsqcap_{i\in\omega}f_i)=\bigsqcap_{i\in\omega}(g\odot f_i)$, 
$(\bigsqcap_{i\in\omega}f_i)\odot h=\bigsqcap_{i\in\omega}(f_i\odot h)$, and
$\overline{F}(\bigsqcap_{i\in\omega}f_i)=\bigsqcap_{i\in\omega}(\overline{F} f_i)$.
%(i.e.\ composition of arrows and applying $\overline{F}$ preserves the greatest lower bound).
%(We denote this order by $\sqsubseteq$.)
\item 
%For each object $X\in\mathbb{C}$, 
The lifting $J(!_X)$ of the unique arrow to 
$1$ %the final object 
is the largest element of $\Kl(T)(X,1)$.\!\!
 \label{asm:existenceWeaklyFinalCoalgTop}
%
%\item The functor  $\overline{F}$ is locally monotone. 
%%and continuous (i.e.\ preserves the greatest lower bound). 
%\label{asm:existenceWeaklyFinalCoalgFMonotone}
%
\item 
The functor $J$ lifts the limit in Assumption~(\ref{asm:existenceWeaklyFinalCoalgFinalSeq}) to a 2-limit.
% $(Z,(J\gamma_i:Z\kto F^i1)_{i\in\omega})$.
Namely, for any cone $(X,(\pi_i:X\kto F^i1)_{i\in\omega})$ over the sequence 
$1\overset{J!_{F1}}{\kleftarrow} \overline{F}1\overset{JF!_{F1}}{\kleftarrow} \overline{F}^2 1\overset{JF^2!_{F1}}{\kleftarrow} \cdots$,
there uniquely exists 
$l:X\kto Z$ such that $\pi_i=J\gamma_i\odot l$ holds for each $i\in\omega$.
Moreover, if $l':X\kto Z$ satisfies $J\gamma_i\odot l'\sqsubseteq J\gamma_i\odot l$ for each $i\in\omega$, then 
$l'\sqsubseteq l$ holds.
\label{asm:existenceWeaklyFinalCoalgWeak2Lim}
\end{enumerate}
Then $F$ and $T$ constitute an infinitary trace situation with respect to $\sqsubseteq$.
%Then $J\zeta:X\kto\overline{F}Z$ is a weakly final $\overline{F}$-coalgebra.
%\qed
\end{prop}

In more elementary terms,
Assumption~(\ref{asm:existenceWeaklyFinalCoalgWeak2Lim}) of Proposition~\ref{prop:generalconstructionnotop} asserts that: $J$
lifts the limit $Z$; and the lifted limit satisfies a stronger condition
of preserving %``carrying over'' 
the order between cones to the order between
mediating maps. 

Intuitively, Assumptions~\ref{asm:existenceWeaklyFinalCoalgFinalSeq}, \ref{asm:existenceWeaklyFinalCoalgGFPcont} and \ref{asm:existenceWeaklyFinalCoalgWeak2Lim}
 together ensure that we can ``transfer'' the greatest element $J(!_X)$ in $\Kl(T)(X,1)$
(Assumption~\ref{asm:existenceWeaklyFinalCoalgTop}) to the greatest homomorphism from $c$ to $J\zeta$.

%\begin{myproof}
\proof
%{\proof[Proof of Prop.\ \ref{prop:generalconstructionnotop}]
Let $c:X\kto\overline{F}X$ be a $\overline{F}$-coalgebra in $\Kl(T)$.
%We construct the largest homomorphism $\trinf(c):X\kto Z$ from $c$ to $J\zeta$.

\begin{minipageparindent}
\begin{wrapfigure}[3]{r}{0pt}
\vspace*{-\intextsep}
%\vspace{-10cm}
\raisebox{3mm}[0pt][0cm]{
\begin{xy}
 \xymatrix@R=1.5em@C=2.6em{
 {\overline{F}X} \ar@{}[dr]|{} \kar[r]^(.5){\overline{F}f} & \overline{F}1 \kar[d]^{J!_{F1}} \\
 {X} \kar[u]^{c} \kar[r]^(.5){f} & {1}  }
\end{xy}}
\end{wrapfigure}
We first construct a cone $(X,(\alpha_i:X\kto \overline{F}^i1)_{i\in\omega})$ over the sequence $1\overset{J!_{F1}}{\kleftarrow} \overline{F}1\overset{JF!_{F1}}{\kleftarrow} \overline{F}^2 1\overset{JF^2!_{F1}}{\kleftarrow} \cdots$.
To this end, we start with defining an arrow $\alpha_0:X\kto 1$.
% For $(\giry,F)$-systems, the value $\alpha_0(x)(*)$ gives the probability of deadlock when we start from $x$.
Let us define a function $\Psi_c:\Kl(T)(X,1)\to\Kl(T)(X,1)$ by $\Psi_c(f)=J!_{F1}\odot \overline{F}f\odot c$
(see the diagram on the right).
As composition in $\Kl(T)$ and $\overline{F}$'s action on arrows are both monotone (by
 Assumption~(\ref{asm:existenceWeaklyFinalCoalgGFPcont})), $\Psi_c$ is also monotone.
Moreover, as $J!_X$ is the largest element in $\Kl(T)(X,1)$ (Assumption~(\ref{asm:existenceWeaklyFinalCoalgTop})), we have $J!_X\sqsupseteq \Psi_c(J!_X)$.
Therefore by repeatedly applying $\Psi_c$ to the both sides of the inequality,
we  obtain a decreasing sequence $J!_X\sqsupseteq \Psi_c(J!_X)\sqsupseteq \Psi_c^2(J!_X)\sqsupseteq\cdots$.
\end{minipageparindent}
\auxproof
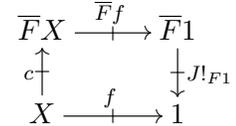
{
\begin{equation*}
\begin{xy}
(0,0)*+{1} = "F01",
(30,0)*+{\overline{F}1} = "F11",
(60,0)*+{\overline{F}^21} = "F21",
(90,0)*+{\overline{F}^31} = "F31",
(-7.5,15)*+{X} = "F0X",
(22.5,15)*+{\overline{F}X} = "F1X",
(52.5,15)*+{\overline{F}^2X} = "F2X",
(82.5,15)*+{\overline{F}^3X} = "F3X",
(12,7.5)*{\sqsupseteq} = "",
(42,7.5)*{\sqsupseteq} = "",
(72,7.5)*{\sqsupseteq} = "",
(-7.5,-15)*+{Z} = "F0Z",
(22.5,-15)*+{\overline{F}Z} = "F1Z",
(52.5,-15)*+{\overline{F}^2Z} = "F2Z",
(82.5,-15)*+{\overline{F}^3Z} = "F3Z",
(12,-7.5)*{=} = "",
(42,-7.5)*{=} = "",
(72,-7.5)*{=} = "",
(112.5,0)*{\cdots}="",
(105,15)*{\cdots}="",
(105,-15)*{\cdots}="",
\kar _{J!_{F1}}"F11";"F01"
\kar _{JF!_{F1}}"F21";"F11"
\kar _{JF^2!_{F1}}"F31";"F21"
\kar _(.1){JF^2!_{F1}} (105,0);"F31"
%\ar @{.} (75,0);(70,0)
%
\kar ^{c}"F0X";"F1X"
\kar ^{\overline{F}c}"F1X";"F2X"
\kar ^{\overline{F}^2c}"F2X";"F3X"
\kar ^(.9){\overline{F}^3c} "F3X";(97.5,15)
%\ar @{.} (65,15);(70,15)
%
\kar ^{J!_X} "F0X";"F01"
\kar ^{JF!_X} "F1X";"F11"
\kar ^{JF^2!_X} "F2X";"F21"
\kar ^{JF^3!_X} "F3X";"F31"
\kar _{J\zeta}"F0Z";"F1Z"
\kar _{JF\zeta}"F1Z";"F2Z"
\kar _{JF^2\zeta}"F2Z";"F3Z"
\kar _(.9){JF^3\zeta} "F3Z";(97.5,-15)
%\ar @{.} (65,-15);(70,-15)
%
\kar _{J!_Z} "F0Z";"F01"
\kar _{JF!_Z} "F1Z";"F11"
\kar _{JF^2!_Z} "F2Z";"F21"
\kar _{JF^3!_Z} "F3Z";"F31"
\end{xy}
%\label{eq:diagramexistenceWeaklyFinalCoalg1}
\end{equation*}
}

By Assumption~(\ref{asm:existenceWeaklyFinalCoalgGFPcont}), their greatest lower bound 
$\bigsqcap_{i\in\omega}\Psi_c^i(J!_X):X\kto 1$ exists;
we write $\Psi_c^{\omega}(J!_X)$ for the greatest lower bound.
Here, as composition of arrows in $\Kl(T)$ and $\overline{F}$'s action on arrows are both $\omega^{\text{op}}$-continuous, 
$\Psi_c$ is also  $\omega^{\text{op}}$-continuous.
Therefore by the Kleene fixed point theorem, $\Psi_c^{\omega}(J!_X)$ is the greatest fixed point of $\Psi_c$.

Using this 
%$\Psi_X^{\mathfrak{l}}(J!_X)$, 
greatest fixed point,
for each $i<\omega$, 
we define an arrow $\alpha_i:X\kto \overline{F}^i1$ %(for each $i<\omega$) 
inductively as follows:
\begin{equation}
%\alpha_0=\bigsqcap_{i\in\omega} J!_{F^i 1}\odot \alpha'_i ,  \;\;\;\;\;
\alpha_0=\Psi_c^{\omega}(J!_X)  \;\;\;\text{and}\;\;\;
\alpha_{i+1}=\overline{F}\alpha_i\odot c\,.
\label{eq:existenceWeaklyFinalCoalgdefcone}
\end{equation}
Then we can prove $\alpha_i=\overline{F}^iJ!_{F1}\odot \alpha_{i+1}$ 
for all $i\in\omega$ by induction on $i$
%inductively 
as follows. 
For $i=0$, we have:
\begin{align*}
J!_{F1}\odot\alpha_1
&=J!_{F1}\odot\overline{F}(\Psi_c^{\omega}(J!_X)) \odot c & (\text{by definition}) \\
&=\Psi_c(\Psi_c^{\omega}(J!_X)) & (\text{by definition}) \\
&= \Psi_c^{\omega}(J!_X) & (\text{$\Psi_c^{\omega}(J!_X)$ is a fixed point}) \\
&=\alpha_0 & (\text{by definition}).
\end{align*}
For the step case,
assume that $\alpha_i=\overline{F}^iJ!_{F1}\odot \alpha_{i+1}$.
Applying $\overline{F}$ and composing $c$ from the right, we have $\alpha_{i+1}=\overline{F}^{i+1}J!_1\odot \alpha_{i+2}$.
Hence we have $\alpha_i=\overline{F}^iJ!_{F1}\odot \alpha_{i+1}$ for all $i\in\omega$. 
This means that $(X,(\alpha_i:X\kto \overline{F}^i1)_{i\in\omega})$ is a cone over the sequence 
$1\overset{J!_{F1}}{\kleftarrow} \overline{F}1\overset{JF!_{F1}}{\kleftarrow} \overline{F}^2 1\overset{JF^2!_{F1}}{\kleftarrow} \cdots$\,.
Therefore by Assumption~(\ref{asm:existenceWeaklyFinalCoalgWeak2Lim}), 
there exists a unique mediating arrow $l:X\kto Z$ 
from the cone $(X,(\alpha_i)_{i\in\omega})$ to the $(Z,(J\gamma_i)_{i\in\omega})$, on the one hand.
%that satisfies the conditions in Assumption~\ref{asm:existenceWeaklyFinalCoalgWeak2Lim}. 
%
\begin{equation*}
\begin{xy}
(0,0)*+{1} = "F01",
(30,0)*+{\overline{F}1} = "F11",
(60,0)*+{\overline{F}^21} = "F21",
(90,0)*+{\overline{F}^31} = "F31",
(-7.5,15)*+{X} = "F0X",
(22.5,15)*+{\overline{F}X} = "F1X",
(52.5,15)*+{\overline{F}^2X} = "F2X",
(82.5,15)*+{\overline{F}^3X} = "F3X",
(12,7.5)*{=} = "",
(42,7.5)*{=} = "",
(72,7.5)*{=} = "",
(-7.5,-15)*+{Z} = "F0Z",
(22.5,-15)*+{\overline{F}Z} = "F1Z",
(52.5,-15)*+{\overline{F}^2Z} = "F2Z",
(82.5,-15)*+{\overline{F}^3Z} = "F3Z",
(12,-7.5)*{=} = "",
(42,-7.5)*{=} = "",
(72,-7.5)*{=} = "",
(112.5,0)*{\cdots}="",
(105,15)*{\cdots}="",
(105,-15)*{\cdots}="",
\kar _{J!_{F1}}"F11";"F01"
\kar _{JF!_{F1}}"F21";"F11"
\kar _{JF^2!_{F1}}"F31";"F21"
\kar _(.1){JF^3!_{F1}} (105,0);"F31"
%\ar @{.} (75,0);(70,0)
%
\kar ^{c}"F0X";"F1X"
\kar ^{\overline{F}c}"F1X";"F2X"
\kar ^{\overline{F}^2c}"F2X";"F3X"
\kar ^(.9){\overline{F}^3c} "F3X";(97.5,15)
%\ar @{.} (65,15);(70,15)
%
\kar ^{\alpha_0} "F0X";"F01"
\kar ^{\overline{F}\alpha_0} "F1X";"F11"
\kar ^{\overline{F}^2\alpha_0} "F2X";"F21"
\kar ^{\overline{F}^3\alpha_0} "F3X";"F31"
\kar _{J\zeta}"F0Z";"F1Z"
\kar _{JF\zeta}"F1Z";"F2Z"
\kar _{JF^2\zeta}"F2Z";"F3Z"
\kar _(.9){JF^3\zeta} "F3Z";(97.5,-15)
%\ar @{.} (65,-15);(70,-15)
%
\kar _{J!_Z} "F0Z";"F01"
\kar _{JF!_Z} "F1Z";"F11"
\kar _{JF^2!_Z} "F2Z";"F21"
\kar _{JF^3!_Z} "F3Z";"F31"
\kar @/_4mm/@{-->} _{l} "F0X";"F0Z"
\end{xy}
%\label{eq:diagramexistenceWeaklyFinalCoalg1}
\end{equation*}
On the other hand, 
$J\zeta^{-1}\odot \overline{F}l \odot c$ is also a mediating arrow
from $(X,(\alpha_i)_{i\in\omega})$ to $(Z,(J\gamma_i)_{i\in\omega})$.
Indeed, for all $i\in\omega$, we have:
%Here, for each $i\in\omega$, we have:
\begin{align*}
J\gamma_i\odot \bigl(J\zeta^{-1}\odot \overline{F}l \odot c\bigr)
&= \overline{F}^{i+1}J!_{F1}\odot\overline{F}J\gamma_{i}\odot \overline{F}l \odot c & (\text{$\zeta$ is a mediating arrow}) \\
&= \overline{F}^{i+1}J!_{F1}\odot \overline{F}\alpha_{i} \odot c & (\text{$l$ is a mediating arrow}) \\
&= \overline{F}^{i+1}J!_{F1}\odot \alpha_{i+1}  & (\text{by definition of $\alpha_{i+1}$}) \\
&= \alpha_{i}  & (\text{$(X,(\alpha_i)_{i\in\omega})$ is a cone}) &\,.
\end{align*}
%Therefore 
%$J\zeta^{-1}\odot \overline{F}l \odot c$ is also a mediating arrow
%from $(X,(\alpha_i)_{i\in\omega})$ to $(Z,(J\gamma_i)_{i\in\omega})$.
Hence by the uniqueness of the mediating arrow, we have $l=J\zeta^{-1}\odot \overline{F}l\odot c$ and
$l$ is a homomorphism from $c$ to $j\zeta$.

To conclude the proof, we have to show that  $l$ the largest homomorphism from $c$ to $J\zeta$.
%is the greatest fixed point of $\Phi_X$.
Let $g:X\kto Z$ be a homomorphism from $c$ to $J\zeta$.
We construct a cone $(X,(\beta_i:X\kto \overline{F}^i1)_{i\in\omega})$ over the sequence 
$1\overset{J!_{F1}}{\kleftarrow} \overline{F}1\overset{JF!_{F1}}{\kleftarrow} \overline{F}^2 1\overset{JF^2!_{F1}}{\kleftarrow} \cdots$
by $\beta_i=J\gamma_i\odot g$.
Then %for each $i\in\omega$, 
we can prove that $\beta_i\sqsubseteq \alpha_i$ for all $i\in\omega$ by induction on $i$ as follows.

For $i=0$, we have:
\begin{align*}
\Psi_c(\beta_0)
&=J!_{F1}\odot \overline{F}J\gamma_0\odot \overline{F}g\odot c & (\text{by definition}) \\
%&=J!_{F1}\odot \overline{F}J\gamma_0\odot J\zeta\odot \Phi_X(g) & (\text{by definition}) \\
&=J!_{F1}\odot \overline{F}J\gamma_0\odot J\zeta\odot g & (\text{$g$ is a homomorphism}) \\
&=J!_{F1}\odot JF\gamma_0\odot J\zeta\odot g & (\text{$\overline{F}$ is a lifting of $F$}) \\
&=J\gamma_0\odot g & (\text{$1$ is a final object in $\mathbb{C}$}) \\
&=\beta_0 & (\text{by definition})&.
\end{align*}
Therefore $\beta_0$ is a fixed point of $\Psi_c$.
As $\alpha_0=\Psi_c^{\omega}(J!_X)$ is the greatest fixed point of $\Psi_c$, we have $\beta_0\sqsubseteq \alpha_0$.

Assume $\beta_i\sqsubseteq \alpha_i$. Then 
\begin{align*}
\beta_{i+1}
&= J\gamma_{i+1}\odot g & (\text{by definition}) \\
&= JF\gamma_i\odot J\zeta \odot g & (\text{$\zeta$ is a mediating arrow}) \\
&= \overline{F}J\gamma_i\odot J\zeta \odot g & (\text{$\overline{F}$ is a lifting of $F$}) \\
&= \overline{F}J\gamma_i\odot \overline{F}g \odot c & (\text{$g$ is a homomorphism}) \\
&= \overline{F}\beta_i \odot c & (\text{$(X,(\beta_i)_{i\in\omega})$ is a cone}) \\
&\sqsubseteq \overline{F}\alpha_i \odot c & (\text{by the induction hypothesis and that $\overline{F}$ is monotone)}) \\
&= \alpha_{i+1} & (\text{by definition})&\,.
\end{align*}

Hence $\beta_i\sqsubseteq \alpha_i$ holds for all $i \in\omega$.
This implies $J\gamma_i\odot g \sqsubseteq J\gamma_i\odot l$ for all $i\in\omega$.
As  $(Z,(J\gamma_i:Z\kto F^i1)_{i\in\omega})$ is a 2-limit (Assumption~(\ref{asm:existenceWeaklyFinalCoalgWeak2Lim})), 
we have $g\sqsubseteq l$.

\begin{equation*}
\begin{xy}
(0,0)*+{1} = "F01",
(30,0)*+{\overline{F}1} = "F11",
(60,0)*+{\overline{F}^21} = "F21",
(90,0)*+{\overline{F}^31} = "F31",
(-20,20)*+{X} = "F0X",
(10,20)*+{\overline{F}X} = "F1X",
(40,20)*+{\overline{F}^2X} = "F2X",
(70,20)*+{\overline{F}^3X} = "F3X",
(-20,-15)*+{Z} = "F0Z",
(10,-15)*+{\overline{F}Z} = "F1Z",
(40,-15)*+{\overline{F}^2Z} = "F2Z",
(70,-15)*+{\overline{F}^3Z} = "F3Z",
(5,13)*+{Z} = "F0Z2",
(35,13)*+{\overline{F}Z} = "F1Z2",
(65,13)*+{\overline{F}^2Z} = "F2Z2",
(95,13)*+{\overline{F}^3Z} = "F3Z2",
(5,-7.5)*{=} = "",
(35,-7.5)*{=} = "",
(65,-7.5)*{=} = "",
(-2,10)*{\sqsupseteq} = "",
(28,10)*{\sqsupseteq} = "",
(58,10)*{\sqsupseteq} = "",
(88,10)*{\sqsupseteq} = "",
(-20,3)*{\sqsupseteq} = "",
(105,0)*{\cdots} = "",
(85,20)*{\cdots} = "",
(85,-15)*{\cdots} = "",
(102,13)*{\cdots} = "",
\kar _{J!_{F1}}"F11";"F01"
\kar _{JF!_{F1}}"F21";"F11"
\kar _{JF^2!_{F1}}"F31";"F21"
\kar _(.1){JF^3!_{F1}} (100,0);"F31"
%\ar @{.} (105,0);(100,0)
%
\kar ^{c}"F0X";"F1X"
\kar ^{\overline{F}c}"F1X";"F2X"
\kar ^{\overline{F}c}"F2X";"F3X"
\kar ^(.9){\overline{F}c} "F3X";(80,20)
%\ar @{.} (80,20);(85,20)
%
%\kar _{J!_X} "F0X";"F01"
%\kar _{JF!_X} "F1X";"F11"
%\kar _{JF^2!_X} "F2X";"F21"
%\kar _{JF^3!_X} "F3X";"F31"
%
\kar _{\alpha_0} "F0X";"F01"
\kar _{\overline{F}\alpha_0} "F1X";"F11"
\kar _{\overline{F}^2\alpha_0} "F2X";"F21"
\kar _{\overline{F}^3\alpha_0} "F3X";"F31"
\kar _{J\zeta}"F0Z";"F1Z"
\kar _{JF\zeta}"F1Z";"F2Z"
\kar _{JF^2\zeta}"F2Z";"F3Z"
\kar _(.9){JF^3\zeta} "F3Z";(80,-15)
%\ar @{.} (80,-15);(85,-15)
%
\kar ^(.2){J\zeta}"F0Z2";"F1Z2"
\kar ^(.2){JF\zeta}"F1Z2";"F2Z2"
\kar ^(.2){JF^2\zeta}"F2Z2";"F3Z2"
%\kar _(.9){JF^3\zeta} "F3Z2";(80,-15)
%\ar @{.} (80,-15);(85,-15)
%
\kar _{J!_Z} "F0Z";"F01"
\kar _{JF!_Z} "F1Z";"F11"
\kar _{JF^2!_Z} "F2Z";"F21"
\kar _{JF^3!_Z} "F3Z";"F31"
\kar ^{J!_Z} "F0Z2";"F01"
\kar ^{JF!_Z} "F1Z2";"F11"
\kar ^{JF^2!_Z} "F2Z2";"F21"
\kar ^{JF^3!_Z} "F3Z2";"F31"
\kar ^(.8){g} "F0X";"F0Z2"
\kar ^(.8){\overline{F}g} "F1X";"F1Z2"
\kar ^(.8){\overline{F}^2g} "F2X";"F2Z2"
\kar ^(.8){\overline{F}^3g} "F3X";"F3Z2"
\kar @/_4mm/@{-->} _{l} "F0X";"F0Z"
\kar  @/^4mm/^{g} "F0X";"F0Z"
\end{xy}
%\label{eq:diagramexistenceWeaklyFinalCoalg1}
\end{equation*}

 Therefore $l$ is the largest homomorphism from $c$ to $J\zeta$.
\qed
%
%\qed
%\end{myproof}

%To prove Theorem~\ref{thm:inftrsitgiry}
Now we show that polynomial $F$ and $T=\giry$ constitute an infinitary trace situation.
To this end, we have to check that
 polynomial $F$ and $T=\giry$ satisfy the assumptions in
 Proposition~\ref{prop:generalconstructionnotop}. The most nontrivial is
 Assumption~(\ref{asm:existenceWeaklyFinalCoalgWeak2Lim}); there we rely
 on results in~\cite{schubert09terminalcoalgebras} 
% on Kolmogorov's consistency theorem, 
 for the fact that a limit is
 lifted to a limit. That the latter is indeed a 2-limit is not hard,
 exploiting suitable monotonicity. %Details are found in the following theorem.
 %Lem.~\ref{lem:satassumptionprob}.

\begin{thm}\label{thm:inftrsitgiry}
The combination of polynomial $F$ and $T=\giry$ constitute an infinitary
 trace situation. % (Definition~\ref{def:infiniteTraceSituation}).
%A functor $\FXi$ and $\giry$ constitute an infinite trace situation.
%with respect to the order in Def.~\ref{def:orderenrichment}. 
 %\qed
\end{thm}

\proof
We show that 
%a polynomial functor 
$F$ 
%on $\Meas$ 
and 
%a sub-Giry monad 
$\giry$ %on $\Meas$ 
satisfy Assumptions~(\ref{asm:existenceWeaklyFinalCoalgFinalSeq})--%, 
%(\ref{asm:existenceWeaklyFinalCoalgDistLaw}), (\ref{asm:existenceWeaklyFinalCoalgGFPcont}), 
% (\ref{asm:existenceWeaklyFinalCoalgTop}) and 
(\ref{asm:existenceWeaklyFinalCoalgWeak2Lim}) in
  Proposition~\ref{prop:generalconstructionnotop}.

It is known that Assumption~(\ref{asm:existenceWeaklyFinalCoalgFinalSeq}) is satisfied~\cite{schubert09terminalcoalgebras}.

It is also known that a distributive law $\lambda:F\giry\Rightarrow \giry F$ exists~\cite{cirstea10genericinfinite}.
Therefore Assumption~(\ref{asm:existenceWeaklyFinalCoalgDistLaw}) is satisfied.

Now we prove that  Assumption~(\ref{asm:existenceWeaklyFinalCoalgGFPcont}) is satisfied.
Assume that a family $(f_i:(X,\sigalg_X)\kto (Y,\sigalg_Y))_{i\in\omega}$ of Kleisli arrows  constitutes a decreasing sequence.
We can define their greatest lower bound $\bigsqcap_{i\in\omega}f_i:X\kto Y$ in a pointwise manner:
%namely, 
for $x\in X$ and $A\in\sigalg_Y$, 
\begin{equation*}
(\bigsqcap_{i\in\omega}f_i)(x)(A)=\lim_{i\to\infty}(f_i(x)(A))\,.
\end{equation*}
It is easy to see that polynomial $F$ preserves this pointwise greatest lower bound. 
Measurability of $\bigsqcap_{i\in\omega}f_i$ and $\omega^{\op}$-continuity of Kleisli composition can be 
proved in a similar manner to the proof of~\cite[Proposition~9]{brengosMP14behaviouralequivalences}.
%
%It remains to prove that $\bigsqcap_{i\in\omega}f_i$ is a measurable function from $(X,\sigalg_X)$ to $\giry(Y,\sigalg_Y)$.
%Moreover, we also have to prove the local continuity of composition. 
%They can be proved in the similar manner to the proof of~\cite[Prop.~9]{brengosMP14behaviouralequivalences}.
%Therefore Assumption~(\ref{asm:existenceWeaklyFinalCoalgGFPcont}) is satisfied.

It is easy to see that Assumption~(\ref{asm:existenceWeaklyFinalCoalgTop}) is satisfied.

Finally, we prove that Assumption~(\ref{asm:existenceWeaklyFinalCoalgWeak2Lim}) is satisfied.
If $F1$ is empty, then the limit $Z$ is also empty and Assumption~(\ref{asm:existenceWeaklyFinalCoalgWeak2Lim}) is satisfied.
Assume that $F1$ is not empty.
It is known that the sub-Giry monad $\giry$ preserves a limit over an $\omega^{\text{op}}$-sequence consisting of standard Borel spaces 
%(see~\cite{doob94measuretheory} for concrete definition) 
and surjective measurable functions~\cite[Corollary~1]{schubert09terminalcoalgebras}.
In the current setting, %, %for all $i\in\omega$, 
%as $\FSigma^i1$ is a countable set with a discrete $\sigma$-algebra, 
$F^i1$ is a standard Borel space for all $i\in\omega$ because:
 $1$ is standard Borel; $\Sigma_n$ is standard Borel for each $n\in\omega$; and 
standard Borel spaces
are closed under 
%is preserved by 
countable coproducts and countable limits~\cite[12.B]{kechris95classicaldescriptive}.
Moreover, it is easy to see that $F^i!_{F1}$ is surjective for each $i\in\omega$ because
$!_{F1}:F1\to 1$ is a surjective function and the polynomial functor $F$ preserves epimorphisms. 
Therefore by~\cite[Corollary~1]{schubert09terminalcoalgebras},
the limit $(Z,(\gamma_i:Z\to F^i1)_{i\in\omega})$ over the final $\omega^{\op}$-sequence 
$1\overset{!_{F1}}{\leftarrow}F1\overset{F!_{F1}}{\leftarrow}F^2 1\overset{F^2!_{F1}}{\leftarrow} \cdots$ is preserved by $\giry$.
This immediately implies that $J:\Meas\to\Kl(\giry)$ preserves the limit. 
It is easy to see that the resulting limit is a 2-limit.
%It is immediate from Lem.~\ref{lem:giry2lim} that assumption~\ref{asm:existenceWeaklyFinalCoalgWeak2Lim} is satisfied.

Hence by Proposition~\ref{prop:generalconstructionnotop},
$F$ and $\giry$ constitute an infinitary trace situation.
\qed
%\end{myproof}
%\end{proof:satassumptionprob}

We will later discuss another pair of a functor and a monad that can also model probabilistic systems in Section~\ref{subsec:subdistmonad}.
The proof that the pair constitutes an infinitary trace situation is very different from the one for polynomial $F$ and $T=\giry$ above,
because the axiomatic results in Proposition~\ref{prop:generalconstructiontop} and Proposition~\ref{prop:generalconstructionnotop} are not applicable.

\subsection{Kleisli Simulations for Probabilistic Systems}\label{subsec:Klsimgiry}
\subsubsection{Forward Simulations}
Soundness of forward simulation, in the current probabilistic setting, 
follows immediately from the 
 the axiomatic development in Lemma~\ref{lem:soundnessFwd}.

\begin{thm}\label{thm:soundnessFwdgiry}
Given  two $(\giry,F)$-systems $\mathcal{X}=(X,s,c)$ and $\mathcal{Y}=(Y,t,d)$,
$\mathcal{X}\fwd\mathcal{Y}$ implies $\trinf(c)\odot s\sqsubseteq \trinf(d)\odot t$.
%\qed
\end{thm}

%thm:soundnessFwdgiry
%\newtheorem*{proof:soundnessFwdgiry}{Proof of Thm.\ \ref{thm:soundnessFwdgiry}}{\itshape}{\rmfamily}
%\begin{proof:soundnessFwdgiry}
%{\proof[Proof of Thm.\ \ref{thm:soundnessFwdgiry}]
%\begin{myproof}
\proof
In a similar manner to the proof of Theorem~\ref{thm:inftrsitgiry},
we can show that $F$ and $\giry$ satisfy the assumptions in Lemma~\ref{lem:soundnessFwd}.
Therefore the statement is immediate from Lemma~\ref{lem:soundnessFwd}.
\qed
%}
%\end{myproof}
%\end{proof:soundnessFwdgiry}

\begin{exa}\label{exa:fwdsimprob}
We define a ranked alphabet $\Sigma=(\Sigma_n)_{n\in\omega}$ by $\Sigma_0=\{a\}$, $\Sigma_2=\{b\}$ and $\Sigma_i=\emptyset$ for each $i\in\mathbb{N}\setminus\{0,2\}$.
We define $(\giry,\FSigma)$-systems $\mathcal{X}=(X,s,c)$ and $\mathcal{Y}=(Y,t,d)$ as follows:
\begin{itemize}
\item $X=(\{x_1,x_2,x_3\},\pow\{x_1,x_2,x_3\})$ and $Y=(\{y_1,y_2\},\pow\{y_1,y_2\})$\,.

\item $s(\{x_1\})=s(\{x_3\})=\frac{1}{2}$, $s(\{x_2\})=0$, $t(\{y_1\})=1$ and $t(\{y_2\})=0$.

\item $c(x_1)(\{a\})=c(x_3)(\{a\})=0$, $c(x_2)(\{a\})=1$, $d(y_1)(\{a\})=0$, $d(y_2)(\{a\})=1$,  
%$c(x_i)(\{(b,x_{i'},x_{i''})\})$ and $d(y_i)(\{(b,y_{i'},y_{i''})\})$ are defined as follows:
\begin{align*}
c(x)(\{(b,x',x'')\})&=\begin{cases}
\frac{1}{2} & \bigl((x,x',x'')\in\{(x_1,x_1,x_2),(x_1,x_3,x_2)\}\bigr) \\
\frac{1}{2} & \bigl((x,x',x'')\in\{(x_3,x_2,x_1),(x_3,x_2,x_3)\}\bigr) \\
0 & (\text{otherwise})
\end{cases}
\quad\text{and}\\
d(y)(\{(b,y',y'')\})&=\begin{cases}
\frac{1}{2} & \bigl((y,y',y'')\in\{(y_1,y_1,y_2), (y_1,y_2,y_1)\}\bigr) \\
0 & (\text{otherwise})\,.
\end{cases}
\end{align*}
\end{itemize}
We can illustrate $\mathcal{X}$ and $\mathcal{Y}$ as follows.
Here $z\!\xrightarrow{a,p}\!\checkmark$ means $c(z)(\{a\})=p$ or $d(z)(\{a\})=p$, 
and $z\!\xrightarrow{b,p}\!{\raisebox{-.4pt}{$\scriptstyle\Box$}} 
{\small\begin{matrix}\rightharpoonup  \\[-2.5mm] \rightharpoondown \end{matrix}}
{\small\begin{matrix}  z_1\\[-1.5mm]  z_2\end{matrix}}
$ means $c(z)(\{(b,z_1,z_2)\})=p$ or $d(z)(\{(b,z_1,z_2)\})=p$.
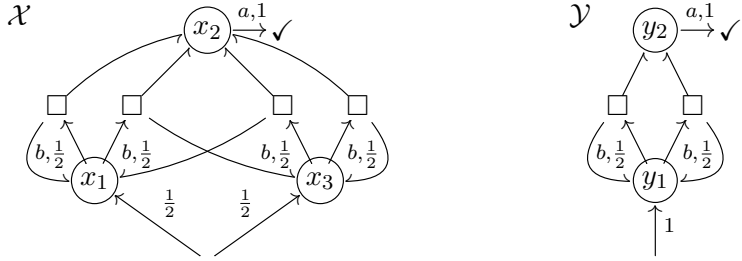
\[
\begin{xy}
(0,22)*{\mathcal{X}}="",
(10,0)*+[Fo]{x_1} = "x1",
(5,10)*{\Box} = "xb1",
(15,10)*{\Box} = "xb2",
(25,20)*+[Fo]{x_2} = "x2",
(35,20)*+{\checkmark} = "xc",
(40,0)*+[Fo]{x_3} = "y1",
(35,10)*{\Box} = "yb1",
(45,10)*{\Box} = "yb2",
%(40,20)*+[Fo]{x_4} = "y2",
%(20,20)*+[Fo]{y_3} = "x22",
%
\ar _(.4){\frac{1}{2}} (24,-10);"x1"*+++[o]{}
\ar ^(.5){b,\frac{1}{2}} "x1"*++{};"xb1"*++[o]{}
\ar _(.5){b,\frac{1}{2}} "x1"*++{};"xb2"*++[o]{}
\ar @/^2mm/ @{-_{>}} "xb1";"x2"*+++[o]{}
\ar @/_7mm/ @{-^{>}} "xb1"*++{};"x1"*+++[o]{}
\ar  @{-_{>}} "xb2";"x2"*+++[o]{}
\ar @/_2mm/ @{-^{>}} "xb2"*++{};"y1"*+++[o]{}
\ar ^(.6){a,1} "x2"*+++{};"xc"*++[o]{}
\ar ^(.4){\frac{1}{2}} (26,-10);"y1"*+++[o]{}
\ar ^(.5){b,\frac{1}{2}} "y1"*++{};"yb1"*++[o]{}
\ar _(.5){b,\frac{1}{2}} "y1"*++{};"yb2"*++[o]{}
\ar  @{-^{>}} "yb1";"x2"*+++[o]{}
\ar @/^2mm/ @{-_{>}} "yb1"*++{};"x1"*+++[o]{}
\ar  @/_2mm/ @{-^{>}} "yb2";"x2"*+++[o]{}
\ar @/^7mm/ @{-_{>}} "yb2"*++{};"y1"*+++[o]{}
%\ar _(.6){a,1} "y2"*+++{};"xc"*++[o]{}
\end{xy}
\qquad\qquad\qquad
\begin{xy}
(0,22)*{\mathcal{Y}}="",
(10,0)*+[Fo]{y_1} = "x1",
(5,10)*{\Box} = "xb1",
(15,10)*{\Box} = "xb2",
(10,20)*+[Fo]{y_2} = "x2",
(20,20)*+{\checkmark} = "xc",
%(20,20)*+[Fo]{y_3} = "x22",
%
\ar _(.4){1} (10,-10);"x1"*+++[o]{}
\ar ^(.5){b,\frac{1}{2}} "x1"*++{};"xb1"*++[o]{}
\ar _(.5){b,\frac{1}{2}} "x1"*++{};"xb2"*++[o]{}
\ar  @{-_{>}} "xb1";"x2"*+++[o]{}
\ar @/_7mm/ @{-^{>}} "xb1"*++{};"x1"*+++[o]{}
\ar  @{-^{>}} "xb2";"x2"*+++[o]{}
\ar @/^7mm/ @{-_{>}} "xb2"*++{};"x1"*+++[o]{}
\ar ^(.6){a,1} "x2"*+++{};"xc"*++[o]{}
%\ar @/^6mm/^(.4){c,\frac{1}{2}} "x22"*++{};"x1"*+++[o]{}
\end{xy}
\qquad
\]
We define a Kleisli arrow $f:Y\kto X$ in $\Kl(\giry)$ as follows:
\[
f(y_1)(\{x\})=\begin{cases} \frac{1}{2} & (x\in\{x_1,x_3\}) \\
 0 & (\text{otherwise}) \end{cases}
 \quad\text{and}\quad
f(y_2)(\{x\})=\begin{cases} 1 & (x=x_2) \\
 0 & (\text{otherwise})\,. \end{cases} 
\]
Then $f$ is a forward simulation from $\mathcal{X}$ to $\mathcal{Y}$.
By Theorem~\ref{thm:soundnessFwdgiry}, we have  $\trinf(c)\odot s\sqsubseteq \trinf(d)\odot t$.
\end{exa}

\subsubsection{Backward Simulations}
\label{subsec:bwdsimprob}
%Next 
We 
turn to
%want to prove the soundness of 
backward simulations.
%Recall that 
Similarly to the nondeterministic setting (Section~\ref{subsubsec:powbwd}),
we have to impose a certain restriction on backward Kleisli simulations to ensure soundness.
By the feature of $\giry$ that composition in $\Kl(\giry)$ is $\omega^{\op}$-continuous---a
feature absent in $\Kl(\pow)$---the  image-finiteness condition is no longer needed.

\begin{defi}[total backward simulation]\label{def:restrictedbwdsimgiry}
Let $\mathcal{X}=(X,s,c)$ and $\mathcal{Y}=(Y,t,d)$ be
 $(\giry,F)$-systems. 
A backward simulation $b:X\kto Y$ 
from $\mathcal{X}$ to $\mathcal{Y}$ is \emph{total} if
$b(x)(Y)=1$ for all $x\in X$.
%If $b$ is total, it is called
%a \emph{T-backward simulation}. 
We write $\mathcal{X}\tbwd\mathcal{Y}$ if there exists a total backward simulation from $\mathcal{X}$ to $\mathcal{Y}$.
\end{defi}

\begin{thm}[soundness of $\tbwd$]\label{thm:soundnesstifbwdgiry}
For  two $(\giry,F)$-systems $\mathcal{X}=(X,s,c)$ and $\mathcal{Y}=(Y,t,d)$,
$\mathcal{X}\tbwd\mathcal{Y}$ implies $\trinf(c)\odot s\sqsubseteq \trinf(d)\odot t$.
\qed
\end{thm}

The proof of Theorem~\ref{thm:soundnesstifbwdgiry} is via the following axiomatic development.

\begin{defi}[total backward simulation, generally]\label{def:restrictedbwdsimnotop}$\!\!\!{}^\dagger$
Let $F$ be an endofunctor and $T$ be a monad on $\mathbb{C}$ that 
satisfy the conditions in Proposition~\ref{prop:generalconstructionnotop} with respect to $\sqsubseteq$. 
%constitute an infinite trace situation with respect to an order $\sqsubseteq$.
%
For  two $(T,F)$-systems $\mathcal{X}=(X,s,c)$ and
 $\mathcal{Y}=(Y,t,d)$,
a \emph{total backward simulation} 
from $\mathcal{X}$ to $\mathcal{Y}$ 
% a \emph{total and image finite backward Kleisli simulation} (\emph{TIF-backward simulation} for short) is 
is a backward simulation $b:X\kto Y$ 
that satisfies the following condition:
\begin{enumerate}
\item 
The arrow $b:X\kto Y$ satisfies $J!_Y\odot b = J!_X$. Here $!_{Y}\colon
      Y\to 1$ is the unique function.%
\label{item:restrictedbwdsim2notop}
%\item One of the following is satisfied.
%\begin{enumerate}
%\item Assumption~\ref{asm:generalconstructiontopGFPcont} in Prop.~\ref{prop:generalconstructiontop} is satisfied. \label{item:restrictedbwdsim3cont}
\end{enumerate}
We write $\mathcal{X}\tbwd\mathcal{Y}$ if there exists a total backward simulation from $\mathcal{X}$ to $\mathcal{Y}$.
\end{defi}

This general total backward simulation satisfies soundness. For its proof
we have to look into the inductive 
 construction of the largest homomorphism in
 Section~\ref{subsec:constructlargesthomgiry} (Proposition~\ref{prop:generalconstructionnotop}).
 
 \begin{lem}\label{lem:soundnessRBwdnotop}$\!\!\!{}^\dagger$
Let $F$ and $T$ be
%We assume that $F$ and $T$ 
as in Proposition~\ref{prop:generalconstructionnotop}.
% wrt.\ $\sqsubseteq$. 
% that constitute an infinite trace situation with respect to an order $\sqsubseteq$.
%We assume that the weakly final coalgebra is obtained by using Prop.~\ref{prop:generalconstructiontop}. 
For two $(T,F)$-systems $\mathcal{X}=(X,s,c)$ and $\mathcal{Y}=(Y,t,d)$, 
$\mathcal{X}\tbwd\mathcal{Y}$ (in the sense of Definition~\ref{def:restrictedbwdsimnotop}) implies $\trinf(c)\sqsubseteq
 \trinf(d)\odot b$.
 Furthermore it follows that  $\trinf(c)\odot s\sqsubseteq \trinf(d)\odot t$.
% \qed
\end{lem}
 
%\newtheorem*{proof:soundnessRBwdnotop}{Proof of Lem.\ \ref{lem:soundnessRBwdnotop}}{\itshape}{\rmfamily}
%\begin{proof:soundnessRBwdnotop}
%{\proof[Proof of Lem.\ \ref{lem:soundnessRBwdnotop}]
%\begin{myproof}
\proof
We prove $\trinf(c)\sqsubseteq \trinf(d)\odot b$ along the construction of $\trinf(c)$ and $\trinf(d)$ 
in the proof of Proposition~\ref{prop:generalconstructionnotop}.

For each $i\in\omega$, we define $\Psi^{i}_c(J!_X):X\kto 1$ and $\Psi^{i}_d(J!_Y):Y\kto 1$ 
as in the proof of Proposition~\ref{prop:generalconstructionnotop}.
We first prove $\Psi^{i}_c(J!_X) \sqsubseteq \Psi^{i}_d(J!_Y)\odot b$ 
for all $i\in\omega$ by  induction on $i$ as follows.
%It is proved by the  induction on $i$.

If $i=0$, by  Assumption~(\ref{item:restrictedbwdsim2notop}) in Definition~\ref{def:restrictedbwdsimnotop}, 
we have $\Psi^0_X(J!_X)=\Psi^0_Y(J!_Y)\odot b$.

Let $i>0$ and
assume that $\Psi^{i-1}_c(J!_X)\sqsubseteq \Psi^{i-1}_d(J!_Y)\odot b$.
Then 
\begin{equation}
\label{eq:lem:soundnessRBwdnotop160107}
\begin{aligned}
\Psi^{i}_c(J!_X)
&= J!_{F1}\odot \overline{F}(\Psi^{i-1}_c(J!_X)) \odot d & (\text{by definition of $\Psi_c$ }) \\
&\sqsubseteq J!_{F1}\odot \overline{F}(\Psi^{i-1}_d(J!_Y))\odot \overline{F}b \odot c  & (\text{by induction hypothesis}) \\
&\sqsubseteq J!_{F1}\odot \overline{F}(\Psi^{i-1}_d(J!_Y))\odot d\odot b & (\text{$b$ is a backward simulation}) \\
&=  \Psi^{i}_d(J!_Y)\odot b & (\text{by definition of $\Psi_d$})&\,.
\end{aligned}
\end{equation}

Therefore we have $\Psi^{i}_c(J!_X)\sqsubseteq \Psi^{i}_d(J!_Y)\odot b$ for all $i\in\omega$.

\begin{equation*}
\begin{xy}
(0,0)*+{1} = "F01",
(20,0)*+{\overline{F}1} = "F11",
(40,0)*+{\overline{F}^21} = "F21",
(60,0)*+{\overline{F}^31} = "F31",
(0,15)*+{Y} = "F0Y",
(20,15)*+{\overline{F}Y} = "F1Y",
(40,15)*+{\overline{F}^2Y} = "F2Y",
(60,15)*+{\overline{F}^3Y} = "F3Y",
(-15,30)*+{X} = "F0X",
(5,30)*+{\overline{F}X} = "F1X",
(25,30)*+{\overline{F}^2X} = "F2X",
(45,30)*+{\overline{F}^3X} = "F3X",
(12,7.5)*{\sqsupseteq} = "",
(32,7.5)*{\sqsupseteq} = "",
(52,7.5)*{\sqsupseteq} = "",
(3,22.5)*{\sqsupseteq} = "",
(23,22.5)*{\sqsupseteq} = "",
(43,22.5)*{\sqsupseteq} = "",
(-5,14)*{=} = "",
(13,18)*{{\scriptstyle=}} = "",
(33,18)*{{\scriptstyle=}} = "",
(53,18)*{{\scriptstyle=}} = "",
%
%(-5,-15)*+{Z} = "F0Z",
%(15,-15)*+{\overline{F}Z} = "F1Z",
%(35,-15)*+{\overline{F}^2Z} = "F2Z",
%(55,-15)*+{\overline{F}^3Z} = "F3Z",
%
%(8,-7.5)*{=} = "",
%(28,-7.5)*{=} = "",
%(48,-7.5)*{=} = "",
%
%(70,-15)*{\cdots} = "",
(75,0)*{\cdots} = "",
(75,15)*{\cdots} = "",
(60,30)*{\cdots} = "",
\kar _{J!_{F1}}"F11";"F01"
\kar _{JF!_{F1}}"F21";"F11"
\kar _{JF^2!_{F1}}"F31";"F21"
\kar _(.1){JF^3!_{F1}} (70,0);"F31"
%\ar @{.} (75,0);(70,0)
%
\kar _(.3){d}"F0Y";"F1Y"
\kar _(.3){\overline{F}d}"F1Y";"F2Y"
\kar _(.3){\overline{F}^2d}"F2Y";"F3Y"
\kar ^(.9){\overline{F}^3d} "F3Y";(70,15)
%\ar @{.} (70,15);(75,15)
%
\kar ^{c}"F0X";"F1X"
\kar ^{\overline{F}c}"F1X";"F2X"
\kar ^{\overline{F}^2c}"F2X";"F3X"
\kar ^(.9){\overline{F}^3c} "F3X";(55,30)
%\ar @{.} (55,30);(60,30)
%
\kar ^{J!_Y} "F0Y";"F01"
\kar ^{JF!_Y} "F1Y";"F11"
\kar ^{JF^2!_Y} "F2Y";"F21"
\kar ^{JF^3!_Y} "F3Y";"F31"
\kar ^{b} "F0X";"F0Y"
\kar ^{\overline{F}b} "F1X";"F1Y"
\kar ^{\overline{F}^2b} "F2X";"F2Y"
\kar ^{\overline{F}^3b} "F3X";"F3Y"
\kar @/_1mm/ _{J!_X} "F0X";"F01"
\kar @/_1mm/ _(.35){JF!_X} "F1X";"F11"
\kar @/_1mm/ _(.35){JF^2!_X} "F2X";"F21"
\kar @/_1mm/ _(.35){JF^3!_X} "F3X";"F31"
%
%\kar _{J\zeta}"F0Z";"F1Z"
%\kar _{JF\zeta}"F1Z";"F2Z"
%\kar _{JF^2\zeta}"F2Z";"F3Z"
%\kar _(.9){JF^3\zeta} "F3Z";(65,-15)
%\ar @{.} (65,-15);(70,-15)
%
%\kar _{J!_Z} "F0Z";"F01"
%\kar _{JF!_Z} "F1Z";"F11"
%\kar _{JF^2!_Z} "F2Z";"F21"
%\kar _{JF^3!_Z} "F3Z";"F31"
\end{xy}
%\label{eq:soundnessbwdsim1}
\end{equation*}
%
%Using $c$ and $d$, % respectively.

Now
let 
%we define cones 
$(X,(\alpha^X_i:X\kto \overline{F}^i1)_{i\in\omega})$ and $(Y,(\alpha^Y_i:Y\kto \overline{F}^i1)_{i\in\omega})$
be cones 
over the sequence
$1\overset{J!_{F1}}{\kleftarrow} \overline{F}1\overset{JF!_{F1}}{\kleftarrow} \overline{F}^2 1\overset{JF^2!_{F1}}{\kleftarrow} \cdots$;
%  $1\kleftarrow \overline{F}1\kleftarrow \overline{F}^2 1\kleftarrow \ldots$ 
they are defined by the equation (\ref{eq:existenceWeaklyFinalCoalgdefcone}) 
in the proof of Proposition~\ref{prop:generalconstructionnotop}.
%
%As we have shown in the proof of Prop.~\ref{prop:generalconstructionnotop}, 
Recall that
 $\trinf(c):X\kto Z$ is the unique mediating arrow from the cone $(X,(\alpha^X_i:X\kto \overline{F}^i1)_{i\in\omega})$ 
to the  2-limit $(Z,(J\gamma_i:Z\kto \overline{F}^i1)_{i\in\omega})$, and 
%Moreover, 
similarly $\trinf(d):Y\kto Z$ is the unique mediating arrow from $(Y,(\alpha^Y_i:Y\kto \overline{F}^i1)_{i\in\omega})$ 
to the same 2-limit. %$(Z,(J\gamma_i:Z\kto \overline{F}^i1)_{i\in\omega})$.
% and therefore 
Note here that for each $i\in\omega$, we have:
\begin{displaymath}
J\gamma_i\odot(\trinf(d)\odot b)  \,=\, (J\gamma_i\odot\trinf(d))\odot b\,=\, \alpha^Y_i\odot b\,.
\end{displaymath}
%Note that
Therefore
$\trinf(d)\odot b:X\kto Z$ is the unique mediating arrow from a cone $(X,(\alpha^Y_i\odot b:X\kto \overline{F}^i1)_{i\in\omega})$ 
to $(Z,(J\gamma_i:Z\kto \overline{F}^i1)_{i\in\omega})$.
%, and therefore 
%$\trinf(d)\odot b$ is a mediating arrow.
% from a cone $(X,(\alpha^Y_i\odot b:X\kto \overline{F}^i1)_{i\in\omega})$
%to $(Z,(J\gamma_i:Z\kto \overline{F}^i1)_{i\in\omega})$.
%Hence by the uniqueness of the mediating arrow, we have $\trinf(c)=\trinf(d)\odot b$.
%By Assumption~\ref{item:restrictedbwdsim4}, $l^Y_{\max}\odot b:X\kto Z$ is the largest mediating arrow from a cone $(X,(\alpha^Y_i\odot b:X\kto \overline{F}^i1)_{i\in\omega})$ 
%to $(Z,(J\gamma_i:Z\kto \overline{F}^i1)_{i\in\omega})$.
%
We prove $\alpha^X_i\sqsubseteq\alpha^Y_i\odot b$ for all $i\in\omega$ by induction on $i$ as follows:
%For all $i\in\omega$, we can prove $\alpha^X_i\sqsubseteq\alpha^Y_i\odot b$ by the induction on $i$ as follows:

For $i=0$, we have:
\begin{align*}
\alpha^X_0&=\bigsqcap_{i\in\omega}\Psi^{i}_c(J!_X) & (\text{by definition}) \\
&\sqsubseteq \bigsqcap_{i\in\omega}(\Psi^{i}_d(J!_Y)\odot b) & 
(\text{by (\ref{eq:lem:soundnessRBwdnotop160107})})\\
%(\text{by the above result}) \\
&=\bigl(\bigsqcap_{i\in\omega}\Psi^{i}_d(J!_Y)\bigr)\odot b & (\text{by Assumption~(\ref{asm:existenceWeaklyFinalCoalgGFPcont}) of Proposition~\ref{prop:generalconstructionnotop}}) \\
&= \alpha^Y_0\odot b & (\text{by definition})&\,.
\end{align*}
%$\alpha^X_0=\Psi^{\omega}_X(J!_X)\sqsubseteq \Psi^{\omega}_Y(J!_Y)\odot b=\alpha^Y_0\odot b$

Let $i>0$ and assume that $\alpha^X_{i-1}\sqsubseteq\alpha^Y_{i-1}\odot b$. Then
\begin{align*}
\alpha^X_i 
&= \overline{F}\alpha^X_{i-1}\odot c & (\text{by definition}) \\
&\sqsubseteq \overline{F}\alpha^Y_{i-1}\odot\overline{F}b\odot c & (\text{by induction hypothesis and the monotonicity of $\overline{F}$}) \\
&\sqsubseteq \overline{F}\alpha^Y_{i-1}\odot d  \odot b & (\text{$b$ is a backward simulation}) \\
&= \alpha^Y_i \odot b & (\text{by definition})&.
\end{align*}

%Moreover, by the monotonicity of $\overline{F}$'s action on arrows and composition in $\Kl(T)$, 
%we can inductively prove that $\alpha^X_i\sqsubseteq\alpha^Y_i\odot b$ for all $i\in\omega$.
Therefore we have $\alpha^X_i\sqsubseteq\alpha^Y_i\odot b$ for all $i\in\omega$.
As $(Z,(J\gamma_i:Z\kto \overline{F}^i1)_{i\in\omega})$ is a  2-limit, this implies $\trinf(c)\sqsubseteq \trinf(d)\odot b$.

\auxproof{
\begin{equation*}
\begin{xy}
(0,0)*+{1} = "F01",
(20,0)*+{\overline{F}1} = "F11",
(40,0)*+{\overline{F}^21} = "F21",
(60,0)*+{\overline{F}^31} = "F31",
(0,15)*+{Y} = "F0Y",
(20,15)*+{\overline{F}Y} = "F1Y",
(40,15)*+{\overline{F}^2Y} = "F2Y",
(60,15)*+{\overline{F}^3Y} = "F3Y",
(-15,30)*+{X} = "F0X",
(5,30)*+{\overline{F}X} = "F1X",
(25,30)*+{\overline{F}^2X} = "F2X",
(45,30)*+{\overline{F}^3X} = "F3X",
(12,7.5)*{=} = "",
(32,7.5)*{=} = "",
(52,7.5)*{=} = "",
(3,22.5)*{\sqsupseteq} = "",
(23,22.5)*{\sqsupseteq} = "",
(43,22.5)*{\sqsupseteq} = "",
(-5,14)*{\sqsubseteq} = "",
(13,18)*{{\scriptstyle\sqsubseteq}} = "",
(33,18)*{{\scriptstyle\sqsubseteq}} = "",
(53,18)*{{\scriptstyle\sqsubseteq}} = "",
(-7,-7)*{{\scriptstyle\sqsubseteq}} = "",
(-5,-15)*+{Z} = "F0Z",
(15,-15)*+{\overline{F}Z} = "F1Z",
(35,-15)*+{\overline{F}^2Z} = "F2Z",
(55,-15)*+{\overline{F}^3Z} = "F3Z",
(8,-7.5)*{=} = "",
(28,-7.5)*{=} = "",
(48,-7.5)*{=} = "",
(70,-15)*{\cdots} = "",
(75,0)*{\cdots} = "",
(75,15)*{\cdots} = "",
(60,30)*{\cdots} = "",
\kar _{J!_{F1}}"F11";"F01"
\kar _{JF!_{F1}}"F21";"F11"
\kar _{JF^2!_{F1}}"F31";"F21"
\kar _(.1){JF^3!_{F1}} (70,0);"F31"
%\ar @{.} (75,0);(70,0)
%
\kar _(.3){d}"F0Y";"F1Y"
\kar _(.3){\overline{F}d}"F1Y";"F2Y"
\kar _(.3){\overline{F}^2d}"F2Y";"F3Y"
\kar ^(.9){\overline{F}^3d} "F3Y";(70,15)
%\ar @{.} (70,15);(75,15)
%
\kar ^{c}"F0X";"F1X"
\kar ^{\overline{F}c}"F1X";"F2X"
\kar ^{\overline{F}^2c}"F2X";"F3X"
\kar ^(.9){\overline{F}^3c} "F3X";(55,30)
%\ar @{.} (55,30);(60,30)
%
\kar ^{\alpha^Y_0} "F0Y";"F01"
\kar ^{\overline{F}\alpha^Y_0} "F1Y";"F11"
\kar ^{\overline{F}^2\alpha^Y_0} "F2Y";"F21"
\kar ^{\overline{F}^3\alpha^Y_0} "F3Y";"F31"
\kar ^{b} "F0X";"F0Y"
\kar ^{\overline{F}b} "F1X";"F1Y"
\kar ^{\overline{F}^2b} "F2X";"F2Y"
\kar ^{\overline{F}^3b} "F3X";"F3Y"
\kar @/_1mm/ _{\alpha^X_0} "F0X";"F01"
\kar @/_1mm/ _(.35){\overline{F}\alpha^X_0} "F1X";"F11"
\kar @/_1mm/ _(.35){\overline{F}^2\alpha^X_0} "F2X";"F21"
\kar @/_1mm/ _(.35){\overline{F}^3\alpha^X_0} "F3X";"F31"
\kar _{J\zeta}"F0Z";"F1Z"
\kar _{JF\zeta}"F1Z";"F2Z"
\kar _{JF^2\zeta}"F2Z";"F3Z"
\kar _(.9){JF^3\zeta} "F3Z";(65,-15)
%\ar @{.} (65,-15);(70,-15)
%
\kar _{J!_Z} "F0Z";"F01"
\kar _{JF!_Z} "F1Z";"F11"
\kar _{JF^2!_Z} "F2Z";"F21"
\kar _{JF^3!_Z} "F3Z";"F31"
\kar @/_8mm/ @{-->} _{\trinf(c)} "F0X";"F0Z"
\kar @/_2mm/ @{-->} _{\trinf(d)} "F0Y";"F0Z"
\end{xy}
%\label{eq:soundnessbwdsim2}
\end{equation*}
}

The last claim follows from $b$'s condition on initial states.
\qed
%}
%\end{myproof}
%\end{proof:soundnessRBwdnotop}

\begin{lem}\label{lem:whenrestrictedgiry}
In Definition~\ref{def:restrictedbwdsimnotop}, if $T=\giry$ and $F$ is a polynomial functor, 
Assumption~(\ref{item:restrictedbwdsim2notop}) is satisfied if $b(x)(Y)=1$ for each $x\in X$.
\end{lem}

%\newtheorem*{proof:whenrestrictedgiry}{Proof of Prop.\ \ref{prop:whenrestrictedgiry}}{\itshape}{\rmfamily}
%\begin{proof:whenrestrictedgiry}
%\begin{myproof}
\proof
%%\noindent{\bf \ref{item:whenrestrictedgiry2}.}
%Next, let $T=\giry$ and $F=\FXi$.
We assume that $b(x)(Y)=1$ for each $x\in X$. %prove that Assumption~\ref{item:restrictedbwdsim2notop}.
By the definition of multiplication $\mu^{\giry}$ of the sub-Giry monad (see Definition~\ref{def:variousmonads}), for  $x\in X$,
we have:
\begin{equation*}
\bigl(J!_Y\odot b\bigr)(x)(1)=b(x)\bigl(!_Y^{-1}(1)\bigr)=b(x)(Y)=1=J!_X(x)(1)\,.
\end{equation*}
Therefore Assumption~(\ref{item:restrictedbwdsim2notop}) is satisfied.
\qed
%\end{myproof}
%\end{proof:whenrestrictedgiry}

%\newtheorem*{proof:soundnesstifbwdgiry}{Proof of Thm.\ \ref{thm:soundnesstifbwdgiry}}{\itshape}{\rmfamily}
%\begin{proof:soundnesstifbwdgiry}
{\proof[Proof of Theorem \ref{thm:soundnesstifbwdgiry}]
In Lemma~\ref{lem:whenrestrictedgiry} we prove that a total backward
 simulation
in the specific sense of Definition~\ref{def:restrictedbwdsimgiry} is also 
a total backward
 simulation in the general sense of Definition~\ref{def:restrictedbwdsimgiry}.
 Therefore Lemma~\ref{lem:soundnessRBwdnotop} yields trace inclusion.
% It is immediate from  Lem.~\ref{lem:soundnessRBwdtop} and that
% if $T=\pow$ and $F=\FSigma$, for a backward simulatio $b:X\kto Y$,
% Assumption~\ref{item:restrictedbwdsim2} In Def.~\ref{def:restrictedbwdsimtop} is satisfied if $b(x)\neq\emptyset$ for each $x\in X$, 
% while Assumption~\ref{item:restrictedbwdsim3} is satisfied if $b(x)$ is finite for each $x\in X$ (Lem.~\ref{lem:whenrestrictedpow}).
\qed}
%\end{proof:soundnesstifbwdgiry}

\begin{exa}\label{exa:bwdsimprob}
We continue Example~\ref{exa:fwdsimprob}.
The arrow $f:Y\kto X$ in Example~\ref{exa:fwdsimprob} is also a total backward simulation from $\mathcal{Y}$ to $\mathcal{X}$.
By Theorem~\ref{thm:soundnesstifbwdgiry}, this implies
$\trinf(c)\odot s\sqsupseteq \trinf(d)\odot t$, 
and therefore together with the forward simulation in Example~\ref{exa:fwdsimprob}, 
we have $\trinf(c)\odot s= \trinf(d)\odot t$.
\end{exa}

\subsection{Forward Partial Execution for Probabilistic Systems}\label{subsec:FPEgiry}
%We show that FPE can be used to aid discovery of forward and backward
% simulations,
%also in the current probabilistic setting.
%

\begin{thm}\label{thm:propertiesFPEgiry}
Let $F$ be a polynomial functor on $\Meas$.
For $(\giry,F)$-systems $\mathcal{X}=(X,s,c)$ and $\mathcal{Y}=(Y,t,d)$, the following  hold.
\begin{enumerate}[itemsep=1ex]
\item
\begin{enumerate}
\item (soundness of FPE for forward simulation)  $\mathcal{X}_{\FPE}\fwd\mathcal{Y}$ implies $\trinf(c)\odot s\sqsubseteq \trinf(d)\odot t$.
\label{item:soundnessfpefwdgiry}
\item (adequacy of FPE for forward simulation) $\mathcal{X}\fwd\mathcal{Y}$ implies $\mathcal{X}_{\FPE}\fwd\mathcal{Y}$.
\label{item:adequacyfpefwdgiry}
\end{enumerate}
\label{item:safpefwdgiry}
\item
\begin{enumerate}
\item (soundness of FPE for backward simulation)  $\mathcal{X}\tbwd\mathcal{Y}_{\FPE}$ implies $\trinf(c)\odot s\sqsubseteq \trinf(d)\odot t$.%
\label{item:soundnessfpebwdtopgiry}
\item (adequacy of FPE for backward simulation) 
$\mathcal{X}\tbwd\mathcal{Y}$ implies
      $\mathcal{X}\tbwd\mathcal{Y}_{\FPE}$, assuming that:
$d(y)(FY)=1$ for all $y\in Y$. \qed
%  the following hold.
% \begin{enumerate}
% \item 
% $d(y)\neq\emptyset$ for all $y\in Y$.
% \label{item:restrictedfpe2topgiry}
% \qed
% \end{enumerate}
\label{item:adequacyfpebwdtopgiry}
\end{enumerate}
\label{item:safpebwdgiry}
\end{enumerate}
\end{thm}

The item~(\ref{item:safpefwdgiry}) for forward simulations follows
immediately from Theorem~\ref{thm:soundnessFPEfwd}. For the relationship to
backward simulations, 
%we develop another general result.
we develop the following general result that
 can be proved in a similar manner to the proof of Theorem~\ref{thm:propertiesFPEpow}.

% The proof of soundness and adequacy of FPE for forward simulation can be 
% proved using axiomatic result in Thm.~\ref{thm:soundnessFPEfwd}.
% Those of FPE for backward simulation is derived from the following axiomatic result.

\begin{thm}[FPE and backward simulation]%[correctness of FPE for restricted bwd.\ sim.]
\label{thm:soundnessFPEbwdnotop}$\!\!\!{}^\dagger$
Let $F$ be an endofunctor and $T$ be a monad on $\mathbb{C}$
that satisfy the conditions in Proposition~\ref{prop:generalconstructionnotop}
 (hence those in Lemma~\ref{lem:soundnessRBwdnotop}). 
Let $\mathcal{X}=(X,s,c)$ and $\mathcal{Y}=(Y,t,d)$ be $(T,F)$-systems. % and $Y=Y_1+Y_2$.
\begin{enumerate}
\item (soundness for backward simulation)  $\mathcal{X}\tbwd\mathcal{Y}_{\FPE}$ implies $\trinf(c)\odot s\sqsubseteq \trinf(d)\odot t$.
\label{item:soundnessfpebwdnotop}
\item (adequacy for backward simulation) 
$\mathcal{X}\tbwd\mathcal{Y}$ implies
      $\mathcal{X}\tbwd\mathcal{Y}_{\FPE}$, 
assuming that:
the coalgebra $d:Y\kto \overline{F}Y$ satisfies 
$J!_{FY}\odot d = J!_Y$. 
\qed
% if the following conditions are satisfied.
% \begin{enumerate}
% \item 
% The coalgebra $d:Y\kto \overline{F}Y$ satisfies 
% $J!_{FY}\odot d = J!_Y$. 
% \label{item:restrictedfpe2notop}
% \qed
% \end{enumerate}
\label{item:adequacyfpebwdnotop}
\end{enumerate}
\end{thm}

%\proof
%This can be proved in a similar manner to the proof of Thm.~\ref{thm:propertiesFPEpow}.
%\qed

\auxproof{
%\newtheorem*{proof:propertiesFPEgiry}{Proof of Thm.\ \ref{thm:propertiesFPEgiry}}{\itshape}{\rmfamily}
%\begin{proof:propertiesFPEgiry}
{\proof[Proof of Theorem \ref{thm:propertiesFPEgiry}]
The statement (\ref{item:safpefwdgiry}) is immediate from Theorem~\ref{thm:soundnessFPEfwd}.
In a similar manner to Lemma~\ref{lem:whenrestrictedgiry},
we can prove the statement (\ref{item:safpebwdgiry}) using Theorem~\ref{thm:soundnessFPEbwdnotop}.
\qed}
%\end{proof:propertiesFPEgiry}
}

%\subsection{Coincidence between Coalgebraic Infinitary Trace Semantics and Automata-theoretic Semantics}
\subsection{Coincidence between Automata-theoretic and Coalgebraic Infinitary Trace Semantics}
\label{subsec:automcharagiry}
%\begin{itemize}
%\item Kolmogorov \to Carath\'eodory~\cite{ashD2000probability}
%\item Branching process~\cite{harris64theorybranching}
%\item Reachability in MC~\cite{baier08principlesof}
%\end{itemize}
In this section we again give a sanity-check result like in Section~\ref{subsec:automcharapow}.
Namely, we show a coincidence between
 coalgebraic infinitary trace semantics that is defined in Section~\ref{subsec:constructlargesthomgiry},
 %the previous section, 
 and
 infinitary %probabilistic tree 
 language that is defined using automata-theoretic terms.
 Although its statement is a sanity-check, the result requires somewhat delicate treatment of
 measure-theoretic structures.
 We first describe the definition of the latter.
% The latter is characterized using the notion of \emph{branching processes} 
% (see e.g.~\cite{harris64theorybranching})---a kind of Markov process.

In this section, for simplicity, we assume that both the state space $(X,\sigalg_X)$ and
all components $(\Sigma_n,\sigalg_{\Sigma_n})$ in the ranked alphabet $\Sigma=\bigl((\Sigma_n,\sigalg_{\Sigma_n})\bigr)_{n\in\omega}$ are countable sets with the discrete $\sigma$-algebras.
%We note that 
It is not difficult to generalize the results in this section for automata 
labeled with a general standard Borel ranked alphabet $\Sigma$. 

\begin{defi}
Let $\Sigma=\bigl((\Sigma_n,\pow\Sigma_n)\bigr)_{n\in\omega}$ be a standard Borel ranked alphabet such that all $\Sigma_n$ are countable sets %and 
equipped with the discrete $\sigma$-algebras. % for all $n\in\omega$.
%For a ranked alphabet $\Sigma=(\Sigma_n)_{n\in\omega}$ where $\Sigma_n$ is countable for each $n\in\omega$, 
A \emph{$\Sigma$-labeled probabilistic tree automaton} is a $(\giry,\FSigma)$-system $\mathcal{X}=(X,s,c)$ 
where $X$ is a countable set equipped with the discrete $\sigma$-algebra.
\end{defi}
%Here, differently from \S\ref{sec:powersetmonad},
%the state space is restricted to finite. 

For a given $\Sigma$-labeled probabilistic tree automaton $\mathcal{X}=(X,s,c)$,
we now define its automata-theoretic (infinitary) language.
It is defined as a probability measure $\langinf(\mathcal{X})$ on a set $\Treeinf(\Sigma)$ of infinitary trees.
The definitions are all as usual.
%To this end, we first fix a $\sigma$-algebra on $\Treeinf(\Sigma)$ as follows.
%It is given using a notion of \emph{cylinder sets} (see e.g.~\cite{ashD2000probability}).

\begin{defi}
For a standard Borel ranked alphabet $\Sigma=((\Sigma_n,\pow\Sigma_n))_{n\in\omega}$, we define a set 
$\mathcal{S}_{\Sigma}\subseteq \pow(\Treeinf(\Sigma))$ %over $\Sigma$ is defined 
by
$%\begin{equation*}
\mathcal{S}_{\Sigma}=\{\cyl(t)\mid k\in\omega, t\in\Tree^k(\Sigma)\}
$, %\end{equation*}
where $\cyl(t)$ is from Definition~\ref{def:treeNotions}.
A $\sigma$-algebra $\sigalg_{\infty}$ on $\Treeinf(\Sigma)$ is
%defined as
 the smallest $\sigma$-algebra that contains $\mathcal{S}_{\Sigma}$.
%$\{\cyl(t)\mid k\in\omega, t\in\Treeinf^k(\Sigma)\}\subseteq  \pow(\Treeinf(\Sigma))$.
%$\mathcal{S}_{\Sigma}$.
%follows.
%\begin{equation*}
%\sigalg_{\Treeinf(\Sigma)}=\sigma_{\Treeinf(\Sigma)}()
%\end{equation*}
%Here, for $A\subseteq X$, $\sigma_X(A)$ denotes the smallest $\sigma$-algebra that contains $A$.
\end{defi}

%we employ the notion of \emph{branching process} (see e.g.~\cite{harris64theorybranching}).
To define a probability measure $\langinf(\mathcal{X})$ on a measurable space $(\Treeinf(\Sigma),\sigalg_{\infty})$, 
we have to fix a value $\langinf(\mathcal{X})(A)$ for each  $A\in\sigalg_{\infty}$. 
%In fact,
As is standard,
 by Carath\'eodory's extension theorem (see e.g.~\cite{ashD2000probability}),
it suffices to fix a value $\langinf(\mathcal{X})(A)$ for all cylinders $A=\cyl(t)$ in a ``compatible'' manner.

%However, before that, 
To this end,
we first review the notion of \emph{branching process} (see e.g.~\cite{harris64theorybranching}).
It is used to fix the value $\langinf(\mathcal{X})(\Treeinf(\Sigma))$ 
(note that $\Treeinf(\Sigma)$ %=\cyl(\empseq)$).
is a cylinder set induced by the $0$-prefix tree). 
Intuitively the value is  probability with which the probabilistic automaton does not abort.
%We characterize these values using the notion of \emph{branching process} (see e.g.~\cite{harris64theorybranching}).
%We first fix a value $\lang(\mathcal{X})(\cyl(\empseq))$ where 

\begin{defi}%[branching process]
A \emph{branching process} is a pair $\Delta=(\Gamma,\tau)$ consisting of 
a finite set $\Gamma$ of \emph{types}, and
a \emph{transition function} $\tau: \Gamma\times \Gamma^*\to [0,1]$
such that $\sum_{\alpha\in  \Gamma^*} (x,\alpha)=1$ for all $x\in\Gamma$. %, and

A branching process $\Delta$ and an \emph{initial process} $x_0\in\Gamma$ give rise to a 
Markov chain $\mathcal{M}_{\Delta,x_0}$ such that:
the state space is a set $\Gamma^*$ of \emph{population of processes}; 
the transition function $\tau_{\mathcal{M}}:\Gamma^*\times\Gamma^*\to[0,1]$ is given by 
\begin{displaymath}
\tau_{\mathcal{M}}\bigl(\langle x_0\ldots x_{n-1}\rangle,\beta\bigr)=\sum_{\substack{\beta_0,\ldots,\beta_{n-1}\in\Gamma^* \\ \text{s.t. }
\beta=\beta_0\ldots\beta_{n-1}}}
\prod_{1\leq i\leq n-1}\tau(x_i,\beta_i); %\;\;\;\;\;\text{and}
\end{displaymath}
and the initial state is a singleton tuple $\langle x_0\rangle$.
Here juxtaposition $\beta_0\ldots \beta_{n-1}$ denotes the concatenation of tuples.
%$\hookrightarrow\subseteq \Gamma\times\Gamma^*$ of \emph{transition rules}, 
%a \emph{transition distribution} $\text{Prob}:\hookrightarrow\to[0,1]$ such that $\sum_{\alpha\in \Gamma^*}\text{Prob}(x,\alpha)=1$ for all $x\in\Gamma$, and
%a initial process $x_0\in\Gamma$.
For 
%a branching process $\Delta=(\Gamma,\tau)$ and 
a pair of types $x_0,x\in \Gamma$, 
the \emph{probability of reaching $x$ from $x_0$} is the value $\text{Reach}(\Delta,x_0,x)\in[0,1]$ with which
a population 
%where a type
that has a type
 $x$ in it
 %is appearing 
 is reached in $\mathcal{M}_{\Delta,x_0}$.
\end{defi}
%
%Here, we give an intuitive semantics of branching processes.
%Formal definition can be found in~\cite{harris64theorybranching}, for example.
%A branching process $\Delta=(\Gamma,\tau)$ and an \emph{initial process} $x_0\in\Gamma$ 
%give rise to a discrete-time a Markov chain $\mathcal{M}_{\Delta,x_0}$.
%Its state space is given by $\Gamma^*$ where $\alpha\in\Gamma^*$ can be regarded as a population of processes.
% where there are $\alpha(x)$ processes of type $x$ for each  $x \in\Gamma$.
Intuitively, %in a branching process,
%we start from an initial population $\langle x_0\rangle$ in $\Gamma^*$ that means there is only one process of type $x_0$. 
in every transition of a branching process, each process in the population gives birth to child processes randomly.
The probability that a process $x$ gives birth to children represented by a population $\alpha\in\Gamma^*$
is given by $\tau(x,\alpha)$.

From a given $\Sigma$-labeled probabilistic tree automaton $\mathcal{X}$, 
we can obtain a branching process $\Delta_{\mathcal{X}}$ by 
adding a new process $\bot$ that means aborting of the system, and by
``forgetting'' the labels on transitions.
%
%Formal definition can be found in Def.~\ref{def:skeleton}.

\begin{defi}\label{def:skeleton}
For a $\Sigma$-labeled probabilistic tree automaton  $\mathcal{X}=(X,s,c)$,
its \emph{skeleton} is a branching process $\Delta_{\mathcal{X}}=(\Gamma_{\mathcal{X}},\tau_{\mathcal{X}})$ where
%Here 
%types is defined by 
$\Gamma_{\mathcal{X}}=X+\{\bot\}$
and
%A transition function 
$\tau_{\mathcal{X}}$ 
is defined as follows.
\begin{equation*}
\tau_{\mathcal{X}}(x,\alpha)=\begin{cases}
%s(*)(y) & (x=\bullet, \alpha=\langle y\rangle) \\
\sum_{a\in\Sigma_n} c(x)\bigl(\{(a,x_0,\ldots,x_{n-1})\}\bigr)  & (x\in X, \alpha=\langle x_0,\ldots,x_{n-1}\rangle\in X^*) \\
%c(x)(\checkmark)  & (x\in X, \alpha=\langle\checkmark\rangle) \\
1-\sum_{\beta\in X^*} \tau(x,\beta)  & (x\in X, \alpha=\langle\bot\rangle) \\
1 & (x\in\{\bot\}, \alpha=\langle \bot\rangle) \\ 
0 & (\text{otherwise})\,
\end{cases}
\end{equation*}
%An initial state is given by $x_0=\bullet$.
\end{defi}

Now we are ready to define a value $\langinf(\mathcal{X})(\cyl(t))$ for each prefix tree $t$.
In particular, the value $\langinf(\mathcal{X})(\Treeinf(\Sigma))$ is defined as 
the probability with which the state $\bot$ is not reached  from $x$ in the skeleton $\Delta_{\mathcal{X}}$.

\begin{prop}\label{prop:existencesemantics}
Let $\mathcal{X}=(X,s,c)$ be a $\Sigma$-labeled probabilistic tree automaton. 
For a state $x\in X$, a natural number $k\in\omega$ and a $k$-prefix tree $t\in \Tree^k(\Sigma)$, 
we define the value $\nu_x(t)\in[0,1]$ by induction on $k$ as follows.
\begin{itemize}
\item If $k=0$, then 
$%\begin{equation}
\nu_x(t)
%=\nu(\Treeinf(\Sigma))
=1-\text{Reach}(\Delta_{\mathcal{X}},x,\bot) \,.
$ %\label{eq:def:probtreeautomaton1}%\end{equation}
\item Let $k>0$. If $t=(D,l)$, $l(\empseq)=a\in\Sigma_n$ and $t_i$ is the $i$-th subtree of $t$, then:
%If $t=(D,l)\in \Treeinf^{k+1}(\Sigma)$
%where $l(\empseq)=a\in\Sigma_n$ and $t_i$ is the $i$'th subtree of $t$,
%then 
%Then
\begin{equation*}
\nu_x(t)=\sum_{x_0,\ldots,x_{n-1}\in X} \Bigl(c(x)\bigl(\{(a,x_0,\ldots,x_{n-1})\}\bigr)\cdot \prod_{i=0}^{n-1} \nu_{x_i}(t_i)\Bigr)\,.
%\nu_x(t)=\int_{(x_0,\ldots,x_{n-1})\in X^n} \bigl(\prod_{i=0}^{n-1} \nu_{x_i}(t_i)\bigr)d(c_a(x))\,.
%\label{eq:def:probtreeautomaton2}
\end{equation*}
%Here, $l(\empseq)=a\in\Sigma_n$, and
%$t_i$ is the $i$'th subtree of $t$.
%and $c_a(x)$ is a probability measure on $X^n$ that is defined by $c_a(x)(A)=c(x)(\{a\}\times A)$.
%, and $a_i=l_i(\empseq)$.
\end{itemize}
Then for each $x\in X$, there exists 
a unique probability measure $\langinf(\mathcal{X},x)$ on $(\Treeinf(\Sigma),\sigalg_{\infty})$ such that %such that 
$\langinf(\mathcal{X},x)(\cyl(t))=\nu_x(t)$.
%The function $\lang(\mathcal{X},x)$ in Def.~\ref{def:probtreeautomaton} satisfies an equality in Lem.~\ref{lem:uniqueextensionmeas}.
%Therefore extension to $\sigalg_{\infty}$ uniquely exists.
%\qed
\end{prop}

This proposition is proved using 
Carath\'eodory's extension 
%Kolmogorov's consistency 
theorem and the following ``compatibility'' lemma.

\begin{lem}\label{lem:existencesemanticslem}
In Proposition~\ref{prop:existencesemantics}, for all $k\in\omega$ and $t\in\Tree^k(\Sigma)$, we have:
\begin{equation}
\label{eq:existencesemanticslem}
\sum_{\substack{s\in \Tree^{k+1}(\Sigma) \\\text{s.t. } t\treeprefix s}}\nu_x(s)=\nu_x(t)\,.
%\hfill\qed
\end{equation}
\end{lem}

%\newtheorem*{proof:existencesemanticslem}{Proof of Lem.\ \ref{lem:existencesemanticslem}}{\itshape}{\rmfamily}
%\begin{proof:existencesemanticslem}
%\begin{myproof}
\proof
%By Kolmogorov's consistency theorem (see~\cite{kechris95classicaldescriptive}), 
%it suffices to prove that for all $k\in\omega$ and $t\in\Treeinf^k(\Sigma)$, $\sum_{s\in \Treeinf^{k+1}(\Sigma), t\treeprefix s}\nu_x(s)=\nu_x(t)$ holds.
For a $(k+1)$-prefix tree $s=(D_s,l_s)\in \Tree^{k+1}(\Sigma)$, 
we write $a_{s}$ for $l_s(\empseq)$ and $n_s$ for the arity of $a_s$ (i.e.\ $a_s\in \Sigma_{n_s}$).
We prove the
equation~(\ref{eq:existencesemanticslem}) %the given equation 
by induction on $k$. 

If $k=0$, as $t$ and all the subtrees of $s$ except for $s$ itself are 
$0$-prefix 
%empty
trees, we have:
\allowdisplaybreaks[2]\begin{align*}
&\sum_{\substack{s\in \Tree^{k+1}(\Sigma) \\\text{s.t. } t\treeprefix s}}\nu_x(s) \\
&= \sum_{s\in \Tree^{1}(\Sigma)}\;\;\;\sum_{x_0,\ldots,x_{n_s-1}\in X} c(x)\bigl(\{(a_s,x_0,\ldots,x_{n_s-1})\}\bigr)\cdot \prod_{i=0}^{n_s-1} \nu_{x_i}(s_i) \\
& \mypushright{\text{(by definition of $\nu_x(s)$)}} \\
&=\sum_{n=0}^{\infty}\sum_{a\in\Sigma_n}\sum_{x_0,\ldots,x_{n-1}\in X}c(x)\bigl(\{(a,x_0,\ldots,x_{n-1})\}\bigr)\cdot \prod_{i=0}^{n-1} \nu_{x_i}(s_i)\\
%&\mypushright{(\text{as $1$-prefix tree $s$ consists of one node})} \\
%
&= \sum_{n=0}^{\infty}\sum_{a\in\Sigma_n} \sum_{x_0,\ldots,x_{n-1}\in X}c(x)\bigl(\{(a,x_0,\ldots,x_{n-1})\}\bigr)\cdot \prod_{i=0}^{n-1}(1-\Reach(\Delta_{\mathcal{X}},x_i,\bot)) \\
& \mypushright{(\text{$s_i\in\Tree^0(\Sigma)$ for each $i$})} \\
%& \mypushright{(\text{by definition of $\nu_x(t)$ where $t\in\Treeinf^0(\Sigma)$})} \\
%&= \sum_{n=0}^{\infty}\sum_{x_0,\ldots,x_{n-1}\in X} \sum_{a\in\Sigma_n} c(x)(\{(a,x_0,\ldots,x_{n-1})\})\cdot \prod_{i=0}^{n-1}(1-\Reach(\Delta_{\mathcal{X}},x_i,\bot)) \\
%&\mypushright{(\text{as all elements are nonnegative})} \\
&=\sum_{n=0}^{\infty} \sum_{x_0,\ldots,x_{n-1}\in X} \tau_{\mathcal{X}}(x,\langle x_0,\ldots,x_{n-1}\rangle)\cdot \prod_{i=0}^{n-1}(1-\Reach(\Delta_{\mathcal{X}},x_i,\bot)) \\
&\mypushright{(\text{by definition of $\tau_{\mathcal{X}}$ of $\Delta_{\mathcal{X}}$})} \\
&\mypushleft{=1-\Reach(\Delta_{\mathcal{X}},x,\bot)} (\text{by definition of branching process}) \\
&\mypushleft{=\nu_x(t)}  (\text{by definition of $\nu_x(t)$})\,.
\end{align*}
%
%For $k>0$, 
Next, let $k>0$ and assume that  
%the equation~\ref{eq:existencesemanticslem}
$\sum_{v\in \Tree^{k}(\Sigma)\text{ s.t. } u\treeprefix v}\nu_x(v)=\nu_x(u)$ 
holds for all $u\in\Tree^{k-1}(\Sigma)$ and $x\in X$.
Let $t=(D,l)\in\Tree^k(\Sigma)$, $a=l(\empseq)$ and $a\in\Sigma_n$.
%Moreover, let $t_i$ and $s_i$ be the $i$'th subtrees of $t\in\Treeinf^k(\Sigma)$ and $s\in\Treeinf^{k-1}(\Sigma)$, respectively.
Moreover, let $t_i$ be the $i$-th subtrees of $t\in\Tree^k(\Sigma)$. % and $s\in\Treeinf^{k-1}(\Sigma)$, respectively.
Then:
\allowdisplaybreaks[1]
\begin{align*}
&\sum_{\substack{s\in \Tree^{k+1}(\Sigma) \\\text{s.t. } t\treeprefix s}}\nu_x(s) \\
&= \sum_{\substack{s\in \Tree^{k+1}(\Sigma) \\\text{s.t. } t\treeprefix s}}\;\;\sum_{x_0,\ldots,x_{n-1}\in X} \left( c(x)\bigl(\{(a,x_0,\ldots,x_{n-1})\}\bigr)\cdot \prod_{i=0}^{n-1} \nu_{x_i}(s_i) \right) \hspace{3cm} \\
& \mypushright{(\text{by definition of $\nu_x(s)$})} \\
&=\sum_{x_0,\ldots,x_{n-1}\in X} \left(c(x)\bigl(\{(a,x_0,\ldots,x_{n-1})\}\bigr)\cdot\left(\sum_{\substack{s\in \Tree^{k+1}(\Sigma) \\\text{s.t. } t\treeprefix s}} \prod_{i=0}^{n-1} \nu_{x_i}(s_i)\right)\right) \\
&=\sum_{x_0,\ldots,x_{n-1}\in X}\left( c(x)\bigl(\{(a,x_0,\ldots,x_{n-1})\}\bigr)\cdot\left(\sum_{\substack{s_0\in \Tree^{k}(\Sigma)\\ \text{s.t. } t_0\treeprefix s_0}}\cdots\sum_{\substack{s_{n-1}\in \Tree^{k}(\Sigma) \\\text{s.t. }t_{n-1}\treeprefix s_{n-1}}} \prod_{i=0}^{n-1} \nu_{x_i}(s_i)\right)\right) \\
& \mypushright{\textstyle{(\{s\in\Tree^{k+1}(\Sigma)\mid t\treeprefix s\}\cong 
\prod_{i=0}^{n-1}\{s_i\in\Tree^k(\Sigma)\mid t_i\treeprefix s_i\})}} \\
%& \mypushright{(\{s\in\Treeinf^{k+1}(\Sigma)\mid t\treeprefix s\}\cong 
%\prod_{i=0}^{n-1}\{s_i\in\Treeinf^k(\Sigma)\mid t_i\treeprefix s_i\})} \\
%& \mypushright{(\{s\in\Treeinf^{k+1}(\Sigma)\mid t\treeprefix s\}\cong 
%\{s_0\in\Treeinf^k(\Sigma)\mid t_0\treeprefix s_0\}\times \cdots\times
%\{s_{n-1}\in\Treeinf^k(\Sigma)\mid t_{n-1}\treeprefix s_{n-1}\})} \\
%
&=\sum_{x_0,\ldots,x_{n-1}\in X} \left(c(x)\bigl(\{(a,x_0,\ldots,x_{n-1})\}\bigr)\cdot\left(\prod_{i=0}^{n-1} \sum_{\substack{s_i\in \Tree^{k}(\Sigma)\\ \text{s.t. } t_i\treeprefix s_i}}\nu_{x_i}(s_i)\right)\right) \\
& \mypushright{(\text{$s_i$ does not appear in $\nu_{x_j}(s_j)$ if $i\neq j$})} \\
&\mypushleft{= \sum_{x_0,\ldots,x_{n-1}\in X} \Bigl(c(x)\bigl(\{(a,x_0,\ldots,x_{n-1})\}\bigr)\cdot \prod_{i=0}^{n-1} \nu_{x_i}(t_i)\Bigr)} (\text{by induction hypothesis}) \\
&\mypushleft{= \nu_x(t)}  (\text{by definition of $\nu_x(t)$}).
\end{align*}
Therefore $\sum_{s\in \Tree^{k+1}(\Sigma)\text{ s.t. } t\treeprefix s}\nu_x(s)=\nu_x(t)$ holds for all $k\in\omega$ and $t\in\Tree^k(t)$. % and this concludes the proof.
\qed
%\end{proof:existencesemanticslem}
%\end{myproof}

%\newtheorem*{proof:existencesemantics}{Proof of Prop.\ \ref{prop:existencesemantics}}{\itshape}{\rmfamily}
%\begin{proof:existencesemantics}
%\begin{proof}[Proof of Prop.\ \ref{prop:existencesemantics}]
{\proof[Proof of Proposition \ref{prop:existencesemantics}]
%\proof
Immediate from 
Carath\'eodory's extension theorem~\cite{ashD2000probability} 
%Kolmogorov's consistency theorem~\cite{kechris95classicaldescriptive} 
and Lemma~\ref{lem:existencesemanticslem}.
\qed}
%\end{proof}
%\end{proof:existencesemantics}

\begin{defi}\label{def:probtreeautomaton}
Let $\mathcal{X}=(X,s,c)$ be a $\Sigma$-labeled probabilistic tree automaton. 
%
%For each integer $k\in\omega$ and $k$-prefix tree $t\in \Treeinf^k(\Sigma)$, 
%we define a value $\nu(\cyl(t))\in[0,1]$ by an induction on $k$ as follows.
%\begin{itemize}
%\item If $k=0$, then 
%\begin{equation}
%\nu(\cyl(t))=\nu(\Treeinf(\Sigma))=\text{Reach}(\Delta_{\mathcal{X}},\bullet) \,.
%\label{eq:def:probtreeautomaton1}
%\end{equation}
%\item If $k=i+1$ and $t=(D,l)$ where , then 
%\begin{equation}
%\nu(\cyl(t))=\sum_{y_0,\ldots,y_{n-1}\in X} \bigl(c(x)(a,y_0,\ldots,y_{n-1})\cdot \prod_{i=0}^{n-1} \nu(\cyl(t_i))\bigr)\,.
%\label{eq:def:probtreeautomaton2}
%\end{equation}
%Here, $l(\empseq)=a\in\Sigma_n$, 
%$t_i=(D_i,l_i)$ is $\langle i \rangle$'th subtree of $t$, and
%$y_i=l_i(\empseq)$.
%\end{itemize}
%
For a state $x\in X$, the \emph{infinitary language of $\mathcal{X}$ from $x$} is a 
%unique 
probability measure $\langinf(\mathcal{X},x)$ on 
%a measurable space 
$(\Treeinf(\Sigma),\sigalg_{\infty})$ in Proposition~\ref{prop:existencesemantics}. %that is the unique extension of $\nu$ (i.e.\ $\lang(\mathcal{X},x)=\nu^*$).
The \emph{infinitary language of $\mathcal{X}$} is a probability measure $\langinf(\mathcal{X})$ on 
a $(\Treeinf(\Sigma),\sigalg_{\infty})$ that is defined by 
$\langinf(\mathcal{X})(A)=\sum_{x\in X} s(*)(x)\cdot\langinf(\mathcal{X},x)(A)$ for $A\in\sigalg_{\infty}$.
\end{defi}

%The measure $\lang(\mathcal{X},x)$ defined above is well-defined.

%\begin{lem}\label{lem:existencesemantics}
%The function $\lang(\mathcal{X},x)$ in Def.~\ref{def:probtreeautomaton} satisfies an equality in Lem.~\ref{lem:uniqueextensionmeas}.
%Therefore extension to $\sigalg_{\infty}$ uniquely exists.
%\qed
%\end{lem}

The following is the main result of this section.
%---coincidence between
%the automata-theoretic infinitary language that we have just defined %in Def.~\ref{def:probtreeautomaton} 
%and coalgebraic infinitary trace semantics defined in the previous section.
%
%Then we can show that 
%the language defined in Def.~\ref{def:probtreeautomaton} 
%coincides with coalgebraic infinite trace semantics.
\begin{prop}\label{prop:largestsemanticsgiry}
The carrier of a final $\FSigma$-coalgebra in $\Meas$ is isomorphic to $(\Treeinf(\Sigma),\sigalg_{\infty})$,
and 
%Moreover, 
for a $\Sigma$-labeled probabilistic tree automaton $\mathcal{X}=(X,s,c)$,
%For the order in Def.~\ref{def:orderenrichment}, 
%we have 
$\trinf(c)(x)=\langinf(\mathcal{X},x)$ holds for all $x\in X$.
%wrt.\ the order in Def.~\ref{def:orderenrichment}.
Furthermore it follows that $\trinf\odot s(c)(*)=\langinf(\mathcal{X})$.
%\qed
\end{prop}

To prove this proposition, we use the lemma below that 
%It 
states that the unreachable probability of a branching process
can be calculated as the greatest fixed point of a certain function.
It is a 
direct consequence
%immediate corollary 
of the well-known result that 
the reachability probability of a Markov chain 
can be calculated as the least fixed point of a certain function (see e.g.~\cite[Theorem 10.15]{baier08principlesof}).
A generalized statement of the following lemma (one for \emph{branching Markov decision processes}) is given in~\cite{etessamiSY15greatestfixed}.
In the rest of this section, for a vector $\mathbf{v}\in[0,1]^{X}$ and $x\in X$, 
$\mathbf{v}_x$ denotes the $x$-th element of $\mathbf{v}$.

%is calculated as the greatest fixed point of a certain function.
%It is an immediate corollary of the known fact that:
%the reachable probability of a Markov 
%This theorem is proved using the result in~\cite{etessamiSY15greatestfixed}---the unreachable probability of a branching process
%is calculated as the greatest fixed point of a certain function.

\begin{lem} %[\cite{etessamiSY15greatestfixed}] 
\label{lem:BPunreachability}
Let $\Delta=(\Gamma,\tau)$ be a branching process and $y\in \Gamma$.
We define a function $P_y:[0,1]^{\Gamma}\to[0,1]^{\Gamma}$ as follows:
\begin{displaymath}
\bigl(P_y(\mathbf{v})\bigr)_x=\sum_{\substack{n\in\omega, \;x_0,\ldots,x_{n-1}\in\Gamma \text{ s.t.} \\ \text{$y\notin\{x_0,\ldots,x_{n-1}\}$}}} \tau\bigl(x,\langle x_0\ldots x_{n-1}\rangle\bigr)\cdot \prod_{i=0}^{n-1} \mathbf{v}_{x_i}\,.
\end{displaymath}
%Here, $\mathbf{v}\in[0,1]^{\Gamma}$ and $v_{x_i}$ denotes $v(x_i)\in[0,1]$.
%Here, $|\alpha|$ denotes the length of $\alpha\in\Gamma^*$ and $\alpha_i$ denotes the $i$'th letter of $\alpha$.
As $P_y$ is a monotone function, $P_y$ has the greatest fixed point $\mathbf{v}^{y,\max}\in[0,1]^{\Gamma}$. 
Then 

\noindent\begin{minipage}{0.9\hsize}
\begin{displaymath}
1-\Reach(\Delta,x,y)=(\mathbf{v}^{y,\max})_x\,.
%$.
\end{displaymath}
\end{minipage}
\begin{minipage}{0.1\hsize}
\qed
\end{minipage}
\end{lem}

%Using this lemma, Thm.~\ref{thm:largestsemanticsgiry} is proved as follows.

%\newtheorem*{proof:largestsemanticsgiry}{Proof of Thm.\ \ref{thm:largestsemanticsgiry}}{\itshape}{\rmfamily}
%\begin{proof:largestsemanticsgiry}

{\proof[Proof of Proposition \ref{prop:largestsemanticsgiry}]
We define an arrow $\zeta:(\Treeinf(\Sigma),\sigalg_{\infty})\to F_{\Sigma}(\Treeinf(\Sigma),\sigalg_{\infty})$ 
in $\Meas$ in a similar manner to the final $\FSigma$-coalgebra in $\Sets$ (see Proposition~\ref{prop:largestsemanticspow}):
namely, $\zeta(t)=(a,(t_0,\ldots,t_{n-1}))$. % where
(Here 
$t=(D,l)$, $a=l(\empseq)\in\Sigma_n$, 
and for each $i\in\{0,\ldots,n-1\}$, $t_i=(D_i,l_i)$ where 
$D_i=\{\alpha\in\mathbb{N}^*\mid i\alpha\in D\}$ and $l_i(\alpha)=l(i\alpha)$.) % holds.
%In a similar manner to the proof of 
It is easy to see that $\zeta$ is a measurable function.
%Moreover, 
It is also easy to see that $\zeta$ is a final $F_{\Sigma}$-coalgebra: the unique homomorphism from %an arbitrary coalgebra
a coalgebra
is given in the same way as the final $\FSigma$-coalgebra in $\Sets$. %Hence we skip the proof.

Next we show that a function $\langinf(\mathcal{X},\place):X\kto \Treeinf(\Sigma)$ in Definition~\ref{def:probtreeautomaton}
is the largest homomorphism from $c$ to $J\zeta$.
%where $\zeta$ is defined as in Lem.~\ref{lem:finalcoalgmeas}. 
As $X$ is equipped with the discrete $\sigma$-algebra, $\langinf(\mathcal{X},\place)$ is indeed an arrow in $\Meas$.

Let $\mathbf{v}^{\max}\in[0,1]^X$ be the greatest fixed point of a function $P:[0,1]^X\to [0,1]^X$ that is defined as follows
(much like in Lemma~\ref{lem:BPunreachability}):
\begin{equation}
\bigl(P(\mathbf{v})\bigr)_x=\sum_{n=0}^{\infty}\sum_{x_0,\ldots,x_{n-1}\in X} \left(\sum_{a\in\Sigma_n} c(x)\bigl(\{(a,x_0,\ldots,x_{n-1}\})\bigr)\right) \cdot \prod_{i=0}^{n-1} \mathbf{v}_{x_i}\,.
\label{eq:largestsemanticsgiry1}
\end{equation}
Recall that $\langinf(\mathcal{X},x)(\Treeinf(\Sigma))$ is defined by $\langinf(\mathcal{X},x)(\Treeinf(\Sigma))=1-\Reach(\Delta_{\mathcal{X}},x,\bot)$.
Therefore by %Definition~\ref{def:skeleton} and 
Lemma~\ref{lem:BPunreachability},
we have $\langinf(\mathcal{X},x)(\Treeinf(\Sigma))=(\mathbf{v}^{\max})_x$.

We first show that $\langinf(\mathcal{X},\place)$ is a homomorphism. 
By Carath\'eodory's extension
%Kolmogorov's consistency 
theorem,  it suffices to prove the following equation for all $k\in\omega$ and $t\in\Tree^k(\Sigma)$.
%commutativity with respect to cylinder sets:
%namely, we show that 
%$
\begin{displaymath}
\bigl(J\zeta^{-1}\odot \overline{\FSigma}\langinf(\mathcal{X},\place)\odot c\bigr)(x)\bigl(\cyl(t)\bigr)=\langinf(\mathcal{X},x)(\cyl(t))
\end{displaymath}
%$ 
%for all $k\in\omega$ and $t\in\Treeinf^k(\Sigma)$.

We prove this equation by induction on  $k$. %the depth $k$ of $t$.

If $k=0$, then as $\cyl(t)=\Treeinf(\Sigma)$,  we have:
\begin{align*}
&\bigl(J\zeta^{-1}\odot \overline{\FSigma}(\langinf(\mathcal{X},\place))\odot c\bigr)(x)\bigl(\cyl(t)\bigr) \\
%&= \sum_{n=0}^{\infty}\sum_{x_0,\ldots,x_{n-1}\in X} \sum_{a\in\Sigma_n} \left(c(x)\bigl(\{(a,x_0,\ldots,x_{n-1}\})\bigr)\cdot \Bigl(J\zeta^{-1}\odot \overline{\FSigma}\bigl(\lang(\mathcal{X},\place)\bigr)\Bigr)(a,x_0,\ldots,x_{n-1})\bigl(\Treeinf(\Sigma)\bigr)\right) \\
%&\mypushright{(\text{by definition of $\odot$ and $\cyl(t)=\Treeinf(\Sigma)$})} \\
&= \sum_{n=0}^{\infty}\sum_{x_0,\ldots,x_{n-1}\in X} \sum_{a\in\Sigma_n} \left(c(x)\bigl(\{(a,x_0,\ldots,x_{n-1}\})\bigr)\cdot \prod_{i=0}^{n-1}\langinf(\mathcal{X},x_i)\bigl(\Treeinf(\Sigma)\bigr)\right) \hspace{1cm}\\
%&\mypushright{(\text{by definition of $\zeta$ and $\cyl(t)=\Treeinf(\Sigma)$})} \\
&= \sum_{n=0}^{\infty}\sum_{x_0,\ldots,x_{n-1}\in X} \sum_{a\in\Sigma_n} \left(c(x)\bigl(\{(a,x_0,\ldots,x_{n-1}\})\bigr)\cdot \prod_{i=0}^{n-1}(\mathbf{v}^{\max})_{x_i}\right) \\
%&\mypushright{(\text{by definition of $\lang(\mathcal{X},x)$})} \\
&\mypushleft{=\bigl(P(\mathbf{v}^{\max})\bigr)_x} (\text{by definition of $P$}) \\
&\mypushleft{=(\mathbf{v}^{\max})_x} (\text{as $\mathbf{v}_{\max}$ is a fixed point of $P$}), \\
\intertext{on the one hand. On the other hand, we have shown that} 
&\langinf(\mathcal{X},x)\bigl(\cyl(t)\bigr)=(\mathbf{v}^{\max})_x\,.
\end{align*}
Therefore we have $\bigl(J\zeta^{-1}\odot \overline{\FSigma}\langinf(\mathcal{X},\place)\odot c\bigr)(x)(\cyl(t))=\langinf(\mathcal{X},x)(\cyl(t))$ for $t\in\Tree^0(\Sigma)$.

Let $k>0$ and $t=(D,l)\in\Tree^{k+1}(\Sigma)$ where $l(\empseq)=a\in\Sigma_n$. 
Moreover, let $t_i$ be the $i$-th subtree of $t$ where $0\leq i \leq n-1$. 
Then 
\begin{align*}
&\Bigl(J\zeta^{-1}\odot \overline{\FSigma}\bigl(\langinf(\mathcal{X},\place)\bigr)\odot c\Bigr)(x)\bigl(\cyl(t)\bigr) \\
&= \sum_{x_0,\ldots,x_{n-1}\in X} c(x)\bigl(\{(a,x_0,\ldots,x_{n-1})\}\bigr)\cdot \prod_{i=0}^{n-1}\langinf(\mathcal{X},x_i)\bigl(\cyl(t_i)\bigr) 
& \tag*{(\text{by definition of $\zeta$})}\\
&= \langinf(\mathcal{X},x)(\cyl(t)) & \tag*{(\text{by definition of $\langinf(\mathcal{X},x)$})\rlap{\,.}}
\end{align*}
Therefore we have $\bigl(J\zeta^{-1}\odot \overline{F}\langinf(\mathcal{X},\place)\odot c\bigr)(x)(\cyl(t))=\langinf(\mathcal{X},x)(\cyl(t))$ for $t\in\Tree^{k+1}(\Sigma)$.

Hence %$J\zeta^{-1}\odot \overline{F}\lang(\mathcal{X},\place)\odot c=\lang(\mathcal{X},\place)$ holds and 
$\langinf(\mathcal{X},\place)$
is a homomorphism from $c$ to $J\zeta$.

%Next we 
It remains to 
show that $\langinf(\mathcal{X},\place)$ is the largest homomorphism.
Let $g:X\kto \Treeinf(\Sigma)$ be a homomorphism from $c$ to $J\zeta$.
By  monotonicity of the extension of a measure, % on cylinder sets,
it suffices to prove 
%\begin{equation}
%\label{eq:prooflargestsemanticsgiryLargest}
$g(x)\bigl(\cyl(t)\bigr)\leq\langinf(\mathcal{X},x)\bigl(\cyl(t)\bigr)$
%\end{equation}
for all $x\in X$, $k\in\omega$ and $t\in\Tree^k(\Sigma)$.
We prove this %equation~(\ref{eq:prooflargestsemanticsgiryLargest}) 
by induction on $k$.

If $k=0$, then $\cyl(t)=\Treeinf(\Sigma)$. Hence we have:
\begin{align*}
&g(x)\bigl(\cyl(t)\bigr) \\
&\mypushleft{=\bigl(J\zeta^{-1}\odot\overline{\FSigma}g\odot c\bigr)(x)\bigl(\Treeinf(\Sigma)\bigr)}  \text{($g$ is a homomorphism)} \\ 
&= \sum_{n=0}^{\infty}\sum_{x_0,\ldots,x_{n-1}\in X} \left(\sum_{a\in\Sigma_n} c(x)\bigl(\{(a,x_0,\ldots,x_{n-1})\}\bigr)\right) \cdot \prod_{i=0}^{n-1}g(x_i)\bigl(\Treeinf(\Sigma)\bigr)\,. \hspace{2cm}%\\
%& \mypushright{\text{(by the definition of $\overline{\FSigma}$)}} \,.
\end{align*}
Here we define a vector $\mathbf{w}\in[0,1]^{X}$ by $\mathbf{w}_x=g(x)(\Treeinf(\Sigma))$ for each $x\in X$.
The equation above implies that $\mathbf{w}$ 
is a fixed point of $P$ defined in (\ref{eq:largestsemanticsgiry1}).
%Therefore a vector $\mathbf{w}\in[0,1]^{X}$ that is defined by $\mathbf{w}_x=g(x)(\Treeinf(\Sigma))$ is a fixed point of $P$ defined in (\ref{eq:largestsemanticsgiry1}).
As $\mathbf{v}^{\max}$ is the greatest fixed point of $P$, we have $g(x)(\Treeinf(\Sigma))=\mathbf{w}_x\leq(\mathbf{v}^{\max})_x=\langinf(\mathcal{X},x)(\Treeinf(\Sigma))$.

Let $k>0$ and assume that $g(x)(\cyl(s))\leq\langinf(\mathcal{X},x)(\cyl(s))$ holds for all $x\in X$ and $s\in\Tree^{k-1}(\Sigma)$.
Let $t=(D,l)\in\Tree^{k}(\Sigma)$ and $l(\empseq)=a\in\Sigma_n$. 
We write $t_i$ for the $i$-th subtree of $t$. % where $0\leq i \leq n-1$.
Then  %as $g$ is a homomorphism, by the definition of $\overline{\FSigma}$ on $\Kl(\giry)$, and by induction hypothesis, we have
\allowdisplaybreaks[2]
\begin{align*}
&g(x)(\cyl(t)) \\
&\mypushleft{=\bigl(J\zeta^{-1}\odot\overline{\FSigma}g\odot c\bigr)(x)(\cyl(t))}  \text{($g$ is a homomorphism)} \\
&= \sum_{x_0,\ldots,x_{n-1}} \left(c(x)\bigl(\{(a,x_0,\ldots,x_{n-1})\}\bigr)\cdot  \prod_{i=0}^{n-1} g(x_i)(\cyl(t_i))\right)\\
%&\mypushright{\text{(by  definition of $\overline{\FSigma}$ on $\Kl(\giry)$)}} \\
&\leq \sum_{x_0,\ldots,x_{n-1}} \left(c(x)\bigl(\{(a,x_0,\ldots,x_{n-1})\}\bigr)\cdot  \prod_{i=0}^{n-1} \langinf(\mathcal{X},x_i)(\cyl(t_i))\right) \hspace{4cm}\\
&\mypushright{\text{(by induction hypothesis)}} \\
&\mypushleft{=\bigl(J\zeta^{-1}\odot\overline{\FSigma}(\langinf(\mathcal{X},\place)\odot c\bigr)(x)(\cyl(t))} \;\;\;\;\;\text{(by  definition of $\overline{\FSigma}$)} \\
&\mypushleft{= \langinf(\mathcal{X},x)(\cyl(t))}  \text{($\langinf(\mathcal{X},\place)$ is a homomorphism)}\,.
\end{align*}
Therefore we have $g(x)(\cyl(t))\leq\langinf(\mathcal{X},x)(\cyl(t))$ for $t\in\Tree^{k}(\Sigma)$.

Hence $\langinf(\mathcal{X},\place)$ is the largest homomorphism from $c$ to $J\zeta$ and therefore 
  $\trinf(c)=\langinf(\mathcal{X},\place)$.
This immediately implies  $\trinf(c)\odot s(*)=\langinf(\mathcal{X})$.
\qed}
%\end{proof:largestsemanticsgiry}

%
\auxproof{
We show that the language defined in Definition~\ref{def:probtreeautomaton} coincides with coalgebraic infinite trace semantics.
The carrier of the final $\FSigma$-coalgebra in $\Meas$ is given by a set of  infinite trees.
\begin{lem}\label{lem:finalcoalgmeas}
%For $\FSigma$ in Def.~\ref{def:FSigmaMeas}, 
We define an arrow $\zeta:(\Treeinf(\Sigma),\sigalg_{\infty})\to F_{\Sigma}(\Treeinf(\Sigma),\sigalg_{\infty})$ 
in $\Meas$ in the same way as Lemma~\ref{lem:finalcoalgsets}.
%\begin{equation*}
%\zeta(D,l)=(a,(t_0,\ldots,t_{n-1})).
%\end{equation*}
%Here, $a=l(\empseq)\in\Sigma_n$, 
%and for each $i\in\{0,\ldots,n-1\}$, $D_i=\{\alpha\in\mathbb{N}^*\mid i\alpha\in D\}$ and $l_i(\alpha)=l(i\alpha)$.
%
Then $\zeta$ is measurable and moreover,  is a final $F_{\Sigma}$-coalgebra.
\qed
\end{lem}

%The infinite language of a probabilistic tree automaton can be captured by the largest homomorphism.
Then we have the following theorem.
It is proved using the known result in~\cite{etessamiSY15greatestfixed}---the 
unreachability probability of a branching processes can be calculated as
the greatest fixed point of a certain function $P:[0,1]^{\Gamma}\to[0,1]^{\Gamma}$.

\begin{thm}\label{thm:largestsemanticsgiry}
Let $\mathcal{X}=(X,s,c)$ be a $\Sigma$-labeled probabilistic tree automaton.
%For the order in Def.~\ref{def:orderenrichment}, 
For $x\in X$, we have $\trinf(c)(x)=\langinf(\mathcal{X},x)$
with respect to the order in Definition~\ref{def:orderenrichment}.
Moreover, $\trinf\odot s(c)(*)=\langinf(\mathcal{X})$.
\qed
\end{thm}
}

%\subsection{Subdistribution Monad and Infinite Trace Situation}
\subsection{Another Modeling of Probabilistic Branching: Subdistribution Monad}
\label{subsec:subdistmonad}
In the previous sections we used 
the sub-Giry monad $\giry$ 
and 
a (standard Borel) polynomial functor $F$ on
%the category of measurable spaces 
$\Meas$ 
to model probabilistic systems.
In this section, we discuss another pair---a polynomial functor $F$ and the subdistribution monad $\dist$ on $\Sets$---that 
 can also model  probabilistic systems.
For a given set $X\in\Sets$, $\dist X$ is the set 
%$\{p\colon X\to [0,1]\mid \sum_{x\in X}p(x)\le 1\}$ 
of (discrete) subdistributions over $X$.

%Here we give definitions of the subdistribution monad and orders on the homsets of the Kleisli category.
%
\begin{defi}[subdistribution monad]\label{def:subdistmonad}
A \emph{subdistribution monad} is a monad $(\dist,\eta^{\dist},\mu^{\dist})$ on $\Sets$  such that
\begin{itemize}
\item $\dist X=\{p:X\to [0,1]\mid \sum_{x\in X}p(x)\leq 1\}$,
\item $\dist f(p)(y)= \sum_{x\in f^{-1}(y)}p(x)$, 
\item $\eta^{\dist}_{X}(x)(y)=\begin{cases} 1 & (y=x) \\ 0 & (\text{otherwise}),\end{cases}$ and 
\item $\mu^{\dist}_{X}(\Phi)(x)=\sum_{p\in \dist X}\Phi(p)\cdot p(x)$. 
\end{itemize}
\end{defi}

\begin{defi}[order enrichment of $\Kl(\dist)$]\label{def:orderenrichmentdist}
We define an order on $\Kl(\dist)(X,Y)$ by $f\sqsubseteq g$ if and only if $\forall x\in X.\, \forall y\in Y.\, f(x)(y)\leq g(x)(y)$.
\end{defi}

%Let $\Sigma$ be a ranked alphabet.
Next, 
%For a ranked alphabet $\Sigma$, 
we show that $\FSigma$ 
in Definition~\ref{def:FSigma}
%on $\Sets$ 
and the subdistribution monad $\dist$ constitute an infinitary trace situation by giving an explicit definition of the largest homomorphism. % from each $\overline{\FSigma}$-coalgebra $c$ to $J\zeta$.

\begin{prop}\label{prop:satinftracesituationdist}
Let $\Sigma$ be a ranked alphabet and $\FSigma$ be the functor on $\Sets$ defined in Definition~\ref{def:FSigma}.
Then $\FSigma$ and the subdistribution monad $\dist$ constitute an infinitary trace situation.
\end{prop}

%\begin{myproof}
\proof
Let $\zeta:\Treeinf(\Sigma)\to F_{\Sigma}\left(\Treeinf(\Sigma)\right)$ be a final $\FSigma$-coalgebra in $\Sets$
that we defined %have used 
in the proof of Proposition~\ref{prop:largestsemanticspow}.
%We define $\zeta:\Treeinf(\Sigma)\to F_{\Sigma}\Treeinf(\Sigma)$ in $\Sets$ in the same way as 
%the proof of Thm.~\ref{thm:largestsemanticspow}.
%%Lem.~\ref{lem:finalcoalgsets}.
%Then as we have shown in the proof $\zeta$ is a final $\FSigma$-coalgebra. 
For an $\overline{\FSigma}$-coalgebra $c:X\kto\overline{\FSigma}X$, we construct the largest homomorphism $h:X\kto \Treeinf(\Sigma)$ 
from $c$ to $J\zeta$. % as follows.
%Let $t=(D,l)$ be a $\Sigma$-labeled infinite tree. 
%We define a value $h(c)(t)\in[0,1]$ as follows.

To this end, for $x\in X$, an integer $k\in\omega$ and a $k$-prefix tree $t\in \Tree^k(\Sigma)$, 
we first define a value $\xi_x(t^k)\in[0,1]$ by induction on $k$ as follows:
\begin{itemize}
\item for $k=0$, $\xi_x(t)=1$, and
\item for $k>0$, 
%if $t^k=(D^k,l^k)\in \Treeinf^{k}(\Sigma)$ 
%where $l^k(\empseq)=a\in\Sigma_n$ and
%the $i$'th subtree of $t^k$ is $t^k_i$,
\begin{equation}
\xi_x(t)=\sum_{x_0,\ldots,x_{n-1}\in X} \bigl(c(x)(a,x_0,\ldots,x_{n-1})\cdot \prod_{i=0}^{n-1} \xi_{x_i}(t_i)\bigr)\,,
\label{eq:disttreeautomaton2}
\end{equation}
%Here, 
where
$t=(D,l)\in \Tree^{k}(\Sigma)$, 
 $a=l^k(\empseq)\in\Sigma_n$, and
$t_i$ is the $i$-th subtree of $t$.
%Here, $a=l^k(\empseq)\in\Sigma_n$, and
%$t^k_i$ is the $i$'th subtree of $t^k$.
\end{itemize}

For 
%an infinite tree 
$t'\in\Treeinf(\Sigma)$ and $k\in\omega$, 
%we define a \emph{$k$-prefix of $t'$} as a $k$-prefix tree $\prefix_k(t')=(\prefix_k(D'),\prefix_k(l'))$ such that 
%we denote 
let 
$\prefix_k(t')=(\prefix_k(D'),\prefix_k(l'))$ be the unique $k$-prefix tree that is a prefix of $t'$.
%$\prefix_k(D')=\{\alpha\in \bigcup_{i<k}\mathbb{N}^i \;\mid\; \exists \beta\in\mathbb{N}^*.\, \alpha\beta\in D'\}$, and
%$\prefix_k(l')(\alpha)=l'(\alpha)$.
We define $h:X\kto \Treeinf(\Sigma)$ by $h(x)(t)=\lim_{k\to\infty}\xi_x(\prefix_k(t))$.
As $\sum_{n\in\omega}\sum_{a\in\Sigma_n}\sum_{x_0,\ldots,x_{n-1}\in X} c(x)(a,x_0,\ldots,x_{n-1})\leq 1$, 
the sequence $\bigl(\xi_x(\prefix_k(t))\bigr)_{k\in\omega}$ is decreasing with respect to $k$.
Therefore this $h$ is well-defined.
%Therefore we define $h:X\kto \Treeinf(\Sigma)$ by $h(x)(t)=\lim_{k\to\infty}\xi_x(\prefix_k(t))$.

We first show that this $h$ is a homomorphism.
For all $x \in X$, $n\in\omega$ and $t=(D,l)\in\Treeinf(\Sigma)$ such that $l(\empseq)=a\in\Sigma_n$ and $i$-th subtree of $t$ is $t_i$, we have:
%\allowdisplaybreaks[1]
\begin{align*}
&\bigl(J\zeta^{-1}\odot \overline{\FSigma}h\odot c\bigr)(x)(t)\\ %(a,t_0,\ldots, t_{n-1}) \\
&= \sum_{x_0,\ldots,x_{n-1}\in X} c(x)(a,x_0,\ldots,x_{n-1})\cdot \prod_{i=0}^{n-1}h(x_i)(t_i) & (\text{by definition of $\zeta$}) \\
&= \sum_{x_0,\ldots,x_{n-1}\in X} c(x)(a,x_0,\ldots,x_{n-1})\cdot \prod_{i=0}^{n-1}\lim_{k\to\infty}\xi_{x_i}(\prefix_k(t_i)) & (\text{by definition of $h$}) \\
%&= \sum_{x_0,\ldots,x_{n-1}\in X} \lim_{k\to\infty} c(x)(a,x_0,\ldots,x_n-1)\cdot \prod_{i=0}^{n-1}\xi_{x_i}(\prefix_k(t_i)) & \\
&= \lim_{k\to\infty} \sum_{x_0,\ldots,x_{n-1}\in X} c(x)(a,x_0,\ldots,x_{n-1})\cdot \prod_{i=0}^{n-1}\xi_{x_i}(\prefix_k(t_i)) & \\
&= \lim_{k\to\infty} \xi_x(\prefix_k(t)) & (\text{by definition of $\xi$}) \\
&= h(x)(t) & (\text{by definition of $h$})&\,. 
\end{align*}

To conclude the proof, we show that $h$ is the largest homomorphism.
Let $g:X\kto \Treeinf(\Sigma)$ be a homomorphism from $c$ to $J\zeta$. 
We prove $g(x)(t)\leq h(x)(t)$ for all $x\in X$ and $t\in\Treeinf(\Sigma)$.
%To this end, we first prove that for all $k\in\omega$, for all $x\in X$ and $t\in\Treeinf(\Sigma)$,
%$g(x)(t)\leq \xi_x(\prefix_k(t))$ holds by the induction on $k$.
To this end, we first prove $g(x)(t)\leq \xi_x(\prefix_k(t))$ for all $k\in\omega$, $x\in X$ and $t\in\Treeinf(\Sigma)$
by  induction on $k$.

If $k=0$ then for all $x$ and $t$, we have $g(x)(t)\leq 1 =\xi_x(\prefix_k(t))$.

Let $k>0$ and assume that $g(x)(t)\leq \xi_x(\prefix_{k-1}(t))$ for all $x$ and $t$.
Then 
\begin{align*}
&g(x)(t) \\
&= \bigl(J\zeta^{-1}\odot \overline{\FSigma}g\odot c\bigr)(x)(t) & (\text{$g$ is a homomorphism}) \\
&= \sum_{x_0,\ldots,x_{n-1}\in X} c(x)(a,x_0,\ldots,x_{n-1})\cdot \prod_{i=0}^{n-1}g(x_i)(t_i) & (\text{by definition of $\zeta$}) \\
&\leq \sum_{x_0,\ldots,x_{n-1}\in X} c(x)(a,x_0,\ldots,x_{n-1})\cdot \prod_{i=0}^{n-1}\xi_{x_i}(\prefix_{k-1}(t_i)) & (\text{by induction hypothesis}) \\
&= \xi_x(t) & (\text{by definition of $\xi$}) 
\end{align*}
Hence for all $x$ and $t$, we have $g(x)(t)\leq\lim_{k\to\infty}\xi_x(\prefix_{k}(t))=h(x)(t)$.
\qed
%\end{myproof}

In the proof above, we have constructed infinitary traces for $T=\dist$ in concrete terms. 
It is rather different from the axiomatic proofs for $T=\pow$ (Theorem~\ref{thm:inftrsituationpow}) and $T=\giry$ (Theorem~\ref{thm:inftrsitgiry}):
%Another difficulty is that 
It is because infinitary traces for $T=\dist$ does not
follow from either of our general axiomatic results  (Proposition~\ref{prop:generalconstructiontop} or
Proposition~\ref{prop:generalconstructionnotop}).
%---in the proof above %~\S\ref{subsec:subdistmonad}

\vspace{1mm}\hspace*{-\parindent}\begin{minipageparindent}
It is easy to see that there exists $X$ and $Z$ in $\Sets$ such that $\Kl(\dist)(X,Z)$ does not have $\top_{X,Z}$.
Therefore we cannot construct the largest homomorphism by using Proposition~\ref{prop:generalconstructiontop}.
\begin{wrapfigure}[4]{r}{0pt}
%\begin{center}
\begin{xy}
%(0,-20)*{} = "",
(-8,6)*{\mathcal{X}} = "",
(0,0)*+{\circ} = "x0",
\ar (0,-5);"x0"
\ar @(dl,ul)^{p,\frac{1}{2}} "x0";"x0"
\ar @(dr,ur)_{q,\frac{1}{2}} "x0";"x0"
\end{xy}
\vspace{5mm}
%\end{center}
\end{wrapfigure}
%
%Moreover, in fact, we cannot
Neither can we
% construct the largest homomorphism %even by 
% using 
use
 Proposition~\ref{prop:generalconstructionnotop}.
 In fact Assumption~(\ref{asm:existenceWeaklyFinalCoalgWeak2Lim}) fails.
Indeed, let $F$ be an endofunctor on $\Sets$ that is defined by $F(\place)=\{p,q\}\times(\place)$.
Then the limit of the final $\omega^{\op}$-sequence $1\overset{!_{F1}}{\leftarrow}F1\overset{F!_{F1}}{\leftarrow}F^2 1\overset{F^2!_{F1}}{\leftarrow} \cdots$
is given by $(Z,(\gamma_i:Z\to F^i1)_{i\in\omega})$ where $Z=\{p,q\}^{\omega}$ and $\gamma_i(a_0a_1\ldots)=a_0a_1\ldots a_{i-1}$. 
%As $F$ preserves $\omega^{\op}$-limit (see the proof of Thoerem~\ref{thm:satassumptionexception}), the carrier object of the final $F$-coalgebra is given by $Z$.
Hence the carrier of the final $F$-coalgebra $\zeta$ is given by $Z$.
We define $X\in\Kl(\dist)$ and $c:X\kto \overline{F}X$ by $X=\{*\}$ and $c(*)(a,*)=\frac{1}{2}$ where $a\in\{p,q\}$.
It is not so hard to see that the largest homomorphism $\trinf(c):X\kto Z$ from $c$ to $J\zeta$ is given by $\trinf(c)(x)(w)=0$ for each $w\in Z=\{p,q\}^{\omega}$\,.
However, we cannot obtain this $\trinf(c)$ with the procedure in the proof of Proposition~\ref{prop:generalconstructionnotop}.

For each $i\in\omega$, 
%we define $\alpha_i:X\kto\overline{F}^i1$ inductively as follows:
we inductively define $\alpha_i:X\kto\overline{F}^i1$ by $\alpha_0=J!_X$ and $\alpha_{i+1}=\overline{F}\alpha_i\odot c$.
%inductively as follows:
%\begin{equation*}
%\alpha_0=J!_X, \;\;\;\text{and}\;\;\; \alpha_{i+1}=\overline{F}\alpha_i\odot c\,.
%\end{equation*}
It is easy to see that $(X,(\alpha_i)_{i\in\omega})$ is a cone over a sequence 
$1\overset{J!_{F1}}{\kleftarrow} \overline{F}1\overset{JF!_{F1}}{\kleftarrow} \overline{F}^2 1\overset{JF^2!_{F1}}{\kleftarrow} \cdots$.
However, it is also easy to see that there does not exist $f:X\kto Z$ such that $J\gamma_i\odot f=\alpha_i$.
%This means that Assumption~(\ref{asm:existenceWeaklyFinalCoalgWeak2Lim}) is not satisfied.
\end{minipageparindent}
\vspace{-0.1mm}

As a consequence, we can construct the largest homomorphism from $c$ to $J\zeta$ neither by using the construction in 
Proposition~\ref{prop:generalconstructiontop} nor Proposition~\ref{prop:generalconstructionnotop}.
This prevents us from applying the general theories for Kleisli
simulations in Sections~\ref{sec:powersetmonad}--\ref{sec:girymonad}.

The resulting infinitary trace semantics in Proposition~\ref{prop:satinftracesituationdist} has limited use, however, due to
the discrete nature of an arrow $X\to \dist\left(\Treeinf(\Sigma)\right)$.
That is, it assigns a
probability  to  each single tree, and the probability is most of the time
$0$ (see Example~\ref{ex:firstExample}).

%\section{The Lift Monad and a Tree Automaton with Exception}
\section{Systems with Exception}
\label{sec:liftmonad}
In this section, we focus on systems that possibly abort with  exception. 
They are modeled as $(\lift,F)$-systems in the category $\Sets$, 
where $\lift$ is from Definition~\ref{def:variousmonads}.
Here the categorical/axiomatic results in Section~\ref{sec:girymonad} are applicable.
Therefore, much like for $\giry$,  
%we can check trace inclusion by 
forward or total backward simulations (see Section~\ref{subsec:Klsimgiry})
witness trace inclusion.
In this section, we assume that $F$ is a polynomial functor on $\Sets$.

\subsection{Construction of Infinitary Traces}
In this section, 
we prove the same results as Sections~\ref{subsec:constructlargesthompow} and \ref{subsec:constructlargesthomgiry}
for $T=\lift$.
%we show that 
%a polynomial functor $F$ on $\Sets$ 
%%$\FSigma$ on $\Sets$ 
%and the lift monad $\lift$ 
%%satisfy the assumptions in Prop.~\ref{prop:generalconstructionnotop}, hence 
%constitute an infinite trace situation.
%
%To show that  $F$ and $\lift$ constitute an infinite trace situation, 
To this end,
we rely on
Proposition~\ref{prop:generalconstructionnotop}
(but not Proposition~\ref{prop:generalconstructiontop}, since $\lift X$ does
not have the greatest element).	

\begin{thm}\label{thm:satassumptionexception}
The combination of polynomial $F$ and $T=\lift$ constitute an infinitary
 trace situation.
\end{thm}

%\newtheorem*{proof:satassumptionexception}{Proof of Prop.\ \ref{prop:satassumptionexception}}{\itshape}{\rmfamily}
%\begin{myproof}%\begin{proof:satassumptionexception}
\proof
We show that A polynomial functor $F$ 
% functor $F_{\Sigma}$ 
on $\Sets$ and a lift monad $\lift$ satisfy Assumptions~(\ref{asm:existenceWeaklyFinalCoalgFinalSeq})--%, 
%(\ref{asm:existenceWeaklyFinalCoalgDistLaw}), (\ref{asm:existenceWeaklyFinalCoalgGFPcont}), 
% (\ref{asm:existenceWeaklyFinalCoalgTop}), and 
(\ref{asm:existenceWeaklyFinalCoalgWeak2Lim}) in 
 Proposition~\ref{prop:generalconstructionnotop}
with respect to the order in Definition~\ref{def:orderenrichment}. 

It is easy to see that $F$ and $\lift$ on $\Sets$ satisfy the 
Assumptions~(\ref{asm:existenceWeaklyFinalCoalgFinalSeq}) and
%\ref{asm:existenceWeaklyFinalCoalgGFPcont}, 
(\ref{asm:existenceWeaklyFinalCoalgTop}).

It is known that Assumption~(\ref{asm:existenceWeaklyFinalCoalgDistLaw}) is satisfied~\cite[Lemma~2.4]{hasuo07generictrace}.

To prove that Assumption~(\ref{asm:existenceWeaklyFinalCoalgGFPcont}) is satisfied, it suffices to show that for all $x\in X$, 
$\bigsqcap_{i\in\omega}(g_i\odot b)(x)=\bot$ if and only if $(\bigsqcap_{i\in\omega}g_i)\odot b(x)=\bot$.
If $b(x)=\bot$, then we have $\bigsqcap_{i\in\omega}(g_i\odot b)(x)=(\bigsqcap_{i\in\omega}g_i)\odot b(x)=\bot$.
If $b(x)\neq\bot$, we have:
\begin{equation*}
\bigsqcap_{i\in\omega}(g_i\odot b)(x)=\bot
\Leftrightarrow \exists i\in\omega.\, g_i(b(x))=\bot 
\Leftrightarrow \bigsqcap_{i\in\omega}g_i(b(x))=\bot
\Leftrightarrow (\bigsqcap_{i\in\omega}g_i\odot b)(x)=\bot\,.
\end{equation*}
Hence Assumption~(\ref{asm:existenceWeaklyFinalCoalgGFPcont}) is satisfied in both cases.

As a connected limit and a coproduct commute in $\Sets$~\cite{adamekBLR02classificationaccessible}, 
the Kleisli inclusion functor $J:\Sets\to\Kl(T)$ preserves $\omega^{\op}$-limit. % lifts to a limit in $\Kl(\lift)$.
It is easy to see that this limit is a 2-limit. 
Therefore Assumption~(\ref{asm:existenceWeaklyFinalCoalgWeak2Lim}) is satisfied.
%
%Therefore by Proposition~\ref{prop:generalconstructionnotop}, $F$ and $\lift$ constitute an infinitary trace situation.
\qed
%\end{myproof}%\end{proof:satassumptionexception}

\subsection{Kleisli Simulation for Systems with Exception}
%It is not so difficult to see that $\FSigma$ and $\lift$ satisfy assumptions of Lem.~\ref{lem:soundnessFwd}.
It is known that a polynomial $F$ and $\lift$ satisfy the assumptions of Lemma~\ref{lem:soundnessFwd}~\cite{hasuo06genericforward}.
Hence we can use  forward Kleisli simulation to check infinitary trace inclusion between tree automata with exception. 

For an $(\lift,\FSigma)$-system, as we have seen in Theorem~\ref{thm:satassumptionexception}, 
the largest homomorphism can be constructed using Proposition~\ref{prop:generalconstructionnotop}.
Therefore from Lemma~\ref{lem:soundnessRBwdnotop}, we can use total backward simulation (Definition~\ref{def:restrictedbwdsimnotop}) to check infinitary trace inclusion between $(\lift,\FSigma)$-systems.
For $(\lift,\FSigma)$-systems
the totality means the following.
%the sufficient condition for a backward simulation $b$ to satisfy 
%the assumption in Definition~\ref{def:restrictedbwdsimnotop} can be given as follows.

\begin{prop}\label{prop:whenrestrictedlift}
In Definition~\ref{def:restrictedbwdsimnotop}, if $T=\lift$ and 
%$F=\FSigma$, then
$F$ is a polynomial functor, then
%\begin{enumerate}
%\item 
Assumption~(\ref{item:restrictedbwdsim2notop}) is satisfied if $b(x)\neq\bot$ for all $x\in X$. 
%\label{item:whenrestrictedlift2}
%\item Assumption \ref{item:restrictedbwdsim3} is always satisfied. \label{item:whenrestrictedlift3}
%\item Assumption \ref{item:restrictedbwdsim4} is always satisfied. \label{item:whenrestrictedlift4}
%\end{enumerate}
\end{prop}

%\begin{myproof}
\proof
%\noindent{\bf \ref{item:whenrestrictedlift2}.}
Let $*$ be the unique element of the final object $1$.
By the assumption we have $b(x)\neq \bot$.
Therefore we have $* = !_Y(b(x))= J!_Y\odot b(x)$ for all $x\in X$.
%It suffices to show that $J!_Y\odot b(x)=*$ for all $x\in X$.
This concludes the proof.
%
%\noindent{\bf \ref{item:whenrestrictedlift3}.}
%This holds because $\FSigma$ and $\lift$ satisfy Assumption~\ref{asm:existenceWeaklyFinalCoalgGFPcont} in Prop.~\ref{prop:generalconstructionnotop}.
%
%\noindent{\bf \ref{item:whenrestrictedlift4}.}
%This holds because for $T=\lift$ and $F=\FSigma$, $(Z,(J\pi_i:Z\kto \overline{F}^i 1))$ is a (strong) 2-limit (see the proof of Prop.~\ref{prop:satassumptionexception}).
\qed
%\end{myproof}

\subsection{Forward Partial Execution for Systems with Exception}
From Theorem~\ref{thm:soundnessFPEfwd}, soundness and adequacy of FPE for forward simulation hold 
for $(\lift,F)$-systems.

By the construction of the largest homomorphism, soundness and adequacy of FPE for backward simulation hold if 
the simulating automaton satisfies the assumptions in Theorem~\ref{thm:soundnessFPEbwdnotop}(\ref{item:adequacyfpebwdnotop}).
It is easy to see that 
%For a $(\lift,F)$-system, 
the assumptions can be described as follows.

\begin{prop}\label{prop:whenrestrictedfpelift}
If $T=\lift$ and 
$F$ is a polynomial functor,
%$F=\FSigma$,
%\begin{enumerate}
%\item 
the assumption in Theorem~\ref{thm:soundnessFPEbwdnotop}(\ref{item:adequacyfpebwdnotop})
%Assumption \ref{item:restrictedfpe2notop}
 is satisfied if $d(y)\neq\bot$ for each $y\in Y$.
%\item Assumption \ref{item:restrictedfpe3} is always satisfied.
%\item Assumption \ref{item:restrictedfpe4} is always satisfied.
\qed
%\end{enumerate}
\end{prop}

%\subsection{Coalgebraic Infinite Trace Semantics and Automata-theoretic Semantics of Tree Automata with Exception}
\subsection{Coincidence between Automata-theoretic and Coalgebraic Infinitary Trace Semantics}
\label{subsubsec:automcharaexception}
The same results as in Sections~\ref{subsec:automcharapow} and \ref{subsec:automcharagiry}
can be shown also for $T=\lift$.

%From an automata-theoretic perspective,
Automata-theoretically,
an $(\lift,\FSigma)$-system $\mathcal{X}=(X,s,c)$ can be regarded almost as a deterministic automaton
that outputs infinite trees,
except that at each stage of its behavior, 
the transition function $c:X\to\lift\FSigma X$ can output $\bot$ and abort.
In that aborting case, the output of $\mathcal{X}$ is undefined;
this automata-theoretic characterization of $\mathcal{X}$ naturally induces
a function $\langinf(\mathcal{X},\place):X\to\{\bot\}+\Treeinf(\Sigma)$.
It is straightforward to see that this function coincides with the largest homomorphism $\trinf(c):X\to\lift\Treeinf(\Sigma)$
from $c$ to $J\zeta$, which defines coalgebraic infinitary trace semantics. 
%, and
%

\section{Related Work}
In this paper, for a coalgebraic modeling of infinitary traces, 
we followed~\cite{jacobs04tracesemantics} where a \emph{Kleisli} category is used.
In~\cite{jacobs12tracesemantics}, an approach towards a coalgebraic characterization of \emph{finite} traces via 
an \emph{Eilenberg-Moore} category is introduced. 
We used ``Kleisli approach'' because of its
%An advantage of the approach using Kleisli category is a
potential for an extension to B\"uchi or parity automata.
The extension seems difficult for Eilenberg-Moore approach because
the approach is based on (generalized) \emph{determinization} of systems,
and it is well-known that determinization of B\"uchi automata strictly decreases
its expressive power. 

The construction of the largest homomorphism given in
Proposition~\ref{prop:generalconstructionnotop} is based on the one
in~\cite{cirstea10genericinfinite}.
The latter imposes some technical conditions  on a monad $T$, including
 a ``totality'' condition that excludes $T=\pow$ from its instances 
 (while the
 nonempty powerset monad is an instance). Our assumption of lifting to 
a 2-limit (Assumption~(\ref{asm:existenceWeaklyFinalCoalgWeak2Lim}) in
Proposition~\ref{prop:generalconstructionnotop}) is inspired by a condition
in~\cite{cirstea10genericinfinite}, namely that the limit $Z$ is lifted
to a \emph{weak} limit in $\Kl(T)$. It is not the case that
Proposition~\ref{prop:generalconstructionnotop} subsumes the construction 
in~\cite{cirstea10genericinfinite}: the former does not apply to the
nonempty powerset monad (but our Proposition~\ref{prop:generalconstructiontop}
does apply to it).

In~\cite{kerstan13coalgebraictrace}, an explicit description of a
(proper, not weakly) final $\overline{F}$-coalgebra is given 
for
$F\in\bigl\{\,\Sigma\times(\place),\,1+\Sigma\times(\place)\,\bigr\}$
and 
$T\in\{\giry,\Geq\}$. Here $\Geq$ is the \emph{Giry monad} and restricts
$\giry$
to proper, not sub-, distributions.
%four combinations of a 
% two functors $F$ and two monads $T$ on 
% $\Meas$---$F=\Sigma\times(\place)$ and $F=1+\Sigma\times(\place)$, and
% $T=\giry$ (sub-Giry monad) and $T=\Geq$ (strict-Giry monad). 
We do not use their  (proper finality) results for modeling of
infinitary traces, because: 1) if $T=\giry$ then the final coalgebras do
not coincide with the set of infinitary words; and 2) if $T=\Geq$
then language inclusion is reduced to the equality. We are skeptical about the value
of
developing simulation-based methods for the latter degenerate case, one
reason being that 
trace equivalence is often much easier than trace inclusion.
%trace inclusion is often a more difficult problem than trace equivalence.
For example, finite trace inclusion for probabilistic systems is undecidable~\cite{blondel03undecidableproblems} while trace equivalence is decidable~\cite{kiefer11languageequivalence}.

In~\cite{schubert09terminalcoalgebras}, it is shown that:
a limit of an $\omega^{\text{op}}$-sequence consisting of standard Borel
spaces and surjective measurable functions is preserved by
a polynomial functor $F$ 
(where  constants are restricted to  standard Borel spaces), and also by $\giry$.
It is also shown there that such a polynomial functor $F$
preserves standard Borel spaces, and so does $\giry$.
These facts imply the existence of a final $\giry F$-coalgebra in
$\Meas$ for every polynomial functor $F$. Note however that this final
$\giry F$-coalgebra captures (probabilistic) bisimilarity, not trace semantics.

\section{Conclusions and Future Work}\label{sec:conclusion}
We have shown that the technique
 forward and backward Kleisli simulations~\cite{hasuo06genericforward}
 and that of FPE~\cite{urabeH14genericforward,urabeH17matSim}---techniques originally 
developed for witnessing \emph{finite} trace inclusion---are also
 applicable to \emph{infinitary} trace semantics.
%  that are introduced in~\cite{hasuo06genericforward} 
% to witness finite trace inclusion and FPE that was introduced in~\cite{urabeH14genericforward} to aid the simulation can be also used %to witness 
% for
% infinite trace setting.
 We followed~\cite{jacobs04tracesemantics} (and
 also~\cite{cirstea10genericinfinite,kerstan13coalgebraictrace}) to 
characterize infinitary trace semantics in coalgebraic terms, on which we
 established properties of Kleisli simulations such as soundness.
 We developed our theory for three classes of instances: nondeterministic
 systems, probabilistic ones and ones with exception.
 These three turn out to result from two categorical principles that are rather different (see Remark~\ref{rem:differencePandG}).

There are some directions for  future work.
In~\cite{urabeH14genericforward} (and its extended version~\cite{urabeH17matSim}), in addition to FPE, a transformation called \emph{backward partial execution} (BPE) is introduced.
Similarly to FPE, BPE can also aid forward and backward Kleisli simulation for \emph{finite}
 traces in the sense that it satisfy soundness and adequacy.
However, BPE is only defined for word automata (with $T$-branching) and
not generally for $(T,F)$-systems.
Defining BPE categorically and proving its soundness and adequacy with
respect to infinitary traces, possibly restricting to word automata, is one of the future work.

In this paper, we used Kleisli simulation to compare simple automata where
an infinite-depth tree is accepted if it only has an infinite path on the automata.
%However 
More complex automata for infinite-length words have been introduced, 
such as
%for example
%are introduced, e.g.\
B\"uchi automata and parity automata.
Extending the notion of Kleisli simulation so that such  automata with complex accepted conditions
can be compared
is one of the directions of future work.

Another direction is implementation and experiments.
As forward and backward Kleisli simulations in this paper are defined in almost the same way as~\cite{urabeH14genericforward,urabeH17matSim},
we can use the implementation already developed
there.

\subsubsection*{Acknowledgments}
%The authors  are supported by Grants-in-Aid
%No.\ 24680001 \& 15K11984, JSPS.
%
The authors are supported by JST ERATO
HASUO Metamathematics for Systems Design Project (No.\ JPMJER1603), %\pagebreak
and JSPS KAKENHI Grant Numbers 15KT0012, 24680001 \& 15K11984.
Natsuki Urabe is supported by JSPS KAKENHI Grant Number 16J08157.

%
% ---- Bibliography ----
%
%\section*{References}
 \bibliographystyle{alpha} %necessary (bug?)
\bibliography{myref} %

\end{document}